\documentclass[useAMS,usenatbib]{mn2e}

 \usepackage{times}

\usepackage{graphicx}
\usepackage{supertabular}
\usepackage{subfigure}
\usepackage{rotating}
\usepackage{lscape}
\usepackage{epsfig}
\usepackage{amssymb}
\usepackage{myaasmacros}
\usepackage{amsmath}
\usepackage{multicol}
\usepackage{slashbox}

\def\squig{$\sim\!\!$}

\def\arcsec{\hbox{$^{\prime\prime}$}}

\def\deg{\hbox{$^\circ$}}

\def\14{\rm 1.4\,GHz}
\def\27{\rm 2.7\,GHz}
\def\whz1{$\,\rm W\,Hz^{-1}$}
\def\kms1{$\,\rm km\,s^{-1}$}

\def\gsim{\mathrel{\lower0.6ex\hbox{$\buildrel {\textstyle >}
\over {\scriptstyle \sim}$}}}

\def\lsim{\mathrel{\lower0.6ex\hbox{$\buildrel {\textstyle <}
\over {\scriptstyle \sim}$}}}

\title[RLF evolution modelling]{
The luminosity--dependent high--redshift turnover in the steep spectrum
radio luminosity function: clear evidence for downsizing in the
radio-AGN population} 
\author[E. E. Rigby et al.]{E. E. Rigby$^{1,2}$\thanks{E-mail: emmaerigby@gmail.com},
P. N. Best$^{2}$, M. H. Brookes$^{2}$, J. A. Peacock$^{2}$,
J. S. Dunlop$^{2}$, \newauthor H.~J.~A. R\"{o}ttgering$^{3}$,
J. V. Wall$^{4}$, L. Ker$^{2}$ \\ 
$^{1}$School of Physics \& Astronomy, University of Nottingham, University Park, Nottingham, NG7 2RD \\
$^{2}$SUPA\thanks{Scottish Universities Physics Alliance}, Institute for Astronomy, University
  of Edinburgh, Royal Observatory, Edinburgh
  EH9 3HJ, UK \\
$^{3}$Leiden Observatory, P.O. Box 9513, 2300 RA, Leiden, The Netherlands \\
$^{4}$Department of Physics and Astronomy, University of British Columbia, 6224 Agricultural Rd, Vancouver, BC, V6T 1Z1, Canada 
}

\begin{document}


\pagerange{\pageref{firstpage}--\pageref{lastpage}} \pubyear{2002}

\maketitle

\label{firstpage}

\begin{abstract}
This paper presents a new grid--based method for investigating
the evolution of the steep--spectrum radio luminosity function, with the aim of
quantifying the high--redshift cut--off suggested by previous work. To
achieve this, the Combined EIS--NVSS Survey of Radio Sources (CENSORS)
has been developed; this is a 1.4 GHz radio survey, containing 135
sources complete to a flux density of 7.2 mJy, selected from the NRAO VLA
Sky Survey (NVSS) over 6 deg$^2$ of the ESO Imaging Survey (EIS) Patch D. The sample is currently 73\% spectroscopically
complete, with the remaining redshifts estimated via the $K$--$z$ or
$I$--$z$ magnitude--redshift relation.
CENSORS is  combined with additional radio data from the Parkes All--Sky, Parkes Selected Regions, Hercules and VLA COSMOS samples
to provide comprehensive coverage of the radio power vs. redshift
plane. The redshift distributions of these samples, together with
radio source count determinations, and measurements of the local luminosity function, provide the input to the fitting process.  

The modelling reveals clear declines, at $> 3\sigma$ significance, in comoving density at $z > 0.7$ for lower
luminosity sources ($\log P = 25-26$); these turnovers are still present at $\log P > 27$, but move to $z \gsim 3$, suggesting a
luminosity--dependent evolution of the redshift turnover, similar to
the `cosmic downsizing' seen for other AGN populations.   
These results are shown to be robust to the estimated redshift errors and to increases in the spectral index for the highest redshift sources. 

Analytic fits to the best--fitting steep spectrum grid are
provided so that the results presented here can be easily accessed by
the reader, as well as allowing plausible extrapolations outside of the
regions covered by the input datasets.  

\end{abstract}


\section{Introduction}

It has become increasingly apparent in recent years that radio--loud
active galactic nuclei (AGN) play a key role in galaxy evolution; the interplay of their
expanding radio jets and the surrounding intergalactic and
intracluster medium acts to provide part, or possibly all, of the heat required to prevent
both large--scale cluster cooling flows and the continued growth of
massive ellipticals \citep[e.g.][]{fabian06, best06, best07, croton06,
  bower06}. 
Determining the evolution of the radio luminosity function (RLF) is
therefore important for understanding the timescales on which they
impose these effects. Also, since radio--loud AGN are powered by the most massive
black holes, their RLF can be used to investigate the behaviour of the
upper end of the black--hole mass function and hence the build--up of these objects in the early Universe. 

The work of \citet{sandage, osmer, peacock85, schmidt}, and in
particular \citet[][hereafter DP90]{DP90}, has shown that the comoving
number density of both flat and steep--spectrum powerful radio
galaxies, selected at 2.7 GHz, is greater by two to three orders of magnitude at a
redshift of two compared with the present day Universe. This density increase
is expected to peak at some point simply because sufficient time is
needed for their host galaxies to grow into the massive
ellipticals, with correspondingly large central black holes, that are
typically observed for radio--loud AGN \citep[e.g.][]{best98}. This
high--redshift `cut off'  was seen by \citet{peacock85} in the
flat--spectrum population and was also detected in the steep--spectrum
population by DP90 beyond $z \sim 2.5$; but
their results were limited by the accuracy of the photometric
redshifts from their faintest radio--selected sample, so they were not
able to quantify the decline.  It should also be noted that the DP90
work assumed an Einstein--de Sitter cosmology which means that the
high redshift sources in their sample were ascribed lower luminosity than for a $\Omega_{\rm \Lambda} > 0$ cosmology, thus
potentially making a cut--off easier to find.  

Following DP90, \citet{shaver} reported evidence of a sharp cut--off
in space density in their sample of flat--spectrum radio sources. However, this result
was disputed by \citet{jarvis00} who showed that it could be caused by an
increasing curvature of the spectral indices with redshift, and that a
shallower decline was more consistent with the data. A more rigorous
analysis of radio--loud quasars by \citet{wall05}, using a larger
sample, confirmed a decrease in the number density of flat--spectrum sources at $z \gsim 3$.  

A shallow space density decline between $z \simeq 2.5$ and $z \simeq 4.5$ was also
found in low--frequency--selected steep--spectrum sources by
\citet{jarvis01b}, although a constant density
value was also consistent with their data; their sample lacked the
depth needed for firm results. \citet{waddington01} used a deeper
survey and saw evidence that the turnover for lower--luminosity sources appeared
to occur at lower redshift than that of brighter flux--limited samples. They
were also able to discount some of the DP90 models, but their study lacked the
volume needed for better measurements of the space density changes of
powerful radio sources. Indications of a similar luminosity dependence
of the cut--off redshift were also seen for radio--loud
\citet{fanaroff} class I (FRI) sources by \citet{me2}.

The evolution of radio--selected AGN can be linked to the
behaviour of AGN selected in other bands. For example the space
density of optically--selected quasi--stellar objects (QSOs) shows a
strong decrease at $z > 2.1$  \citep{boyle00, fan01, wolfe03, fan04}, consistent with that
found for both radio--loud \citep{wall05}, as well as X--ray
selected quasars \citep{hasinger05, silverman05}.  There is also
evidence for a luminosity--dependent redshift cut--off in the optical
and X--ray selected QSO samples  \citep{ueda03, hasinger05,
  richards05, wall08}. Since radio--loud QSOs are thought to correspond to flat spectrum sources
\citep[e.g.][]{barthel, antonucci}, investigating the evolution of the
flat and steep populations can result in a new understanding of the
links between radio--loud (quasars and radio galaxies) and
radio--quiet (QSOs) sources. A full review of radio--loud AGN
  evolution can be found in \citet{zotti09}.   

To properly investigate the evolution of the steep spectrum  RLF it is
clear that a combination of several radio surveys of differing depth
is needed to ensure a broad coverage of the radio luminosity--redshift ({\it P--z}) plane. In
particular, faint radio samples are needed if the $z \gsim 2$ behaviour is to be
determined. This has motivated the development of the 150--object, 1.4
GHz selected, Combined EIS--NVSS Survey of Radio Sources
\citep[CENSORS;][hereafter Paper I]{CENSORS1}, which has been designed
to maximise the information for high--redshift, steep--spectrum, radio sources close to
the break in the RLF.  

In this paper the CENSORS dataset, combined with additional samples,
is used to investigate the nature of the high--redshift evolution of radio sources, via a new grid--based modelling technique in which no prior
assumptions are made about the behaviour of the luminosity
function. This is an improvement on previous investigations which have
either used functional forms, or only considered pure luminosity or
density evolution, or a combination of both (although \citealt{dye10} have recently developed a similar method to study the evolution
of sub--mm galaxies). 

The layout of the paper is as follows: Section \ref{data} describes
both CENSORS and the additional data sets needed; Section \ref{model} presents the modelling technique; 
Section \ref{results} describes the results from the best--fitting model and investigates their robustness; 
finally Section \ref{conc} summarises the findings. Throughout this
paper values for the cosmological parameters of H$_{0} =
70$~km~s$^{-1}$Mpc$^{-1}$, $\Omega_{\rm m} = 0.3$ and $\Omega_{\rm
  \Lambda} = 0.7$ are used and the spectral index, $\alpha$ is defined
as $S_{\nu} \propto \nu^{-\alpha}$.  

\section{Input data}
\label{data}

\begin{figure}
\centering
\includegraphics[scale=0.35, angle=90]{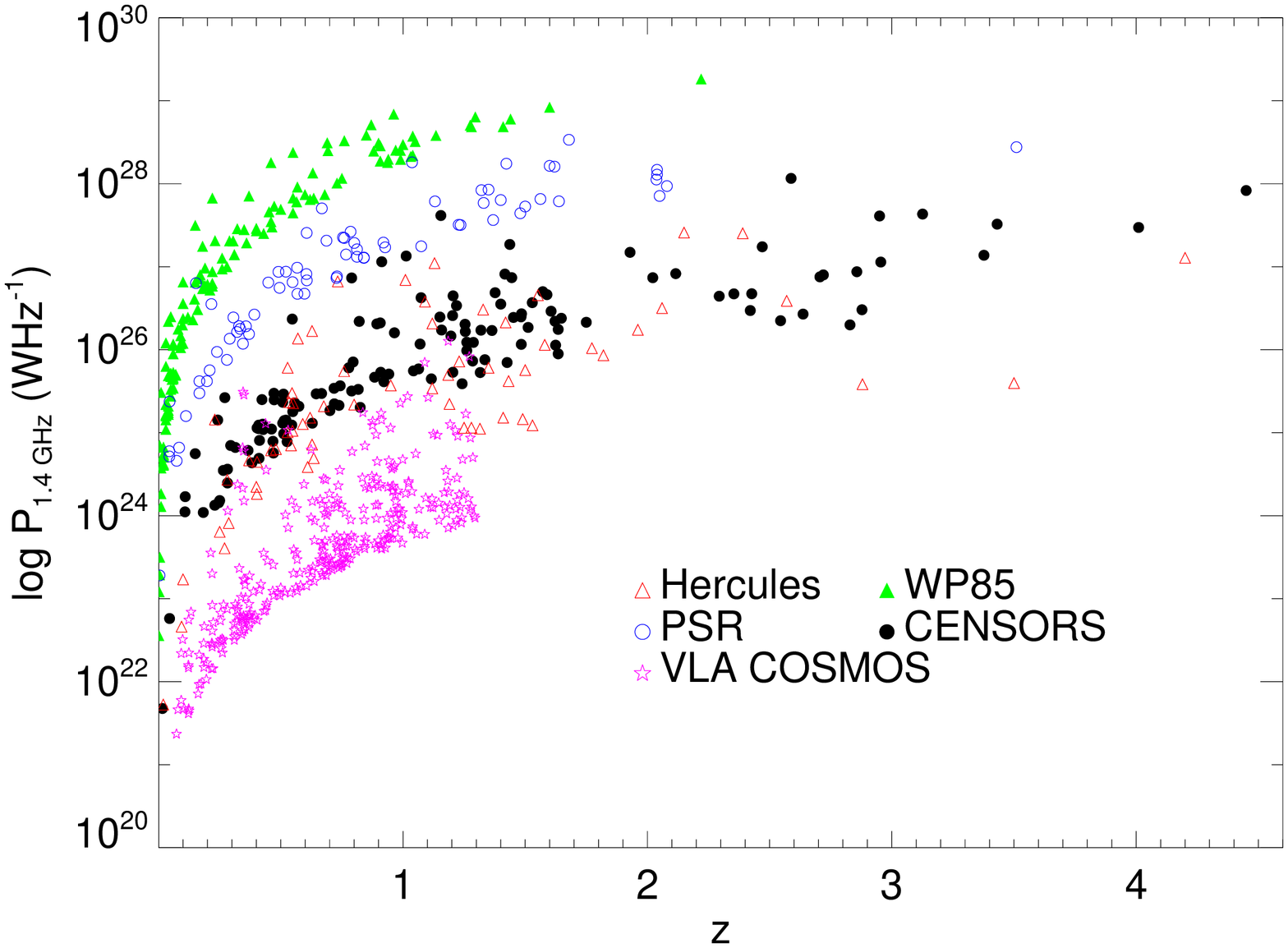}
\caption{The \protect\citet[][WP85] {WP85}, Parkes Selected Regions
(PSR, \protect\citeauthor{Downes86} 1986, \protect\citeauthor{Dunlop89}
  1989), CENSORS, Hercules \protect\citep{waddington01} and VLA
  COSMOS \citep[for $z\leq1.3$ only;][]{smolcic} samples plotted on a radio--luminosity vs. redshift plane to illustrate how they 
efficiently cover a large part of the plane without much
overlap. Radio luminosities were calculated using previously published
spectral indices for the WP85, PSR and Hercules samples; $\alpha =
0.8$ was assumed for the sources in VLA--COSMOS. The spectral indices
for the CENSORS sample are taken from \protect\citet{louise}. The PSR,
WP85 and COSMOS samples are restricted to steep spectrum sources
only. See text for full details of sample selection.} 
\protect\label{pz_plot}
\end{figure}

As discussed above, several data sets are needed to constrain the
radio source cosmic evolution. In addition to the CENSORS sample, therefore, four other radio
samples, along with determinations of the local radio luminosity function, and measurements of the radio source counts are 
used; these are described in this section. Fig. \ref{pz_plot} illustrates the coverage of the {\it P--z} plane obtained using these radio samples.

\begin{table*}
\centering
\caption{\protect\label{sample_table} Key information relating to the
  five radio samples used in the modelling. $N({\rm total})$ gives the
  total number of sources in each sample as used here; $N(z_{\rm spec})$,
  $N(z_{\rm photo})$, $N(z_{\rm est})$ and $N(z_{\rm limit})$ give the number of
  sources with spectroscopic, photometric, estimated (via either the
  $K$--$z$ or $I$--$z$ relations) or lower limit $K$--$z$ redshifts respectively. }
\begin{tabular}{l|c|c|c|c|c|c|c}
\hline
Name & 1.4 GHz Limit (Jy) & Sky Area (sr) & $N({\rm total})$ & $N(z_{\rm spec})$ & $N(z_{\rm photo})$ & $N(z_{\rm est})$ & $N(z_{\rm limit})$\\
\hline
WP85                & 4.0       & 9.81    & 83 & 79 & -- & 4 & --\\
PSR                    & 0.30     & 0.075 & 74   & 41 & -- & 30 & 3 \\
CENSORS         & 0.0072  & 0.0018        & 135 & 99 & -- & 31 & 5 \\
Hercules          & 0.002    & 0.00038        & 64    & 42 & 19 & 1 & 2 \\
VLA-COSMOS  & 0.0001   & 0.00036       & 314  & -- & 314 & -- \\
\hline
\end{tabular}
\end{table*}

\subsection{CENSORS}

The full CENSORS sample contains 150 sources with $S_{\rm 1.4 GHz} > 3.8$ mJy in a 3 by 2 deg field of the ESO Imaging Survey (EIS) Patch D, centred on 09 51 36.0,
$-$21 00 00 (J2000). \citet[Paper I]{CENSORS1} present the radio
data, along with the optical host galaxy identifications, with
additional $K$--band imaging  presented by \citet[Paper II]{CENSORS2}.
Spectroscopic data for a subset of the sample can be found in
\citet[Paper III]{CENSORS3}, which also gives estimated redshifts for
the remainder of the sample, calculated using the {\it K--z} and, for
one source, the {\it I--z} magnitude--redshift relations.  

Since the publication of Papers I--III some small reassessments have
been made to the sample. Subsequent radio data have shown that:
\begin{itemize}
\item  CENSORS 66 and CENSORS 82 are actually the lobes of a FRII
  radio source whose host galaxy is located at 09 50 48.97, $-$21 32
  55.8 (Gendre, priv. comm.), with a $K$--band  magnitude of 17.8
  $\pm$ 0.1 in a 4.5 arcsec diameter aperture ($K = 18.2 \pm 0.1$ when
  corrected to the standard 63.9 kpc diameter aperture; see Paper II
  for details) and $I = 21.7 \pm 0.1$. This  results in an estimated
  {\it K--z} redshift of 
1.40 (calculated using $\log z = 0.0025K^{2} + 0.113K - 2.74$; cf. Paper III); 
\item similarly, CENSORS 84 and CENSORS 85 are also the lobes of an
  extended double radio source, whose host galaxy is located at 09 55
  36.87, $-$21 27 12.5, with a $K$--band  aperture corrected magnitude
  of 13.1 $\pm$ 0.2 and a corresponding estimated $K$--$z$ redshift of
  0.15;  
\item  the host galaxy of CENSORS 64 is located at 09 49 01.60
$-$20 50 00.7 \citep{louise} with $K$ and $I$ band magnitudes of 15.04 $\pm$ 0.08 (when aperture corrected) and 17.44 $\pm$ 0.02
respectively with a corresponding estimated redshift of 0.33. This
compares well to its new spectroscopic redshift  of 0.403 \citep{louise}. 
\end{itemize}
Spectral indices for the sources, calculated using the new radio data,
will be presented in \citet{louise}, but are also included here in the
analysis of Fig. \ref{alpha_dist}.  

Additional $K$--band imaging was obtained for 14 sources using IRIS2
(in service mode) on the Anglo--Australian Telescope. As a result of
this, the host galaxy of CENSORS 69 has now been detected, with a
$K$--band magnitude of 20.1 $\pm$ 0.3 (19.6 $\pm$ 0.3 when 
aperture corrected). New spectroscopic data subsequently showed it to
have a redshift of 4.01 \citep{louise}.  
Finally,  the $K$--$z$ limits presented in Paper II were
discovered to be in error and have been recalculated. The correct
values and corresponding new $K$--$z$ redshift limits, along with the new IRIS2 $K$--band data, can be found in Appendix
\ref{full_cen_table}, within the up--to--date CENSORS dataset.  

Of the original 150 CENSORS sources, 135 are deemed to be complete to
a flux density limit of 7.2 mJy, and it is this subset which is used
in this paper. No other selection is performed, meaning that it
contains both starforming galaxies along with steep-- and flat--spectrum sources. The current host galaxy identification fraction of this
is 96\% with a spectroscopic completeness of 73\%. Table \ref{sample_table} summarises the salient information
for this subsample.

\subsection{The Wall \& Peacock 2.7 GHz radio sample}
\label{wp85_section}

The sample of \citet[hereafter WP85]{WP85} includes the brightest
radio sources at \27 over an area of 9.81\,sr, and is complete to {\rm
  2\,Jy}. The original paper presents redshifts for 171 of the 233 source sample, and a further 20 were added by \citet{WP99}. 
Since then further spectroscopic observations have been published by a
variety of groups, raising the redshift completeness to 98\%. As part
of the back-up programme for CENSORS two of the WP85 sources, 0407-65
and 1308-22, have been observed with FORS2 on the VLT (see Appendix
\ref{morespecs_notcen}). The redshift measured for 0407-65 is in good
agreement with that observed by \citet{labiano}, but the observation of 1308-22 updates the redshift
estimate quoted in \cite{WP99_3}. For the remaining two sources without useful
spectra, the estimated redshifts of \citet{WP85} are used. 
There is an excess of sources in this sample at $z < 0.1$, some of
which may be contaminating starburst galaxies, so a minimum redshift
of 0.1 is imposed here; doing this does not degrade the analysis as
this region of parameter space will be well constrained by the local
radio luminosity function. 

Complete spectral indices are available for this sample making it
straightforward to convert the \27 flux densities to \14; since it is only the
behaviour of the steep--spectrum luminosity function which the model
will assess (as discussed in Section \ref{model}) only steep
spectrum WP85 sources are considered in this. 

In order to use this sample in the modelling of the RLF it needs to be
converted into a sample with an effective \14 flux density limit. To
do this, a flux limit of $S_{\rm 1.4 GHz} = 4$ Jy was adopted,
corresponding to a spectral index $\alpha = 1.06$ for sources at the
\27 limit. It is possible that sources with a steeper spectral index
than this may be below the WP85 sample flux limit at \27 and still have a
\14 flux density above 4 Jy, thus leading to incompleteness in the
sample. A search was therefore carried out for such sources, so they
could be added to the sample.  
In the Northern Hemisphere this is simple as there is a fainter
($S_{\rm 2.7GHz} \ge 1.5$ Jy) sample from \citet{pw81} covering the
WP85 area; this contains one source  with a flux lower than 2 Jy, but
a corresponding 1.4 GHz value above 4 Jy (and $z \geq 0.1$). In the Southern Hemisphere
there are two 408 MHz surveys, the \citet{WP99_14} equatorial sample
and \citet{burgess}, which between them cover the rest of the WP85
survey region. Since these are lower frequency surveys, any such steep
spectrum sources will be bright and can be identified. Searching these
samples reveals two $z \geq 0.1$ steep--spectrum sources brighter than 4 Jy at 1.4
GHz, but not in the WP85 sample. Oddly, calculating the \27 flux
densities of these sources reveals that they should have
been detected by WP85. It is not clear why they were
missed, but it is possible that they have very curved spectra. 
The data for these three `missing' sources are given in Table \ref{missing}, whilst Table \ref{sample_table}
summarises the WP85 sample as used in this paper (including the three sources discussed above); the full
dataset can be found in Appendix \ref{extra_data}.

\begin{table}
\centering
\caption{\protect\label{missing} The additional sources that were
  added into the WP85 sample to correct for incompleteness. The
  $z_{\rm ref}$ gives the reference for the redshift -- G05 for
  \citet{grimes} or BH06 for \citet{burgess}.} 
\begin{tabular}{c|c|c|c|c|c}
\hline
Name & $S_{\rm 2.7 GHz}$& $S_{1.4 \rm GHz}$ & $\alpha$ & $z$ & $z_{\rm ref}$\\
& (Jy) & (Jy) & & &\\
\hline
3C325 & 1.84 & 4.29 & 1.29 & 1.135 & G05\\
\hline
Name & $S_{\rm 408 MHz}$& $S_{1.4 \rm GHz}$ & $\alpha$ & $z$  & $z_{\rm ref}$\\
& (Jy) & (Jy) & &  &\\
\hline
1526--423 & 17.86 & 5.08 & 1.02 & 0.5 & BH06\\
1827--360 & 25.83 & 6.49 & 1.12 & 0.12 & BH06\\
\hline
\end{tabular}
\end{table}

\subsection{The Parkes Selected Regions 2.7 GHz radio sample}
\label{PSR_def}

The Parkes Selected Regions (PSR; \citeauthor{wall68} 1968, \citeauthor{Downes86} 1986,
\citeauthor{Dunlop89} 1989) cover {\rm 0.075\,sr} in six {\rm 6.5\deg}
square fields of view down to a flux density limit of {\rm
  0.1\,Jy} at \27. The updated sample of \cite{Dunlop89} presents
redshifts for 82 of the 178 sources in the sample and subsequent
observations published by a variety of groups result in a further 24
redshifts. Of the remaining 72 sources, 10 have estimated values from
\citet{DP93}, whilst the remainder were estimated using the {\it K--z}
relation as outlined for the CENSORS sources in Paper III. The
$K$--band photometry for the sample \citep{Dunlop89} uses 12.4\arcsec
diameter apertures for most objects. These are large enough such that
aperture corrections are sufficiently small to be ignored. Where a
$K$--band magnitude is not available one is estimated via the relation: $K = -1.1(B - R) + 18.3$ \citep{Dunlop89}. Two sources have been
omitted from this sample due to unclear identification and non--detection. 

The current PSR dataset suffers from some incompleteness below 0.15
Jy, becoming quite large by 0.1 Jy (DP90) so there is also a desire to minimise the contribution of these
faintest sources. The \14 flux density 
limit is therefore set at 0.3 Jy to achieve this. All \27 sources with a \14 flux
density greater than this are included in the final sample; the lowest
of these has a \27 value of 0.14 Jy which should be sufficiently above
the incompleteness limit to avoid problems. 
Table \ref{sample_table} summarises the PSR sample as it used in this paper, whilst the full dataset can be found in Appendix \ref{extra_data}. 

\subsection{The Leiden--Berkeley Deep Survey Hercules sample}

The Leiden--Berkeley Deep Survey covers 5.52 sq. deg. over nine high latitude fields, and was originally based upon multi--colour plates from
the 4 m Mayall telescope at Kitt Peak (\citeauthor{kron80} 1980;
\citeauthor{KooKron82} 1982) alongside Westerbork Synthesis Radio Telescope
1.4\,GHz radio observations \citep{Wind84}. One of these fields, in
the constellation of Hercules, has subsequently been followed up by
\bibpunct[; ]{(}{)}{,}{a}{}{,} \citet[][see also
  \citealt{Wind84b}]{Waddington00,waddington01}, who defined a
complete sample of 64 radio sources (both starforming galaxies and
steep and flat--spectrum objects) with $S_{\rm 1.4 GHz} > 2$ mJy within
1.2 sq. degrees. The spectroscopic redshift completeness of the sample is 66\% [41
redshifts measured by \citealt{waddington01} with one additional value from
\citealt{me1}]. Of the remaining 22 sources, 20 have photometric
redshifts also from \citet{waddington01} but the final two only have
estimated {\it K--z} lower limits due to host galaxy
non--detections. Table \ref{sample_table} summarises the salient
information for the Hercules sample, whilst the full dataset can be
found in Appendix \ref{extra_data}. 
\bibpunct[, ]{(}{)}{,}{a}{}{,} 

\subsection{The AGN subsample of the VLA COSMOS survey}

\citet{smolcic} defined a sample of 601 AGN with z $\leq$ 1.3 in the 2 deg$^2$ COSMOS
field using the multiwavelength imaging available for the region. The
redshift limit was imposed because beyond this the AGN/star--forming
galaxy separation becomes unreliable, leading to possible
contamination of the sample. 
The radio sensitivity varies with position across the field, but over a well--defined area of 1.17 deg$^2$ it is possible to define a clean sample complete to S$_{\rm 1.4 GHz}
\geq$ 100 $\mu$Jy; this contains 314 steep spectrum sources. 
Robust photometric redshifts were calculated for the survey and are
available for all of these sources. Table \ref{sample_table} summarises
the salient information for the VLA COSMOS sample; however, unlike the
other samples, this full dataset is not available in Appendix
\ref{extra_data} as it is still proprietary, and was obtained via
private communication with the authors.  

\subsection{Source counts data}

Any model of RLF evolution must match the measured radio source counts
as a function of flux density and so this comparison will also be used
in the modelling process. The data used for this were taken from
\citet{bondi}, \cite{Seymour04}, \cite{Wind84}, \cite{White97} and
\cite{KW87} to ensure that a sufficient range of flux densities (0.05 mJy -- 94 Jy) was covered.  The White et al. counts are limited to $S_{\rm 1.4 GHz} > 2$
mJy only, as they note that they are incomplete below this. The full
set of \14 source counts used in the modelling can be found in Appendix
\ref{extra_data} and the counts themselves are shown in Fig. \ref{counts_comp}, within the discussion of the modelling results (Section \ref{data_comp}).

\subsection{Local radio luminosity functions}

The local radio luminosity function (LRLF) for AGN was measured by
\cite{Sadler02} using the 2dF galaxy redshift survey (2dFGRS;
\citeauthor{2dFGRS} 1999, \citeauthor{Colless01} 2001), 
by Best et al. \citeyearpar{bestlrlf, BestLRLF2},  using the Sloan Digital Sky Survey
(SDSS; \citeauthor{York00} 2000), and by \citet{mauch} using the 6 degree
Field Galaxy Survey \citep[6dFGS DR2][]{jones04}. 
\cite{Sadler02} compiled their LRLF from a sample of all 2dFGRS, NVSS
selected radio galaxies with $z \leq 0.3$. 
Note that while the 2dFGRS LRLF includes points at $\log
P_{1.4\rm{\,GHz}} = 25.9 \text{~and~} 26.3$  (as converted to
the cosmology used here) these are not included due to the small number (1) of sources in each band. 
\cite{BestLRLF2} use the seventh data release of the SDSS, combined with
the NVSS and FIRST 1.4\,GHz radio surveys (using a similar process to
that carried out by \citet{bestlrlf} for the second data release, but
now with an independent normalisation), to produce a sample of 9168
radio sources, with a median redshift of \squig~~0.1.  Similarly the 
\citet{mauch} LRLF was calculated using the 7824 NVSS radio sources
contained in the second incremental data release of the 6dFGS; the median redshift for their sample is 0.043. 
In addition, 95\% confidence upper limits are added at the highest radio luminosities
($\log P_{1.4{\rm \,GHz}} > 27.0$) as no sources were detected at
these luminosities in the three datasets. The LRLF data used in the modelling can be found in Appendix \ref{extra_data}.

\section{Modelling technique}
\label{model}

Unlike previous attempts to model the evolution of the RLF
\citep[e.g.][etc.]{DP90, jarvis01b, willott01}, no assumptions are made about the shape of the luminosity
functions here.  
Instead, they are determined by allowing the densities,
$\rho$, at various points on a $P$--$z$ grid of radio luminosities and
redshifts to each be free parameters and then simply finding the 
best--fitting values to this many--dimensional problem. The $P$--$z$
grid points were chosen so as to allow sensitive 
calculations to be made without having so many parameters involved
that finding a best fit becomes a prohibitively long task (but with
more powerful computers and larger datasets the technique could be
expanded to sample much finer detail). The range of radio luminosities
covered was $\log P_{1.4{\rm\,GHz}} = 19.25 \rm{~to~} 29.25$, equally spread in steps of
0.5 in $\log P_{1.4{\rm\,GHz}}$. Densities were evaluated at the
redshifts 0.1, 0.25, 0.5, 1.0, 2.0, 3.0, 4.0, 6.0. The $\rho$ at any
$(P,z)$ can then be interpolated from the nearest four grid points,
except at $z < 0.1$ where the densities are assumed to be constant (see
Section \ref{input_grids} below). 

\begin{figure}
\centering
\includegraphics[scale=0.4, angle=90, trim=0mm 0mm 0mm 25mm]{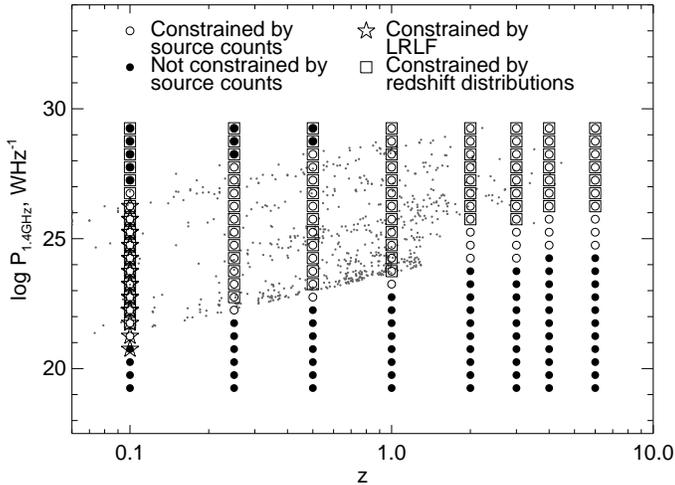}
\caption{\protect\label{cons_fig} The various regions of the $P$--$z$ plane that are constrained by
the redshift distributions, the source counts and the local radio luminosity
function (LRLF). The redshift distribution constraint is determined by
the COSMOS sample at $z \leq 1.3$ and the Hercules sample above this, as they are the deepest surveys; the positions of the
individual sources in the five samples are shown as grey dots. The source
counts cover the $S$--$z$ grid used in the modelling process, which takes a lower limit of 0.1\,mJy and  an upper limit of 50\,Jy.}
\end{figure}

The $P$--$z$ plane is constrained by the samples described in Section
\ref{data}. This is demonstrated in Fig. \ref{cons_fig}, which
highlights the different regions which are covered by different data types.
The densities that are actually included in the fitting process are
those that are constrained by the faintest redshift
distribution or the source counts; grid points that are unconstrained are excluded. 
As Fig. \ref{cons_fig} shows, high--redshift, low--luminosity sources
do not satisfy this criterion and are therefore not included in the
modelling. The total number of points fitted, and hence the dimensions
of the minimisation process, is 101.  

The \verb1amoeba1 algorithm for downhill simplex minimisation
\citep{downhillsimplex} is used to obtain the best fitting space
densities. It takes as input a set of parameters and a scaling factor and uses
these to construct a geometrical object of $N+1$ points in $N$
dimensions called a simplex. It uses a user--defined function to
calculate the likelihood of each vertex of the simplex, and then reflects, contracts or expands
these vertices until the function is minimised. In order to achieve this best fit in reasonable time,  the algorithm
was run in a multi-stage loop with varying scaling and tolerance for
the first five steps. 
The maximum number of iterations allowed within \verb1amoeba1 for each stage is
set to 3000 and the process ends when the likelihood ratio of successive stages is less than 1.0000001.

The subsections below describe the calculation of the input modelling parameters, along with an explanation of the likelihood calculation used.  

\subsection{Construction of the input density grids}
\label{input_grids}

Three $P$--$z$ grids are used as inputs to the modelling program
containing the source densities for steep--spectrum, flat--spectrum and
star--forming populations separately (the total radio source space
density obviously being the sum of these three at any grid
point). Considering these three grids separately is essential since the five 
different radio source samples used in the modelling include and exclude different populations, as
well as on the good physical grounds that the different populations might well evolve
differently. As the aim of this work is to investigate 
the evolution of the radio galaxy RLF, the star--forming grid is
fixed. Its inclusion is necessary as the low flux densities of the
CENSORS and Hercules samples mean that radio emission from star formation becomes
significant. The flat--spectrum radio--source grid is also held constant since 
the low numbers of this type of source in the input samples do not
allow minimisation; more accurate constraints come from previous
surveys explicitly targetting these sources.  

The star--forming grid is created by evolving the local star--forming galaxy luminosity function of \citet{Sadler02} such that
\begin{equation}
\Phi\left( P,z\right) = 
\begin{cases}
\Phi\left( \frac{P}{\left(1 + z\right)^{2.5}}, 0\right)& z \leqq z_{\rm max},\\
\Phi\left( \frac{P}{\left(1 + z_{\rm max}\right)^{2.5}}, 0\right)& z > z_{\rm max}.
\end{cases}
\end{equation}
where $\Phi$ is the comoving density of radio sources due to star
formation, $P$ and $z$ are the luminosity and redshift respectively,
and $z_{\rm max} $ is the redshift where the space density plateaus. The power law index of 2.5 is taken from
\citet{smolcic09a} who studied star forming galaxies in VLA--COSMOS
out to $z \sim 3$; their value agreed with that previously found by
\citet{Seymour04}. The value $z_{\rm max} = 2$ is adopted, following 
\citet{Blain99}, but its precise value is irrelevant to this work as
the contribution of star forming galaxies at $z > 1$ is negligible at
the flux densities studied.  

Similarly, the starting steep--spectrum grid (which the minimisation then varies) is 
formed by evolving the \citet{Sadler02} local AGN RLF by
$(1+z)^3$ in density. Conversely, the flat--spectrum grid is created by taking
the median value from the results of the seven evolutionary models
presented in DP90, after conversion to the cosmology used here. This
is in broad agreement with the recent results of \citet{wall05}, and
minor variations are not critical given that this population is small
compared to steep--spectrum sources.  

An $S$--$z$ (flux density--redshift) grid is, in many cases, more readily compared with
real data than the $P$--$z$ grid, as it can be converted easily into
source numbers. The modelling code therefore uses the three $P$--$z$
grids to populate three corresponding $S$--$z$ grids containing 120
flux density bins covering the range 0.1\,mJy--50\,Jy at 1.4\,GHz, equally spaced in $\log$~S, and 300
redshift bins covering the range 0--6, equally spaced in $z$. 
This $z$--range is wider than that previously used to
define the $P$--$z$ grid in order to cover the full range of the
radio samples. To account for this, two extra bins -- $z = 0.001$ and
$z=0.05$ -- are inserted into the $P$--$z$ grids. For the steep
spectrum grid, the densities for these additional grid points are
assumed to be constant with redshift, and are therefore set to the
$z=0.1$ values; for the flat--spectrum and star forming grids they are
calculated from the input models. 
For a given $P$--$z$ grid the density at a given point on
the $S$--$z$ grid $\rho(S,z)$, is found by linear interpolation of the density
values from the four surrounding points in the expanded $P$--$z$ grid.  
The total number of sources per steradian in each bin is then given by:   
\begin{equation}
N(S,z) = \rho(S,z) \frac{dV}{dz} d(\log S) dz .
\end{equation}
The star forming and flat--spectrum $S$--$z$ grids are only calculated
once as they are not changed in the minimisation process. The steep
spectrum $S$--$z$ grid is recalculated in each cycle of the
\verb1amoeba1 process as $\rho(S,z)$ changes when $\rho(P,z)$ changes.  
Although the source counts extend to lower flux densities, it is not useful to extend the modelled region because source counts alone
do not provide a sufficient constraint on the radio luminosity functions.

\subsubsection{Spectral index selection}

\begin{figure}
\centering
\hspace{-15mm}
\includegraphics[scale=0.4, angle=90]{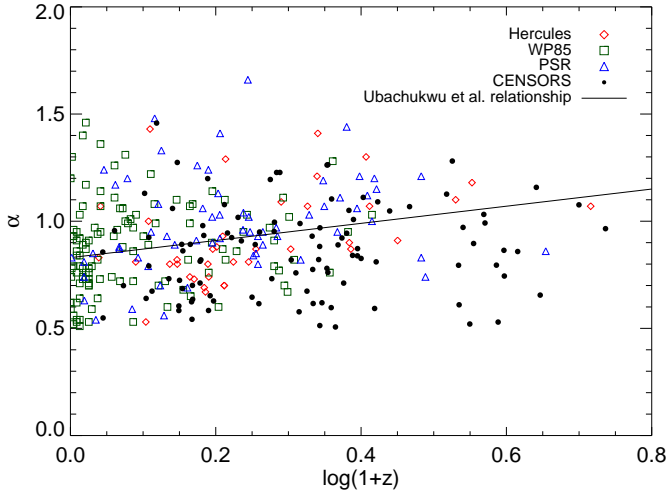} 
\caption{\protect\label{alpha_dist} The changes in spectral index with
  redshift for the steep--spectrum sources in the four radio samples
  with measured $\alpha$ values. Also shown is the relation from
  \citet{ubachukwu} which follows the datapoints reasonably well. The
  spectral indices were calculated using 1.4 GHz to 0.6 GHz, 1.4 GHz
  to 5 GHz, 2.7 GHz to 5 GHz and 1.4 GHz to 325 MHz flux densities for the Hercules, PSR, WP85 and CENSORS samples respectively. } 
\end{figure} 

The calculation of the appropriate luminosity for a given flux density
and redshift bin in the $S$--$z$ grid requires a value for the
spectral index, $\alpha$. For the flat--spectrum and star--forming grids, 
$\alpha$ is assumed to be 0.0 and 0.8  respectively; the small
contribution of these populations means that a more precise value is
not necessary.  
However, the steep--spectrum spectral index may vary with redshift and luminosity, which could significantly alter the results. This $P$--$z$--$\alpha$ degeneracy means that either parameter could be used to give the $\alpha$ variation. Unlike DP90, who adopted a $P$--$\alpha$ relation, it is the redshift dependence which is used here for the steep--spectrum grid; \citet{louise} will investigate these spectral index variations in more detail. 
The default form for this variation is taken from \citet{ubachukwu} who found that the mean spectral index increases with redshift as $\alpha = 0.83 +
0.4\log(1+z)$; as Fig. \ref{alpha_dist} shows, this relation gives a
reasonable approximation to the available steep--spectrum sample
data, with a possible over--estimation of $\alpha$ at high redshift. The CENSORS spectral indices in particular would also be consistent with a  flat relation, but this may be 
because they are low frequency, making them appear flatter than the high frequency WP85 and PSR values. Section \ref{alpha_change} considers the effects other $\alpha$ assumptions, including a constant value, have on the modelling, though it should be noted that reasonable variations in $\alpha$ do not lead to qualitative differences in the results. 

In addition to this, a Gaussian scatter in $\alpha$ of 0.2 is also 
incorporated at each redshift in the steep--spectrum grid to account
for any variations in the value within a redshift bin; this is a reasonable assumption given the spread seen in Fig. \ref{alpha_dist}. In practice
this is implemented by creating 21 versions of the $S$--$z$ grid,
extending to $\pm 2.5\sigma$ (in steps of $0.25\sigma$), which are
each assigned a weight depending on how far they are away from the mean. Redshift bins where $\alpha < 0.5$ are ignored (as sources within them would not be in the steep spectrum sample) and their weight evenly distributed over the remainder. The $P$--$z$ grid densities, $\rho(P,z)$, corresponding to each
$(S,z)$ point are then interpolated onto the bins in these 21 new
$S$--$z$ grids as before; the final grid carried forward into the minimisation
is their weighted sum.
 
\begin{table}
\centering
\caption{\protect\label{comp_table} The grids used for each dataset
  comparison in the modelling process. `steep', `stars' and `flat'
  indicate the grids for steep--spectrum, star--forming and flat--spectrum grids respectively. Note that it is only the steep--spectrum
  grid which is allowed to vary in the modelling process.} 
\begin{tabular}{l|c|c}
\hline
Dataset  & \multicolumn{2}{c|}{Comparison grid} \\
               & $P$--$z$     & $S$--$z$  \\
\hline
VLA--COSMOS &    & steep\\
Hercules &    & steep+stars+flat\\
CENSORS &    & steep+stars+flat\\
PSR &    & steep\\
WP85 &    & steep\\
LRLFs & steep+flat& \\
Source counts & & steep+stars+flat\\
\hline
\end{tabular}
\end{table}

\begin{figure*}
\centering
\includegraphics[scale=0.3, angle=90]{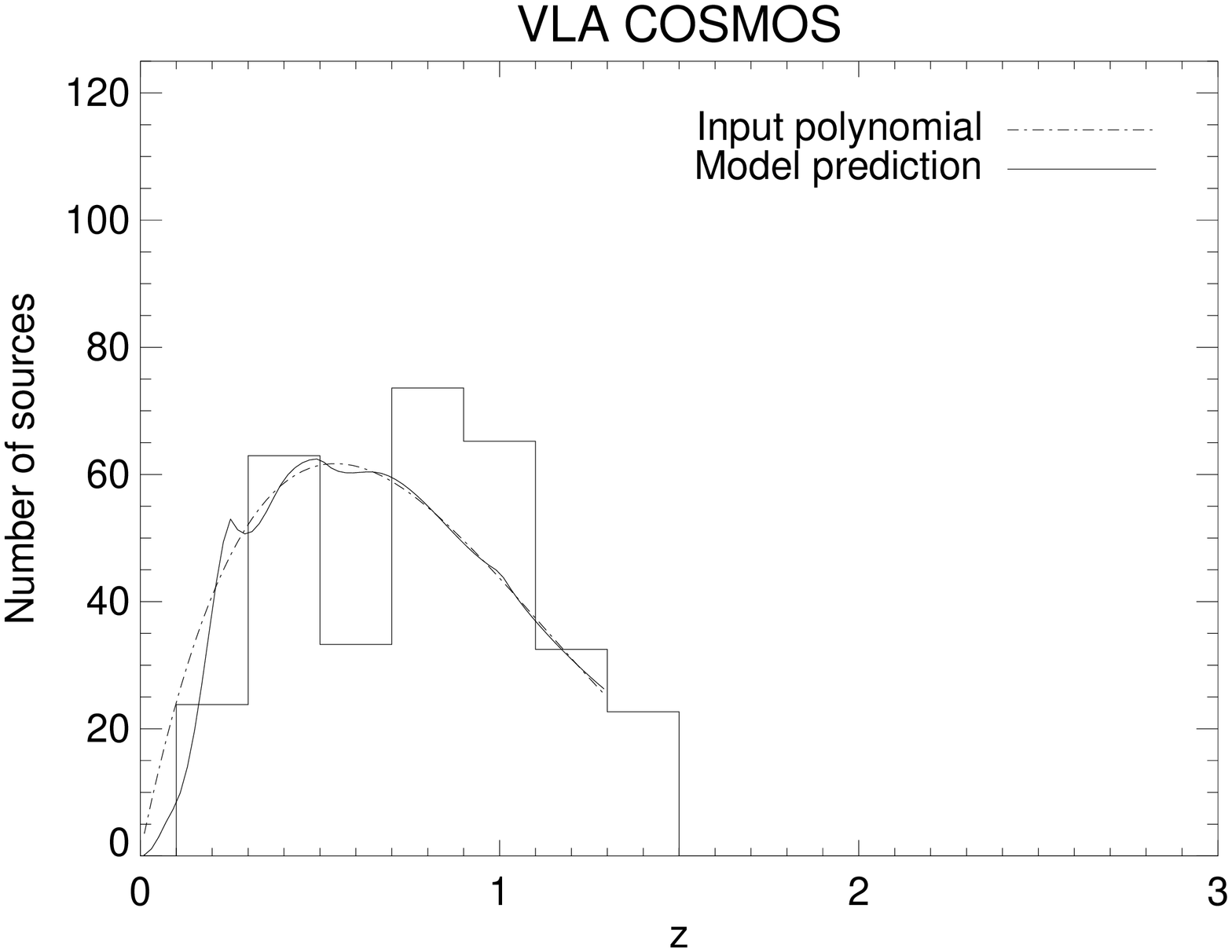}
\includegraphics[scale=0.3, angle=90]{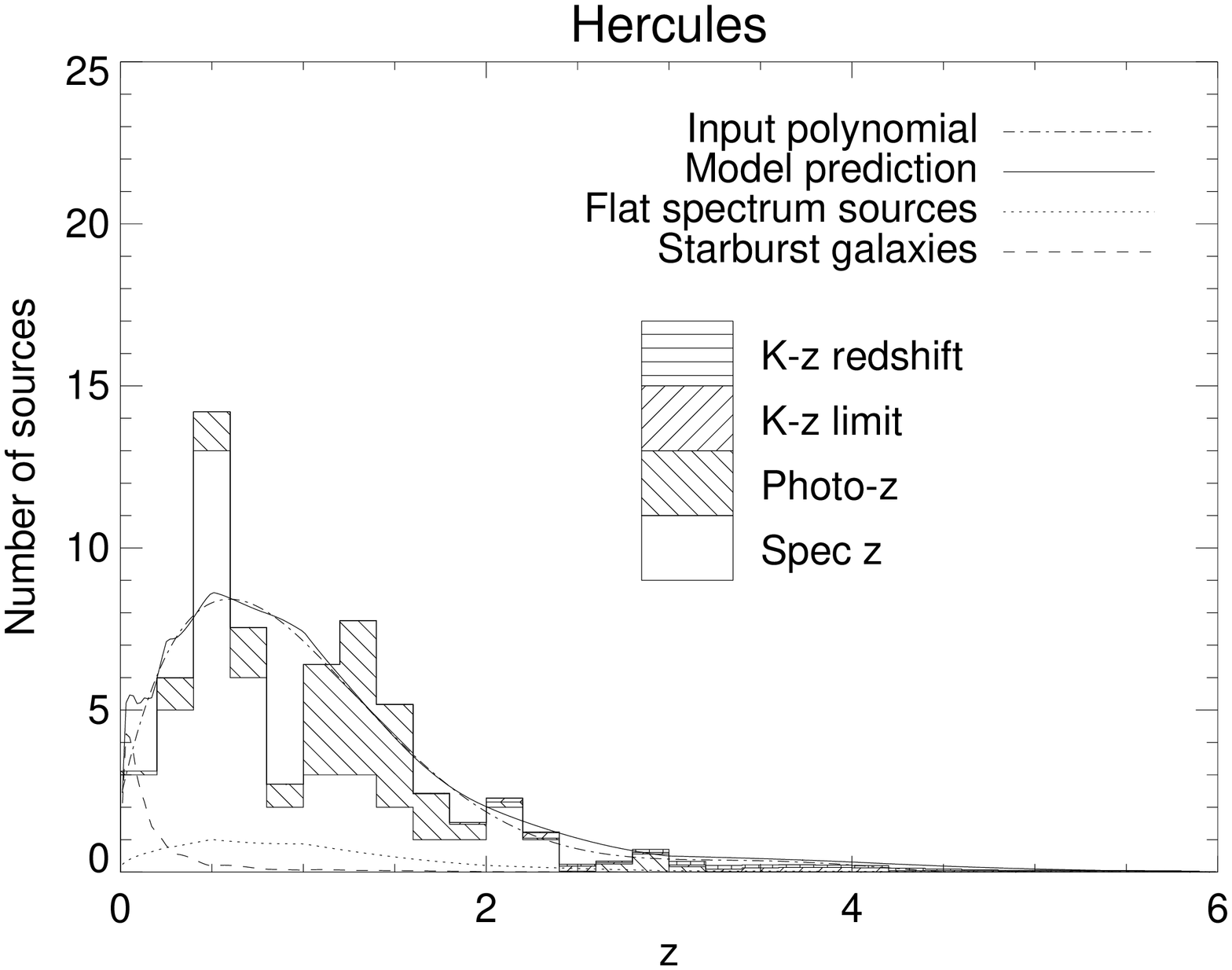}
\includegraphics[scale=0.3, angle=90]{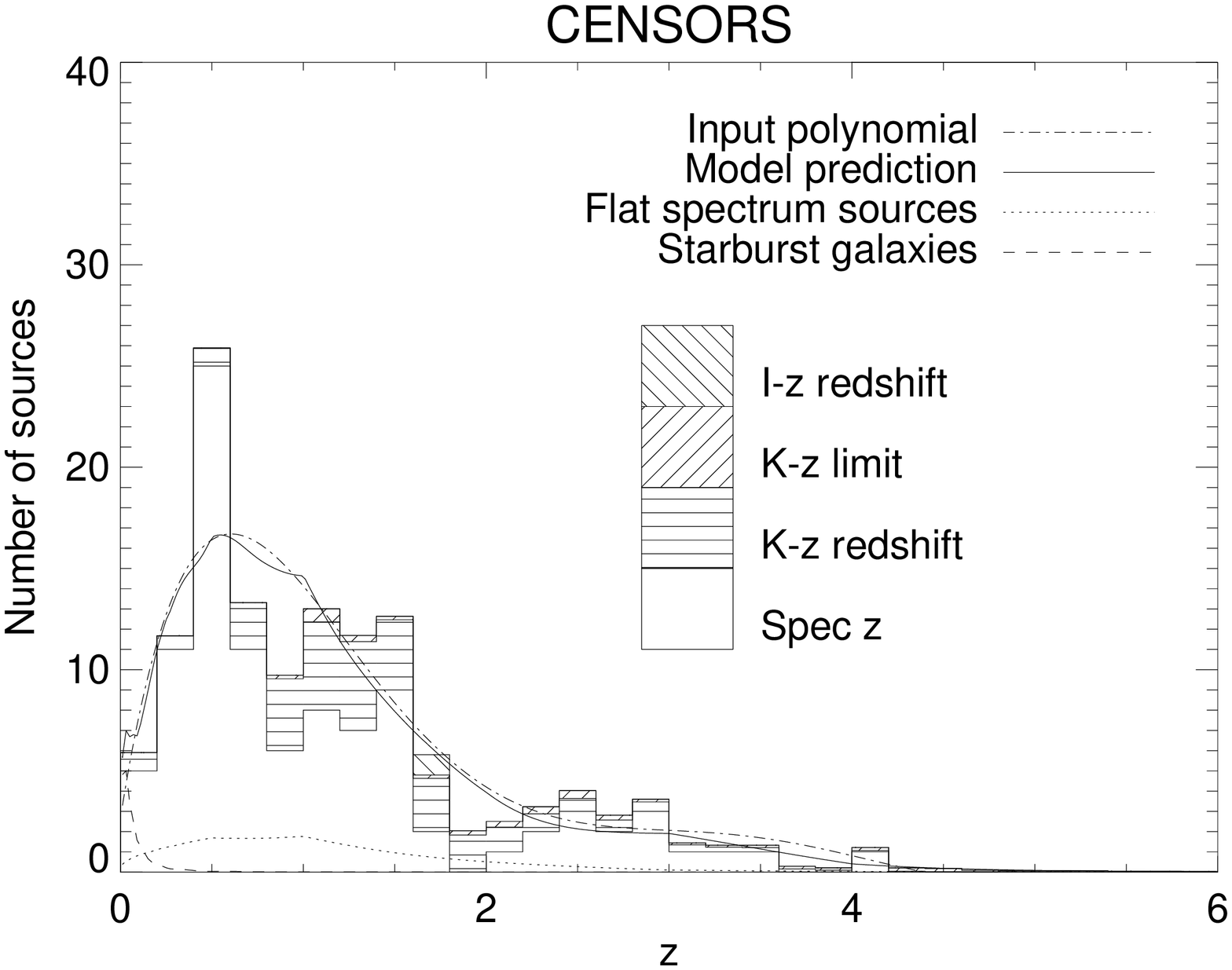}
\includegraphics[scale=0.3, angle=90]{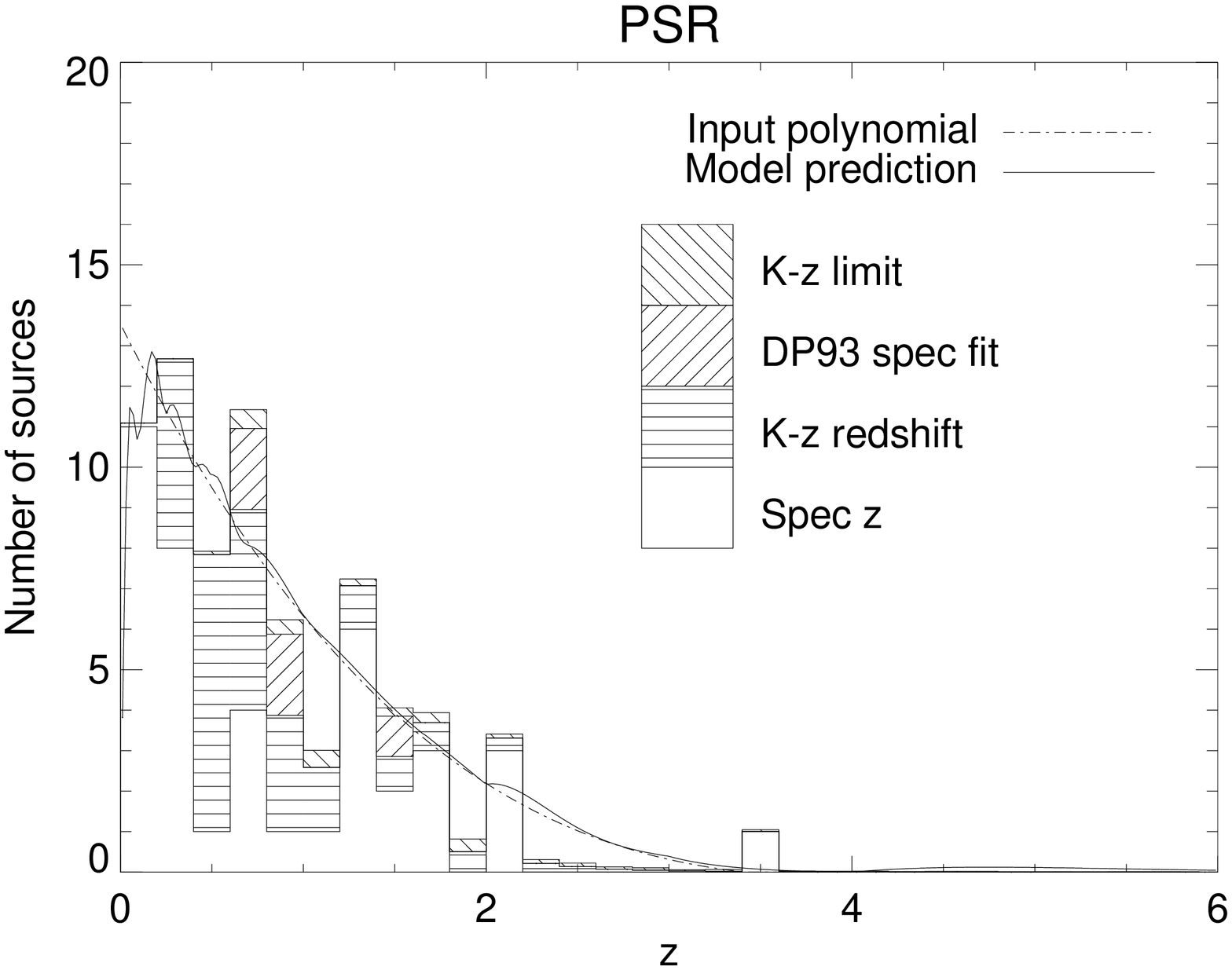}
\includegraphics[scale=0.3, angle=90]{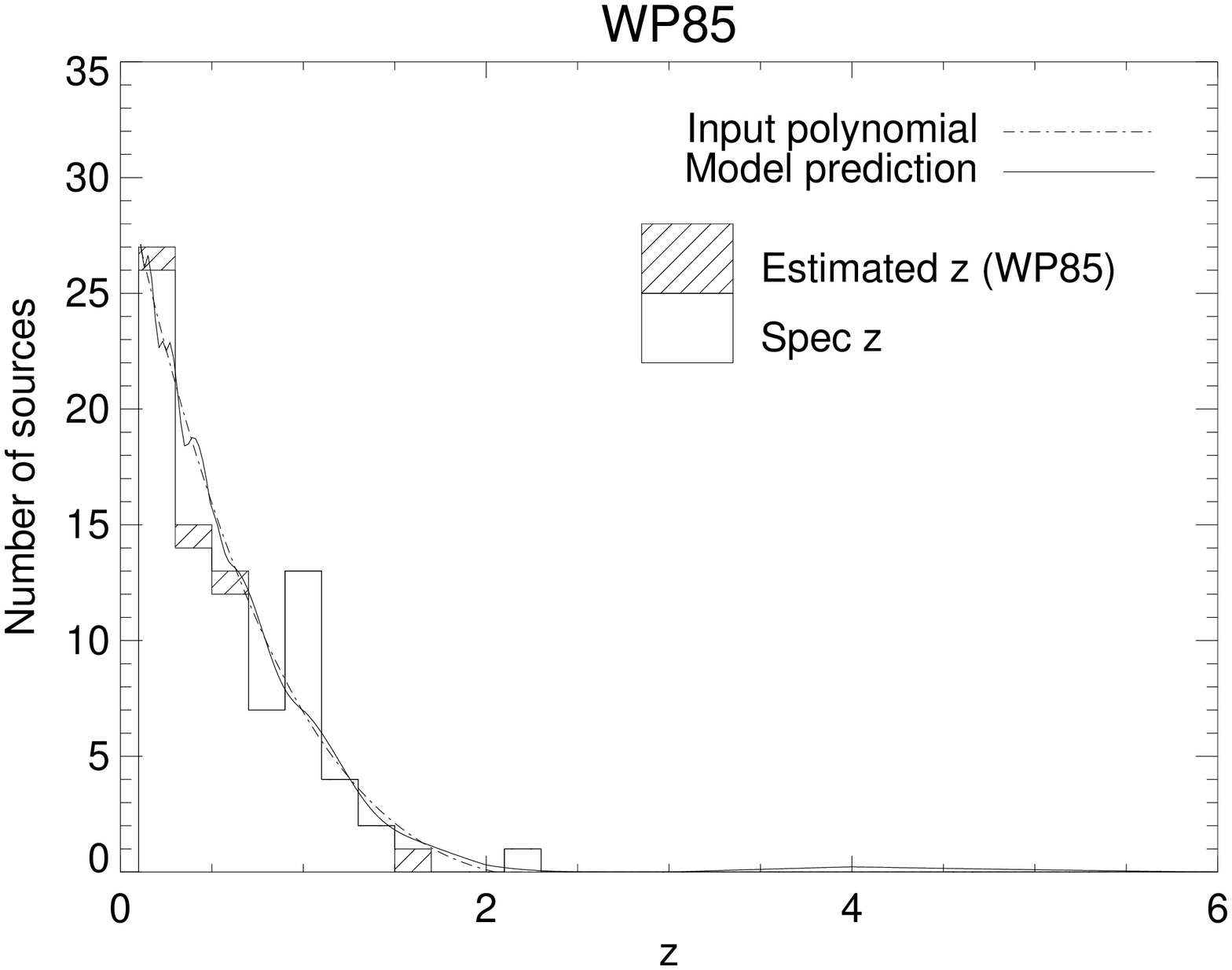}
\caption{\protect\label{z_dist_comp} The redshift distributions for
  the five samples, CENSORS, Hercules, WP85, PSR and VLA--COSMOS,
  included in the modelling. The histograms are plotted with  a binsize of 0.2 in
  $z$ and incorporate the uncertainties arising from the limits and
  estimated redshifts by representing them as weighted Gaussian distributions
  (see Section \protect\ref{comp_pred} for details). The overplotted
  solid lines show the model predictions for each sample, and the bin
  shading shows the contribution from the different redshift
  determinations. For samples which were compared to some combination
  of the three input grids, the contributions from the additional grids
  are also shown separately to illustrate the dominance of the steep
  spectrum population at high redshift. 
  See Table \ref{comp_table} for 
  information on which grids were used for each dataset.} 
\end{figure*}

\subsection{Comparing the model predictions to data}
\label{comp_pred}

The input data from the redshift distributions of the five different radio
source samples described in Section \ref{data}, along with the local RLF
and the observed source counts, now need to be compared to the model grids
described in the previous Section. To do this, the model local RLFs are
simply read from the appropriate ($z=0.1$) column in the $P$--$z$ grid,
and the model redshift distributions, $N(z,S_{\rm limit})$ for some sample
with survey area $A$ and flux limit $S_{\rm limit}$, and source counts,
$N(S)$, are easily calculated from the $S$--$z$ grid using
\begin{equation}
N(z,S_{\rm limit}) = A \times \sum_{S>S_{\rm limit}}{N(S,z)}
\end{equation}
and
\begin{equation}
N(S) = \sum_{z} N(S,z) 
\end{equation}
respectively.  Different datasets are compared to different combinations
of the model grids, depending on their content. For instance, the sum of
the steep and flat spectrum $P$--$z$ grids is fitted to the LRLF data as
they were created using only AGN, but the source counts are compared
against the sum of all three $S$--$z$ grids as they include starforming
galaxies. Table \ref{comp_table} summarises which of the grids are used
for comparison to each dataset.

A $\chi^{2}$--test, $\chi^{2} = \sum_{i}{\frac{(O_{i} - E_{i})^{2}}
   {\sigma^{2}}}$, is used to assess the model predictions for each dataset
in turn. For the source counts and the local radio luminosity function,
the data points and their uncertainties provide the values of $O$ and
$\sigma$. It is impractical, given the large number of data points
involved (and the wish to add more as they become available) to choose
model grid points that exactly match the locations (flux density or
luminosity) of the data, and therefore the values of $E$ are calculated by
interpolating between neighbouring model grid points. The $\chi^{2}$
comparison also includes a term for the upper limits on the high radio
luminosity points in the LRLF; this is done by setting the `data' points
for these luminosities equal to zero and setting the corresponding errors
equal to the upper limit of the LRLF. 

When comparing the model predictions with the redshift distributions, it
is important to note that there are insufficient sources at the
high--redshift end of the redshift distributions (a part of the parameter
space that is of particular interest) to formally allow a $\chi^{2}$--test
to be used. A possible solution to this would be to carry out a
source--by--source likelihood analysis instead, calculated using the
product of the model prediction for the redshift probability distribution
at the redshift of each of the sample sources; this returns the maximum
likelihood value if the data exactly match the model distribution. A
problem with this method, however, is the difficulty of accounting for the
uncertain redshifts present in several of the samples.

An alternative solution is therefore adopted here, by matching the shape
of the redshift distributions using polynomial fits, and comparing these
with the model predictions. The smoothing that this provides at high
redshifts mitigates the issue of small source numbers in the
high--redshift bins. It also helps to lessen anomalous features in the
redshift distributions, such as the drop in source numbers at $z \sim 1.5$
in the CENSORS, Hercules and PSR samples (Fig. \ref{z_dist_comp}).  The
latter arises because of the onset of the `redshift desert', where
spectroscopic redshifts are difficult to obtain: the $K$--$z$ estimates
should fill this gap, but in practice the scatter in the $K$--$z$
relation, combined with the lack of spectroscopic measurements, means
there is a overall bias for redshifts to lie outside of this region.

The polynomials are fitted to histograms with binsizes of 0.1 in $z$,
beginning at $z=0.0$, with the exception of WP85, which, as discussed
previously in Section \ref{wp85_section}, has a minimum redshift of
0.1. Additionally, the VLA COSMOS sample was only defined to $z = 1.3$
(due to lack of AGN/star--formation separation beyond that redshift) so no
fitting is allowed beyond this. Once the value of a polynomial fit reaches zero it is set to zero for all subsequent redshifts, to prevent it returning negative source numbers.\footnote{The polynomial fitting is not done in log space as this appeared to provide a poorer match to the total number of sources in each sample at high redshift.} 
The order of polynomial selected for each
redshift distribution is the one that gives the minimum reduced
chi--squared -- 6 for CENSORS, Hercules and COSMOS, 3 for PSR and 4 for
WP85 -- and these input fits are shown in Fig. \ref{z_dist_comp}.

\begin{figure}
\centering
\includegraphics[scale=0.35, angle=90, clip, trim=4mm 0mm 4mm 10mm]{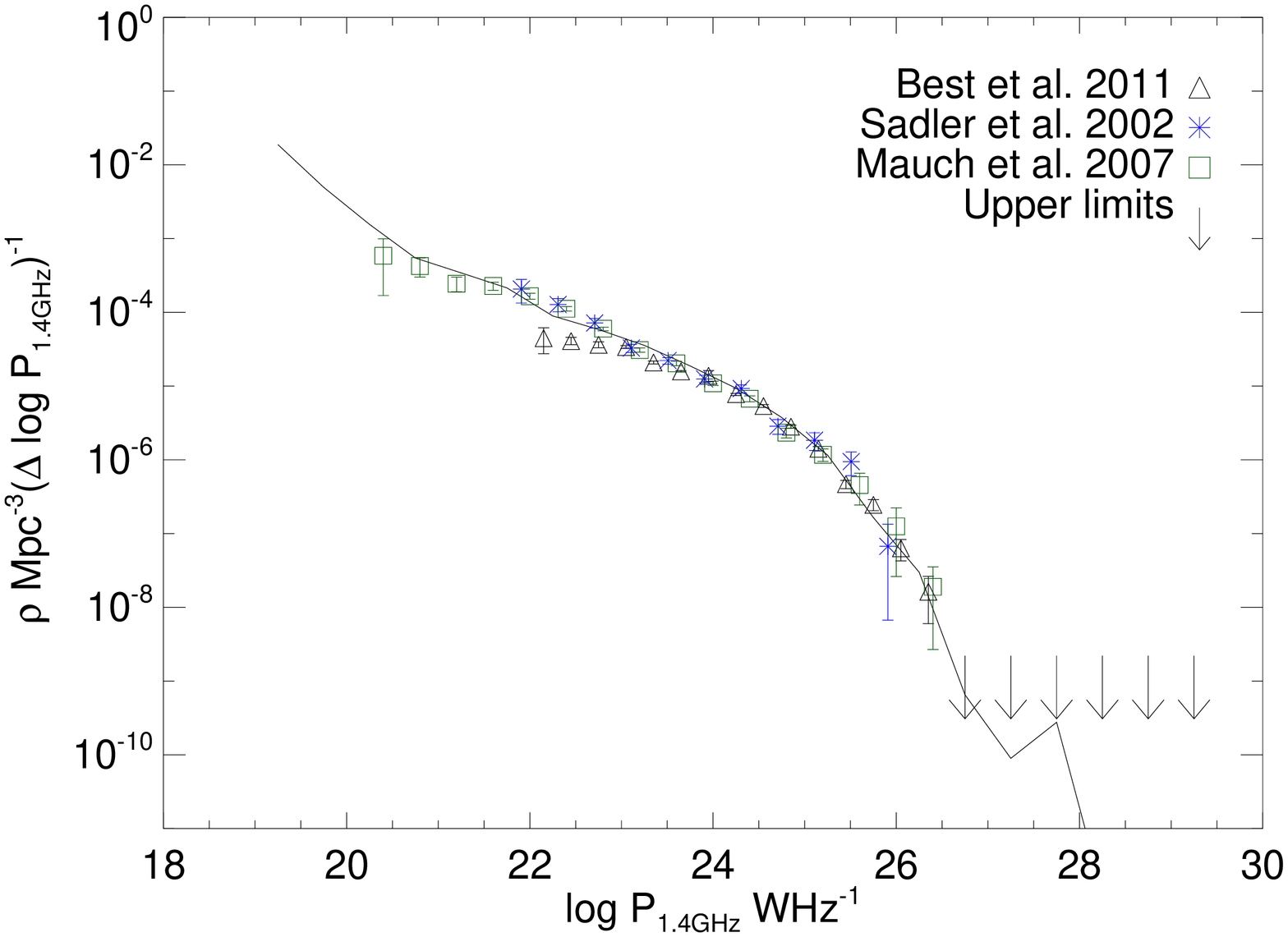}
\caption{\protect\label{lrlf_comp} The data for the three input LRLFs
  used in the modelling process (points) along with the best--fitting
  model LRLF prediction (solid line) from the $z=0.1$ column in the combined steep and flat
  spectrum final $P$--$z$ grids. } 

\centering
\includegraphics[scale=0.35, angle=90, clip, trim=4mm 0mm 4mm 0mm]{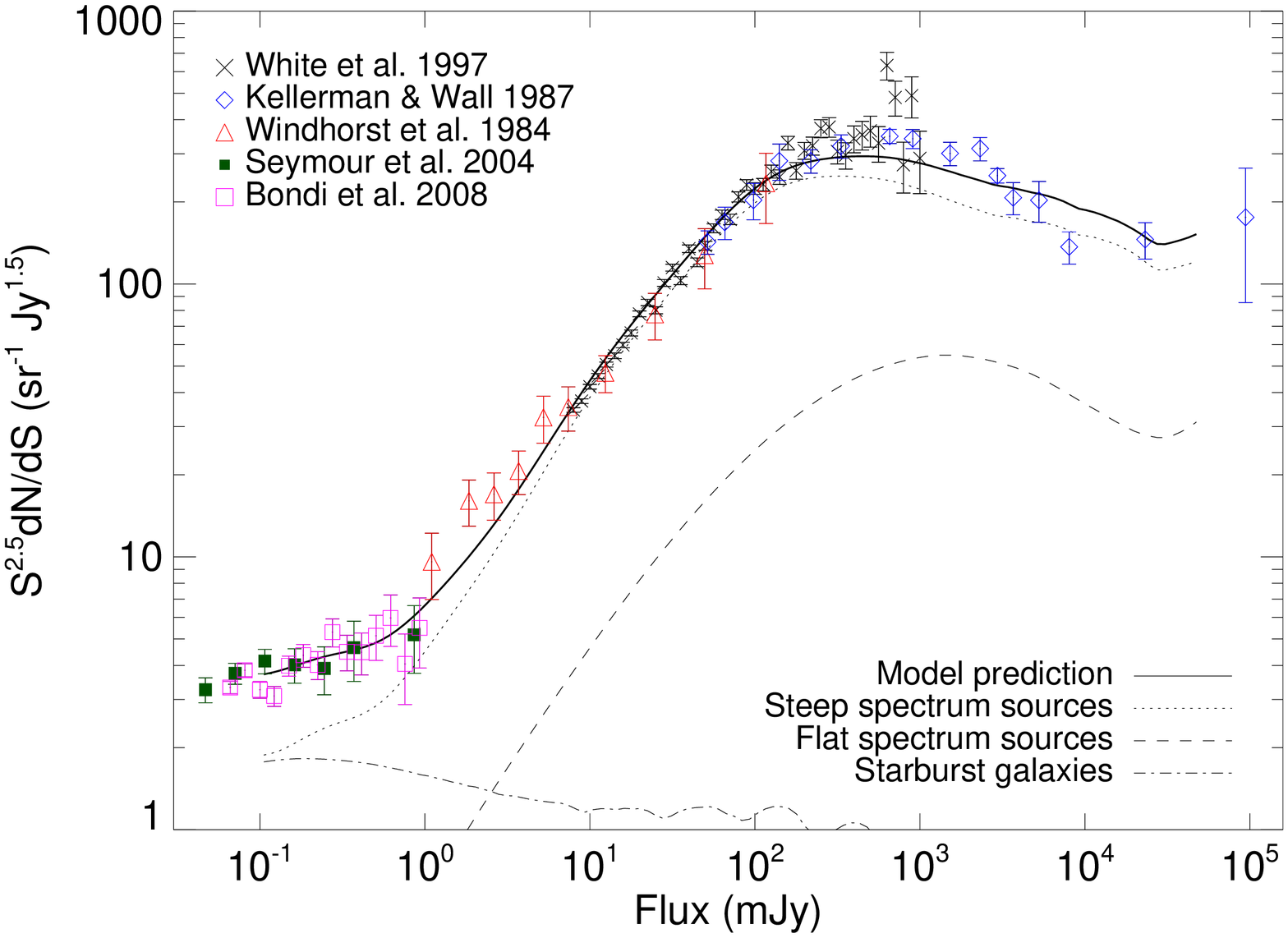}
\caption{\protect\label{counts_comp} The source count data used in the
  modelling process (points) along with the best--fitting prediction
  for the counts from the three final model $S$--$z$ grids (solid
  line). Also shown are the contributions to the model counts for the
  steep--spectrum (dotted line), flat spectrum (dashed line) and
  starforming galaxies (dot--dashed line) separately to illustrate the
  dominance of the steep population.} 

\centering
\includegraphics[scale=0.35, angle=90, clip, trim=4mm 0mm 4mm 10mm]{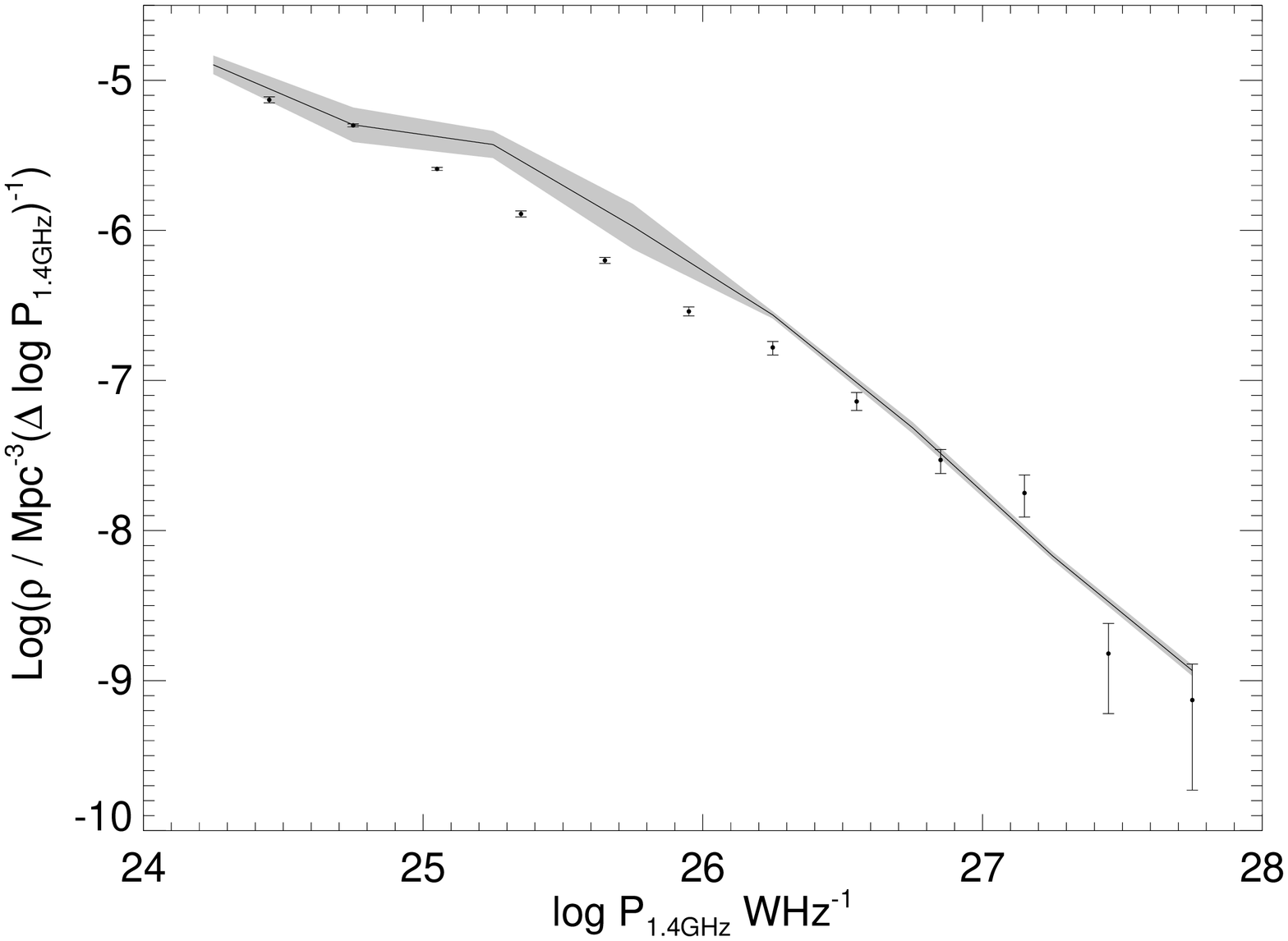}

\caption{\protect\label{rlf_comp} The RLF data points of
  \citet{donoso} compared to the best--fitting model prediction at $z =
  0.5$ from the steep--spectrum grid (solid line). The shaded region
  shows the $5\sigma$ confidence limits of the model.}
\end{figure}

The polynomial fitting takes into account the uncertainty from both
estimated and photometric redshifts present in these datasets, by
representing each source as a Gaussian distribution centred on the given
redshift, with the width equal to either the published photometric
redshift error, or, for the $K$--$z$ estimates, 0.14 in $\log z$ (the
1$\sigma$ spread in the 7C $K$--$z$ relation \citep{Willott03KZ}). The
$K$--$z$ limits can similarly be modelled, but are assigned a width (in
$\log z$) of 0.4 above and 0.1 below the given value to represent their
increased uncertainty (see Paper III for more discussion). This
distribution is then included in the redshift histograms, thus spreading
the source across different redshift bins, depending on the errors. The
effect of these redshift uncertainties on the modelling results are
investigated, in Section \ref{z_change}, by changing these assumptions.

The polynomial fits are then compared to the model predictions using a
$\chi^{2}$--test, with the polynomial values providing the $E$ parameters
in these equations, evaluated at redshifts and flux densities which match
points in the model grid, and the $O$ values given by the model prediction
at those grid points. The corresponding $\sigma$ values are taken as the
uncertainty in the polynomial value, generated from the covariance matrix
of the polynomial fit coefficients.

Using polynomial fits to represent the redshift distributions in this way
has one drawback: although the fitting method should return the correct
minimum point, and thus identify the correct best--fit model grid, the
absolute value of $\chi^{2}$ at that point will not necessarily be a true
measure of the goodness--of--fit of the model. This is because the
polynomial--derived errors are correlated. Any scaling error in $\chi^{2}$
will then lead to a mis-estimate of the uncertainties on the model
predictions (see Section 3.3). To account for this, the $\chi^{2}$ values
derived using the polynomial-fit method for the best-fit model prediction
were compared with `true' $\chi^{2}$ values. The latter were derived by
comparing the best-fit model with the observed source numbers in different
redshift bins, but with all estimated redshifts treated as exact, and the
bin sizes increased to ensure that each redshift bin contained a minimum
of 5 sources; this avoids problems with small number statistics, but means
that resolution is lost at the high--redshift end.  This analysis showed
that the polynomial-fit method provided values of $\chi^{2}$ which were a
factor $f=0.3$ away from the true value. The
$\chi^{2}$ values from the redshift distributions are therefore scaled by
this factor in order to ensure that accurate uncertainties are determined.

The scaled $\chi^{2}$ values are transformed to likelihoods, and combined
to produce an overall measure of the goodness of the fit:
\begin{equation}
L  =  \exp\left(\frac{-\chi_{\rm LRLF}^{2}}{2}\right)
       \exp\left(\frac{-\chi_{\rm counts}^{2}}{2}\right)
       \exp\left(\frac{-f\chi_{\rm z-dists}^{2}}{2}\right), 
\end{equation}
which is then returned to the minimisation routine \verb1amoeba1.

\subsection{Constraint in the $P$--$z$ grid}
\label{errors}

It is not sufficient to find  the best fit to the data without obtaining some measure of the uncertainty associated with each point on the $P$--$z$ grid.
In this section the intrinsic modelling limits offered
for the constraint of the $P$--$z$ grid are presented, whilst the
effect of uncertainty in the input data is discussed later in Sections
\ref{alpha_change} and \ref{z_change}.  

Assuming an ideal data set, uncertainty still arises due to the degeneracy across parameters in the model.
The conditional error for parameter $p$ is given by:
\begin{equation}
\sigma_{\rm cond, p}^{2} = \left(-\frac{\partial^{2} \ln L}{\partial p^{2}}\mid_{\rm peak}\right)^{-1}.
\label{sigma_cond_eqn}
\end{equation}
In practice it is determined by finding the value of the parameter for
which $\ln L(p) - \ln L(\hat{p}) = 0.5$, from which $\sigma = \rho (p)
- \rho (\hat{p})$.  
This is then a measure of the uncertainty in a given parameter whilst holding all other parameters constant.

However the actual uncertainty may be greater when the variation of other parameters is taken into account.
This is the marginalized error and comes from the diagonal of the inverse Hessian matrix, $H$, where:
\begin{equation}
\sigma_{\rm marg,i}^{2} = \left[H\right]_{\rm i,i}^{-1}, 
\end{equation}
for parameter i, where,
\begin{equation}
H_{\rm i,j} = \frac{\partial^{2} \ln  L}{\partial p_{\rm i}\partial p_{\rm j}}.
\end{equation}
This is calculated here via the finite difference approximation (note
that diagonal terms in the Hessian matrix are simply $-1/\sigma_{\rm cond, p}^{2}$). 
It is these marginalized errors that are quoted in the results
presented in the remainder of this paper. In general these are comparable to the conditional errors, but for some grid points they are up to a factor of two higher.

\section{Results and discussion}
\label{results}

This Section presents the best--fitting steep--spectrum $P$--$z$ grid
produced by running the modelling code described in Section
\ref{model} above, and examines how well its predictions agree with
the input datasets. The best--fitting $P$--$z$ grid and its associated
error is given in full in Appendix \ref{fullgrid}.

\subsection{Dataset comparison}
\label{data_comp}

\begin{figure*}
\centering
\includegraphics[scale=0.30, angle=90]{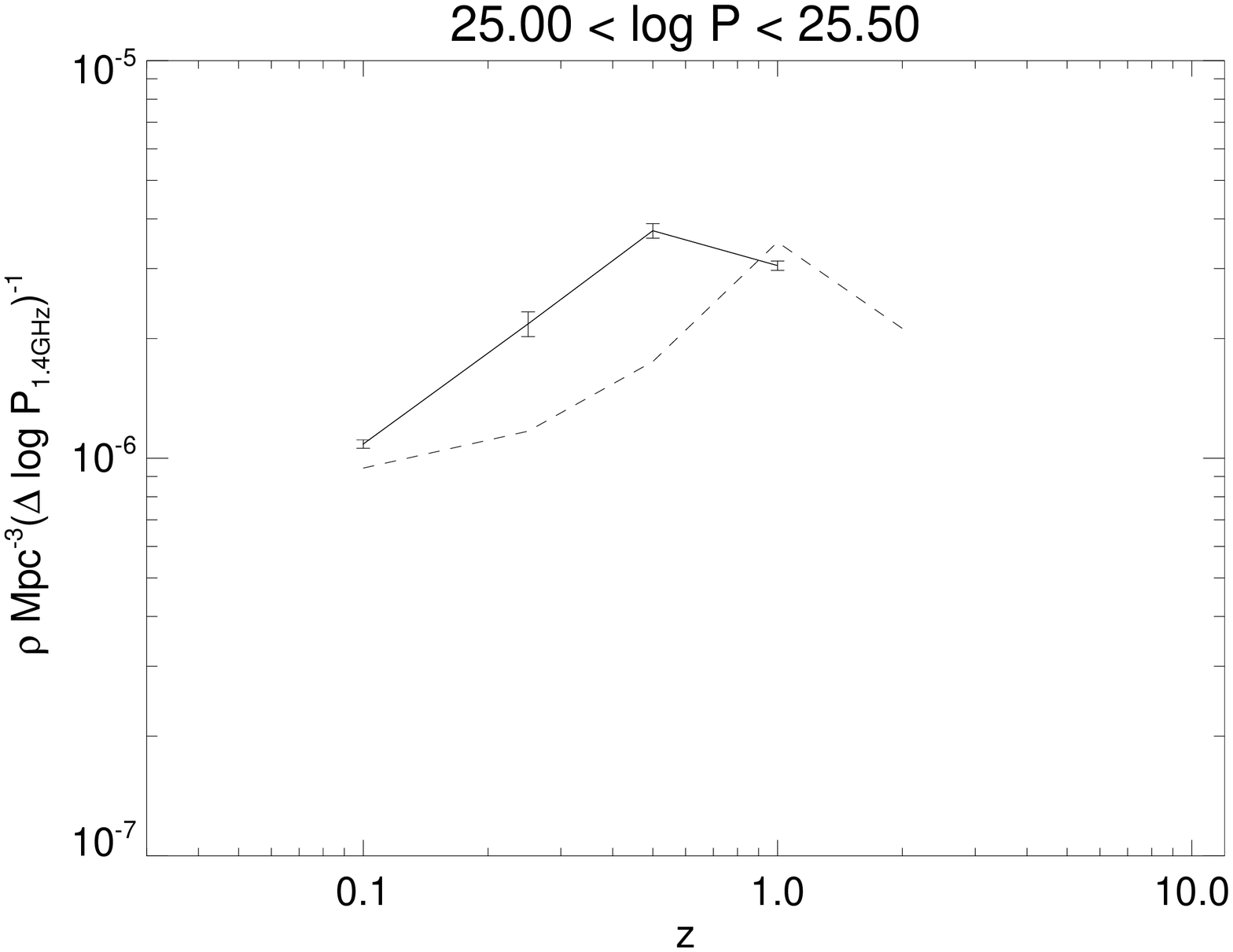}
\includegraphics[scale=0.30, angle=90]{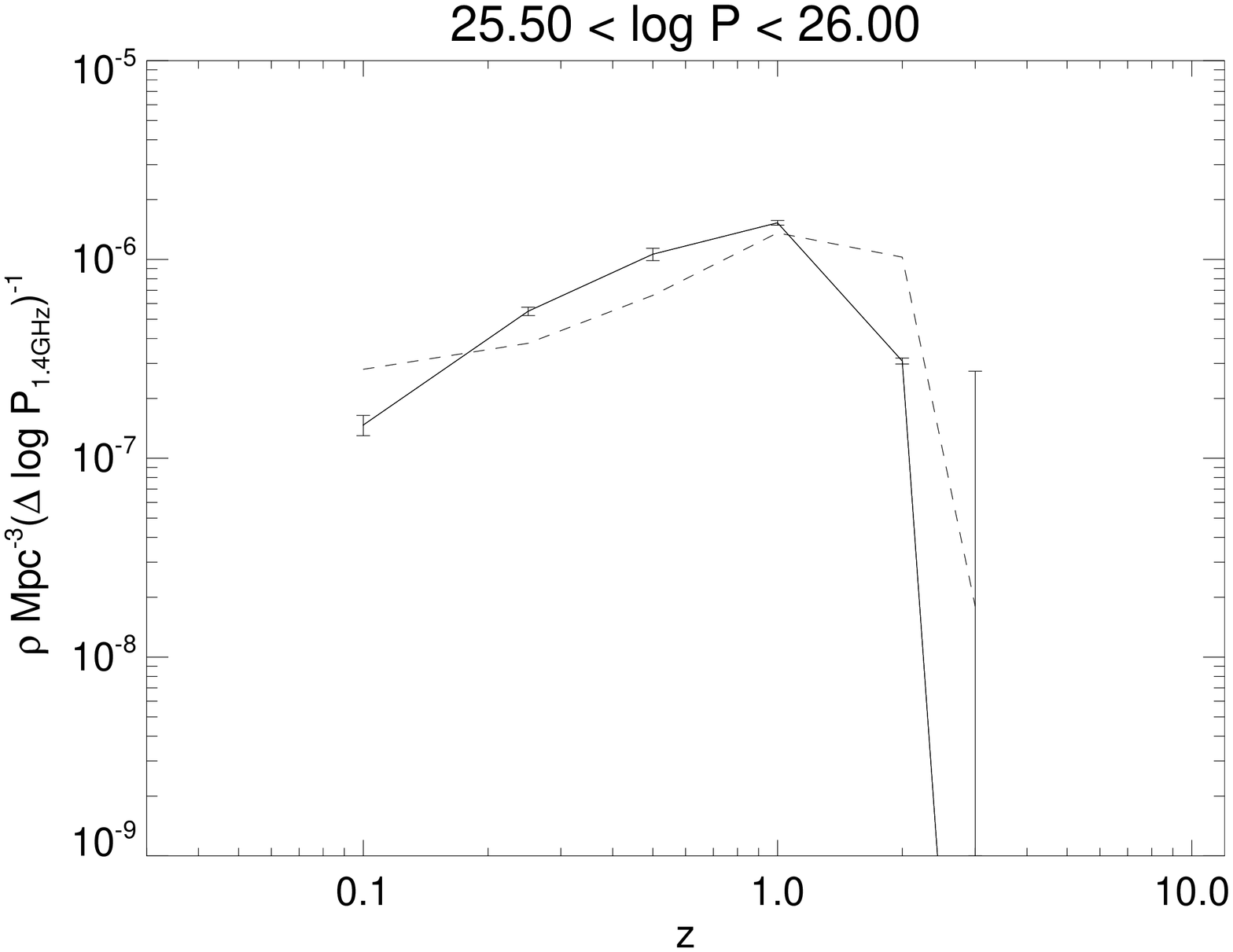}
\includegraphics[scale=0.30, angle=90]{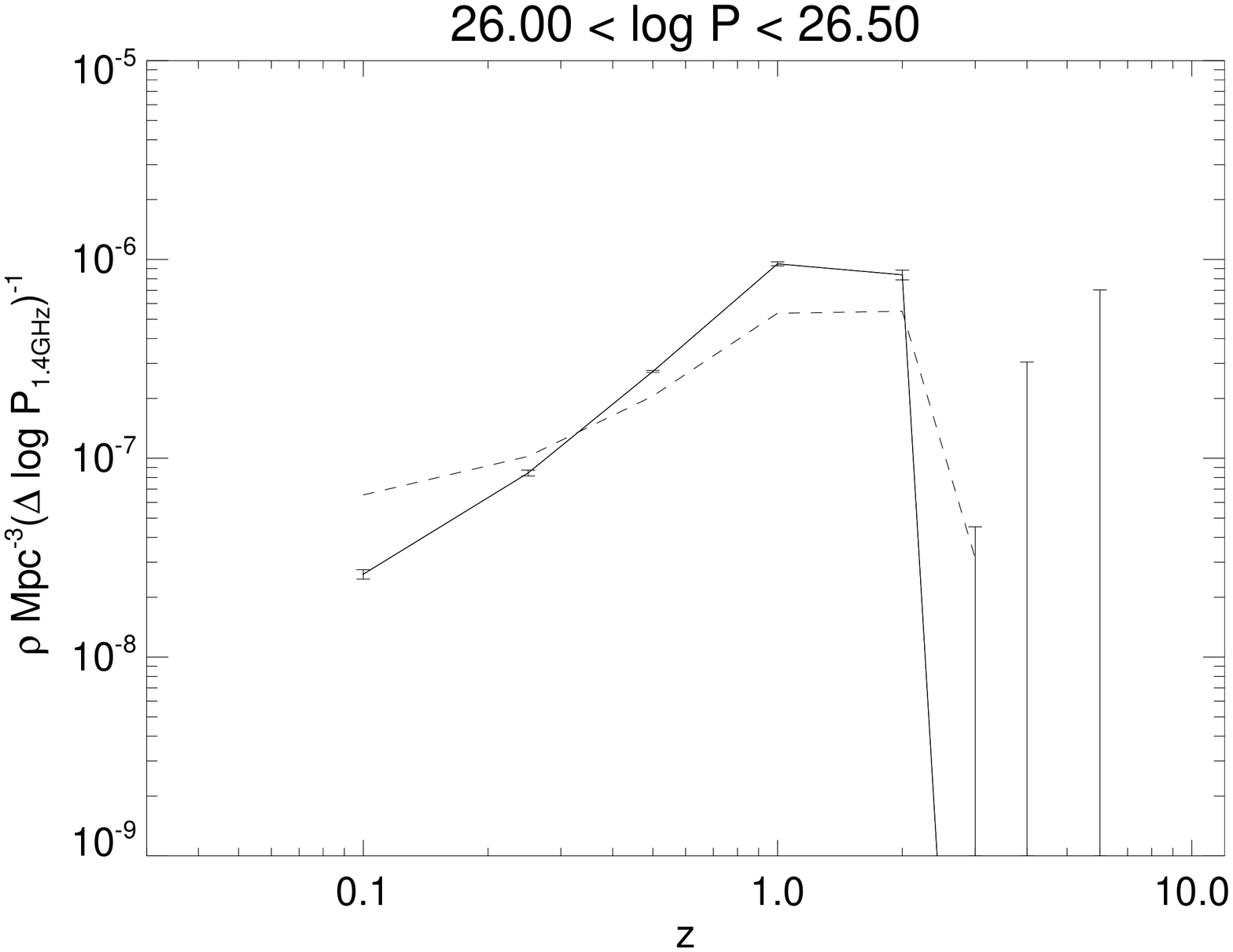}
\includegraphics[scale=0.30, angle=90]{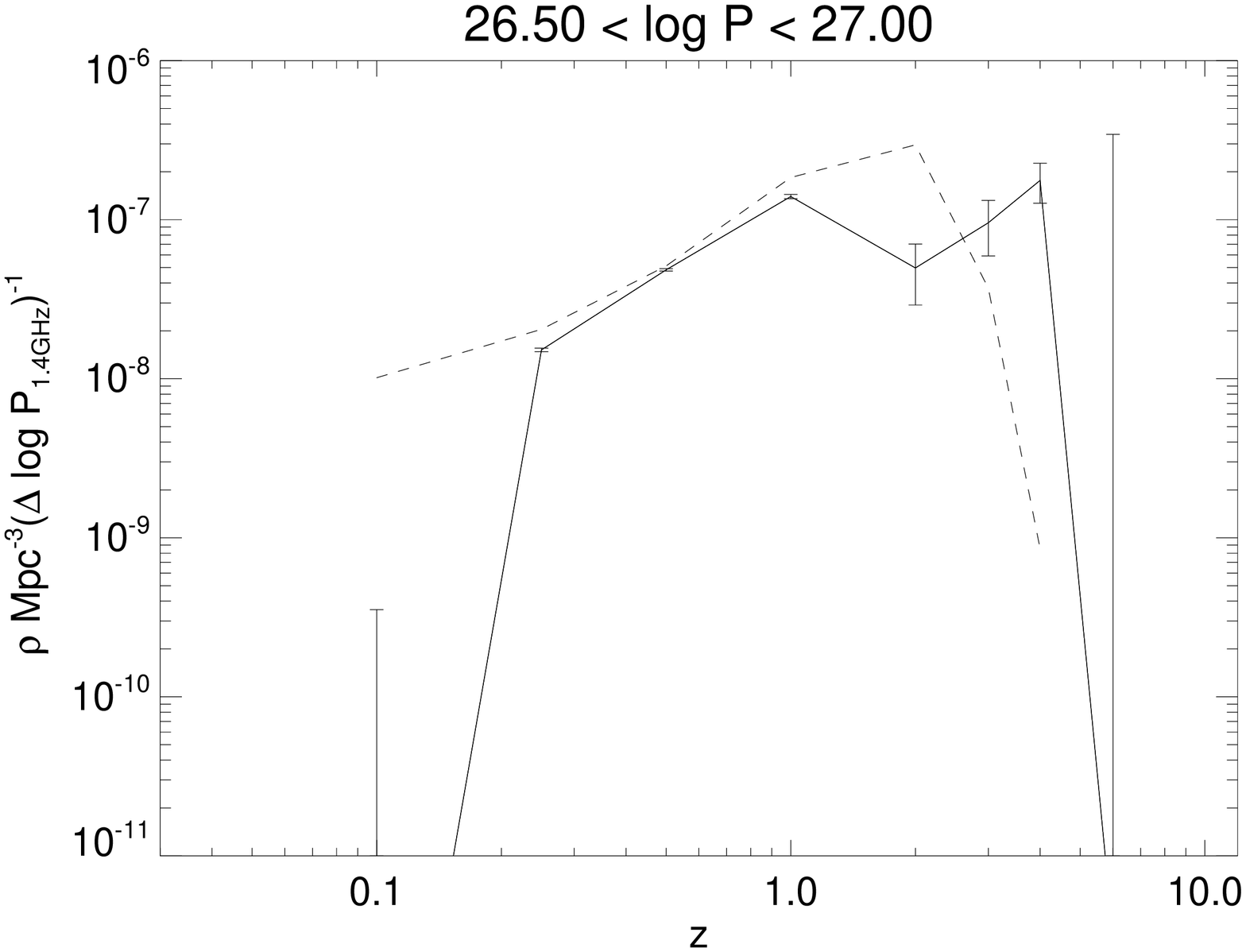}
\includegraphics[scale=0.30, angle=90]{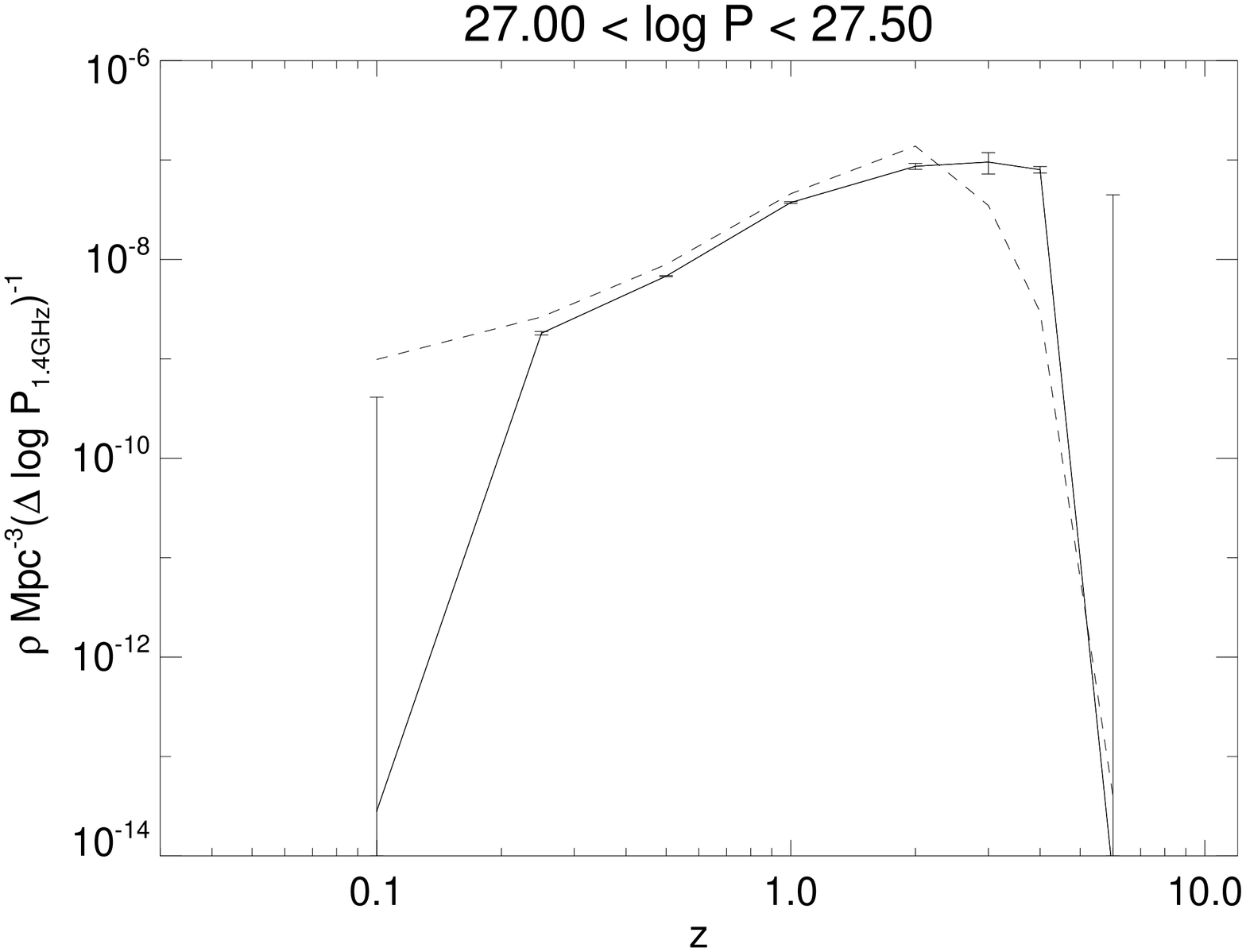}
\includegraphics[scale=0.30, angle=90]{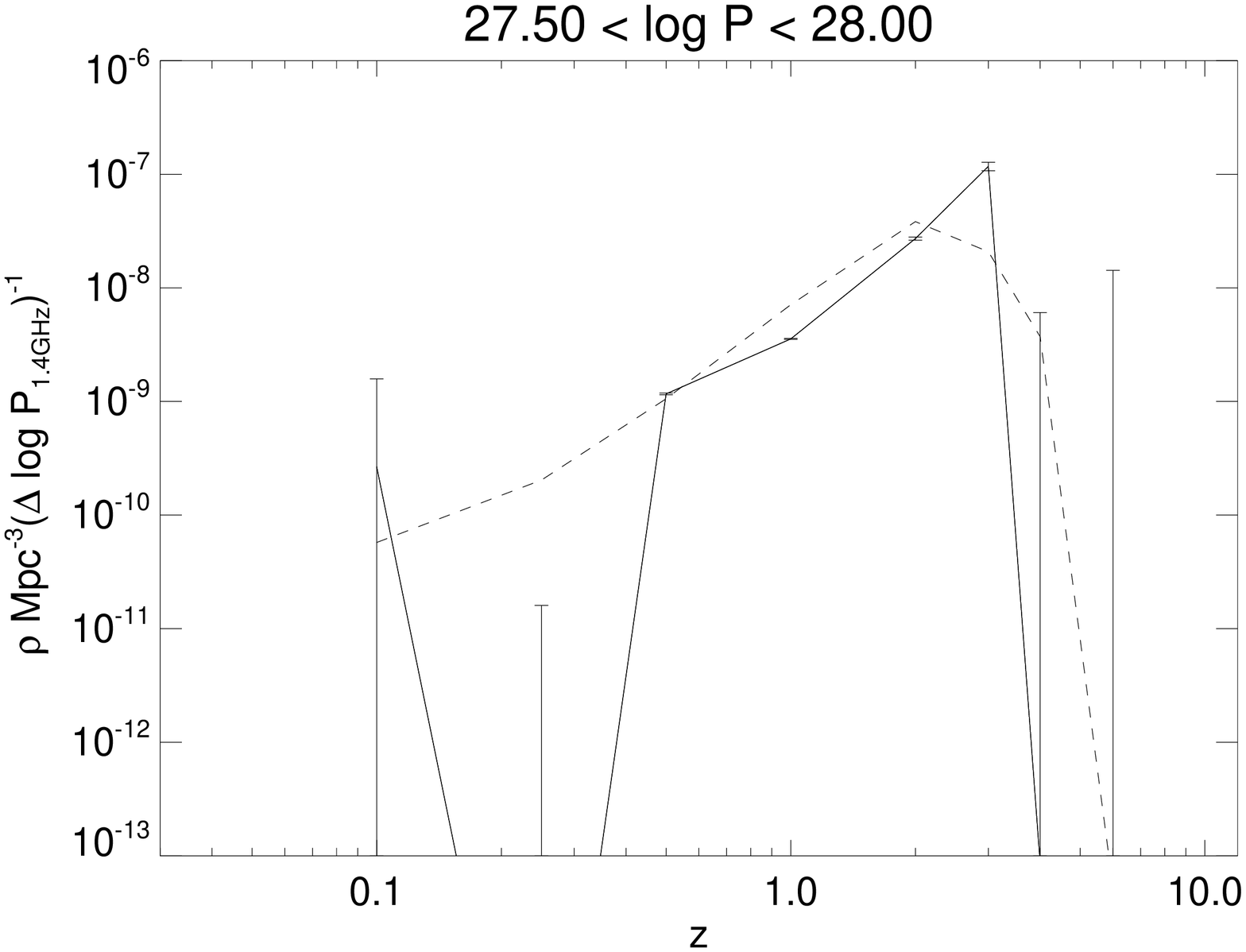}
\caption{\protect\label{indivs_lfs1} The individual model steep--spectrum
  radio luminosity functions vs. redshift from the best--fitting
  $P$--$z$ grid (solid lines). Points are only plotted here if they are constrained
  by at least two of the input datasets (see Fig.
  \protect\ref{cons_fig} for details). Also shown are the median of
  the evolutionary models from DP90 (dashed lines).} 
\end{figure*}

The success of the best-fitting steep spectrum $P$--$z$ grid, combined where
necessary with the unvaried flat spectrum and starburst grids, at fitting
the input data is illustrated in Figs \ref{z_dist_comp} to \ref{counts_comp} for the
five sample redshift distributions, the LRLF,  and the differential source counts respectively.
The agreement with both the LRLF and the source counts is very good -- this is to be
expected for two reasons. The constraint provided by the
LRLFs is tight and particular to specific parameters, meaning that the
model has little freedom to vary it. At the opposite
extreme, there are numerous combinations of densities which will
satisfy the observed counts, as they depend on the summation carried
out across the $P$--$z$ grid to transform it to $S$--$z$. Thus, whilst
fitting these data is essential, on their own they do not provide a particularly
interesting constraint. 
The source counts comparison plot (Fig. \ref{counts_comp}) also
breaks down the model prediction into the different contributions from
the three populations; this shows that the steep--spectrum sources
dominate at the flux densities probed by the current samples, which in turn justifies limiting the fitting to them only at
this point.   

The redshift distributions also derive from the $S$--$z$ grid so they
prevent nonsensical combinations of densities fitting the source
counts, and the stronger constraint that they provide is therefore
essential to obtaining information about the evolution of the
luminosity function. Again the flat and starburst populations are
plotted separately in Fig. \ref{z_dist_comp}  where relevant, showing 
their small contribution, especially at the high redshifts which are
of particular interest here.  

The model predictions for all five samples are generally in good
agreement with the data across the whole of each redshift range. 
The total numbers of sources given by the model are 131.1, 67.7, 74.5, 284.8 and 84.5 
for CENSORS, Hercules, PSR, COSMOS and WP85 respectively; these are well
matched to the actual figures of 135, 64, 74, 314 and 83. 

As a further check on the results, the final model grid can also be
compared to datasets that were not included in the fitting: for
example  the $\bar{z}=0.55$ radio luminosity function
determined by \citet{donoso} using a sample of $\sim$14000 radio--loud
AGN, created by combining the NVSS and FIRST 1.4\,GHz radio surveys
with the MegaZ--LRG catalogue; this comparison is shown in Fig.
\ref{rlf_comp}. The match is reasonable over the luminosity range of 
the data, and the overprediction seen at $25 < \log P < 26$ is likely to
be because Donoso et al. were only considering radio sources associated with luminous red galaxies
in their sample, and therefore miss any bluer radio galaxies (this is also why this sample was not included as an explicit constraint).  

\subsection{Model luminosity functions}
\label{mod_res}
Fig. \ref{indivs_lfs1} shows the behaviour of the individual
best--fitting steep--spectrum luminosity functions with $z$ for $\log P
= 25.25$ to $\log P = 27.75$ inclusive, averaged over bins of 0.5 in $\log P$.
This luminosity range was chosen as it covers the region of the $P-z$
plane with the most constraints as illustrated in Fig.
\ref{cons_fig}; additionally, high redshift points are only plotted if they are
constrained by both the redshift distributions and the source
counts. Also plotted as a comparison is the median of the seven
evolutionary models calculated by DP90 (translated to the cosmology used for this work), which all suggest a mild
high redshift decline in the number density of sources of these powers.

\begin{figure}
\centering
\includegraphics[scale=0.35,angle=90]{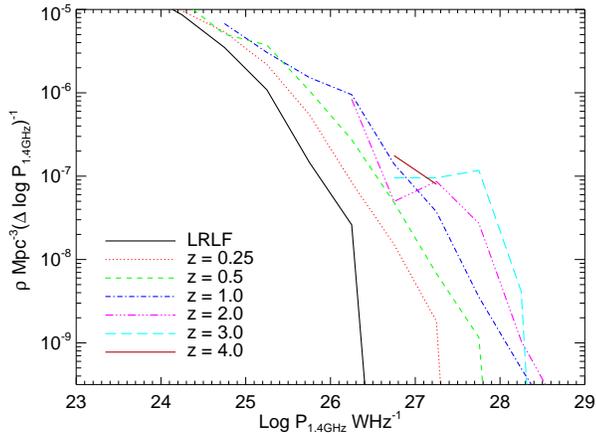}
\caption{\protect\label{combined_lfs} The evolving luminosity
  function. Points are only plotted here if constrained by more than
  one dataset and errors have been ignored for clarity.} 
\end{figure}

The evolving luminosity function (in which the grid densities are
plotted against $\log P$ for different redshifts) is shown in Fig.
\ref{combined_lfs}. As before regions of radio power and redshift are
excluded if they are not constrained by at least two
datasets. Displaying the grid behaviour in this way is a useful
alternative to Fig. \ref{indivs_lfs1} as it gives an overview of the
space density changes with radio power and redshift.  

The low--$z$, $\log P \geq 26.5$ points are discrepantly low (albeit with large error bars) compared to their neighbours. This is a result of the
weaker constraints provided by the upper limits of the LRLF to the
minimisation process (though this region is also constrained by the
source counts and redshift samples). Inspection of Fig.
\ref{lrlf_comp} (the combined LRLF from the steep and flat spectrum
grids) illustrates that the model reaches well below the upper--limits
of the densities in these regions. Similarly, there are some
apparently anomalous high redshift points (e.g.  the `dip' seen at $z
\sim 2$ for $26.5 < \log P < 27$ and the $z=3$ points at $\log P < 26.5$);  Figs \ref{cons_fig} and
\ref{pz_plot} suggest that although these should be constrained by
both the redshift distributions and the source counts, the low
densities are likely to result from a lack of sources in the radio
samples covering this range. However, the `dip' in particular may not just be
the result of low--number statistics as it lies within the `redshift
desert' discussed previously in Section \ref{comp_pred}.

\subsection{The high--redshift turnover}
\label{highz_cut}

Inspection of Figs \ref{indivs_lfs1} shows that decreases in space density are present to some
degree at $z \gsim 0.5$ for all the luminosities considered. At low
powers these declines are clear and occur at $z \sim 1$, but at the
highest powers the densities remain essentially level with no strong
decline, out to $z \sim 4$. The agreement with the DP90 results is
reasonable, though visually there seems to be a trend for the low power cutoffs to be at lower redshifts than DP90, and the high power cutoffs to be at higher redshift. 
However, the range in the DP90 results, and the inability to probe the full distance range for 
these low powers, means it is difficult to draw firm conclusions about the differences. 

The `strength' of the cut--off, $C_{\sigma}$,  between the density at
the peak redshift, $\rho_{\rm peak}$ and that at any of the $n$
subsequent redshift points can be quantified as: 
\begin{equation}
\label{cut_eqn}
C_{\sigma} = \frac{\rho_{\rm peak} - \rho_{\rm peak + n}}{\sqrt(\sigma_{\rm peak}^{2} + \sigma_{\rm peak + n}^{2})} ,
\end{equation}
where $\sigma_{\rm peak}$ and $\sigma_{\rm peak + n}$ are the
corresponding errors on the peak and post--peak space
densities. However, caution is necessary when calculating $C_{\sigma}$
for the results presented here, because of the discrepant, badly
constrained, points discussed in the previous Section. The quantised nature of the grid also means that
it is often relatively flat after the peak, before dropping
sharply.  To minimise these effects, $\rho_{\rm peak+n}$ is taken as
the average of the space densities at redshifts
higher than the peak position. This average is weighted by the
available volume in each redshift bin, and the $z=6$ point is ignored
in all cases due to the lack of supporting data in the redshift distributions.
The error, $\sigma_{\rm peak + n}$, for this
average density is calculated by combining the post--peak space
density errors in quadrature, taking the volume weighting into
account. The results of this calculation are shown in Table
\ref{res_table}; they support the previous observation that the
declines are strong for the faintest powers, but tend to be weaker at
brighter powers.  

\begin{table}
\caption{\protect\label{res_table} The redshift at the grid point at
  which the space density is highest, $z_{\rm grid}$, and the strength
  of the cut--off, $C_{\sigma}$following that point from the average
  of the post--peak densities (excluding the $z=6$ points), using Equation
  \protect\ref{cut_eqn}. Also shown is the peak redshift, $z_{\rm
    fit}$, determined from the polynomial fitting described in Section
  \protect\ref{highz_cut}. The different versions illustrate the
  effects of altering various model parameters; see Section
  \protect\ref{alpha_change} and \protect\ref{z_change} for full details.  
 }
\begin{tabular}{c|c|c|c}
\hline
Radio power range& $z_{\rm grid}$ & $z_{\rm fit}$ & $C_{\sigma}$    \\ 
\hline
\multicolumn{4}{c}{Default: $\alpha = 0.83 + 0.4\log(1+z)$}    \\
\hline
25.00 -- 25.50   &0.5& 0.7&  3.8      \\
25.50 -- 26.00   &1.0& 1.1&  9.6   \\
26.00 -- 26.50   &1.0& 1.4&  5.6      \\
26.50 -- 27.00   &4.0& 2.3&  --      \\
27.00 -- 27.50   &3.0& 2.6&  0.7      \\
27.50 -- 28.00   &3.0& 3.9&  10.0     \\
\hline
\multicolumn{4}{c}{$\alpha = 0.8+0.25z$}     \\
\hline
25.00 -- 25.50&  1.0& 0.8&  --        \\
25.50 -- 26.00&  1.0& 1.2&  2.4   \\
26.00 -- 26.50&  1.0& 1.4&  4.0   \\
26.50 -- 27.00&  2.0& 1.6&  2.1   \\
27.00 -- 27.50&  4.0& 2.6&  --    \\
27.50 -- 28.00&  3.0& 3.2&  5.9   \\
\hline
\multicolumn{4}{c}{$\alpha = 1.5$}     \\
\hline
25.00 -- 25.50&  1.0& 0.7&  --        \\
25.50 -- 26.00&  0.5& 0.7&  1.5    \\
26.00 -- 26.50&  1.0& 1.4&  4.8    \\
26.50 -- 27.00&  1.0& 1.4&  4.0    \\
27.00 -- 27.50&  4.0& 2.8&  --     \\
27.50 -- 28.00&  2.0& 2.3&  11.4    \\
\hline
\multicolumn{4}{c}{$\alpha = 0.8$}     \\
\hline
25.00 -- 25.50&  0.5&  0.7&2.4 \\
25.50 -- 26.00&  1.0&  0.9&7.9 \\
26.00 -- 26.50&  1.0&  1.4&16.1  \\
26.50 -- 27.00&  4.0&  2.5&--   \\
27.00 -- 27.50&  2.0&  1.9&6.7  \\
27.50 -- 28.00&  3.0&  3.8&6.5  \\
\hline
\multicolumn{4}{c}{Average strength from 50 random variations of the redshift limits}     \\
\hline
25.00 -- 25.50&  0.5&  0.7&  1.8   \\
25.50 -- 26.00&  1.0&  1.0&  8.6   \\
26.00 -- 26.50&  1.0&  1.4&  2.6   \\
26.50 -- 27.00&  4.0&  2.1&  --      \\
27.00 -- 27.50&  3.0&  2.6&  0.2   \\
27.50 -- 28.00&  3.0&  4.0&  9.8   \\
\hline
\multicolumn{4}{c}{Uncertain $z$ at $+1\sigma$}     \\
\hline
25.00 -- 25.50&  1.0&  0.8&  --      \\
25.50 -- 26.00&  1.0&  1.3&  6.5    \\
26.00 -- 26.50&  1.0&  1.4&  6.7   \\
26.50 -- 27.00&  2.0&  1.8&  5.9    \\
27.00 -- 27.50&  3.0&  2.5&  4.4    \\
27.50 -- 28.00&  3.0&  3.6&  15.5    \\
\hline
\multicolumn{4}{c}{Uncertain $z$ at $-1\sigma$}    \\
\hline
25.00 -- 25.50&  0.5&  0.7&  3.5    \\
25.50 -- 26.00&  1.0&  1.2&  3.1    \\
26.00 -- 26.50&  1.0&  1.3&  5.0    \\
26.50 -- 27.00&  1.0&  2.3&  2.4    \\
27.00 -- 27.50&  3.0&  2.1&  1.1    \\
27.50 -- 28.00&  3.0&  3.8&  4.6    \\
\hline
\end{tabular}
\end{table}

\begin{figure}
\centering
\includegraphics[scale=0.35,angle=90]{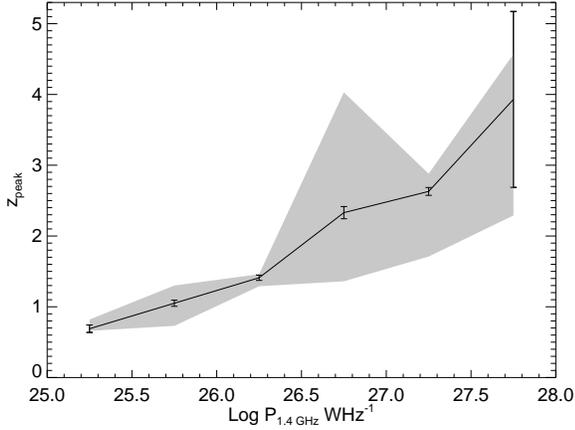}
\caption{\protect\label{zpeaks} 
An illustration of the changes in peak
  redshift with radio power for the best--fitting steep--spectrum
  grid (solid line). The error bars show the uncertainty in the
  polynomial fits and the shaded region represents the range in
  results which come from repeating this 
  process for the additional grids in Sections \protect\ref{alpha_change} and \protect\ref{z_change}.  
}
\end{figure}

Fig. \ref{indivs_lfs1} and Table \ref{res_table} both show an
apparent luminosity--dependence of the peak redshift, $z_{\rm peak}$,
but the wide redshift bins in the $P$--$z$ grid means the position of
the peak is ambiguous. For a better estimate, polynomials (generally
of order 2, but  also of order 3 where necessary)  were fitted to the
model steep--spectrum $\rho$--$z$ distributions, for various radio power bins, with the aim of roughly
parametrizing this $z_{\rm peak}$ vs $\log P$ relation. Fig.
\ref{zpeaks} shows the results of this for the best--fitting steep
spectrum grid, with error bars showing the uncertainties in the
polynomial fit. Also shown is  the range in $z_{\rm peak}$ found from
repeating this fitting for the different spectral indices and redshift
limits discussed in Sections \ref{alpha_change} and
\ref{z_change}. This is not a rigorous analysis but it does illustrate
the general increase in $z_{\rm peak}$ over the radio luminosity range
probed.  

\subsubsection{The effect of the spectral index}
\label{alpha_change}

The creation of the $S$--$z$ grid, used in the modelling for the
data from the five samples and source counts, requires a value for the
spectral index, $\alpha$.  The choice of $\alpha$ is complicated by
the spectral curvature seen in some radio--loud sources
\citep[e.g.][]{laing}, which can result in an increase in the spectral
indices at higher redshift (see \citealt{jarvis00} for a discussion of
this effect for flat--spectrum sources); as Fig. \ref{alpha_dist}
shows, this effect is seen in the radio samples used here, albeit with
a large scatter. Ignoring this may lead to under--estimation of the
high--$z$ space density (and increase the significance of any density
cut--off) since the steepest spectrum sources are missed.  In the modelling results presented in the previous Section attempts
were made to take this into account by using the $\alpha$--$z$
relation from \citet{ubachukwu} in the creation of the $S$--$z$
grid. This provides a reasonable match to the data (Fig.
\ref{alpha_dist}) but it is important to investigate the effect a
different choice has on the high--redshift behaviour of the model
RLFs. Changing the assumed value of $\alpha$ will move sources into different radio power bins and hence strengthen some turnovers, and weaken others. 

\begin{figure}
\includegraphics[scale=0.30, angle=90]{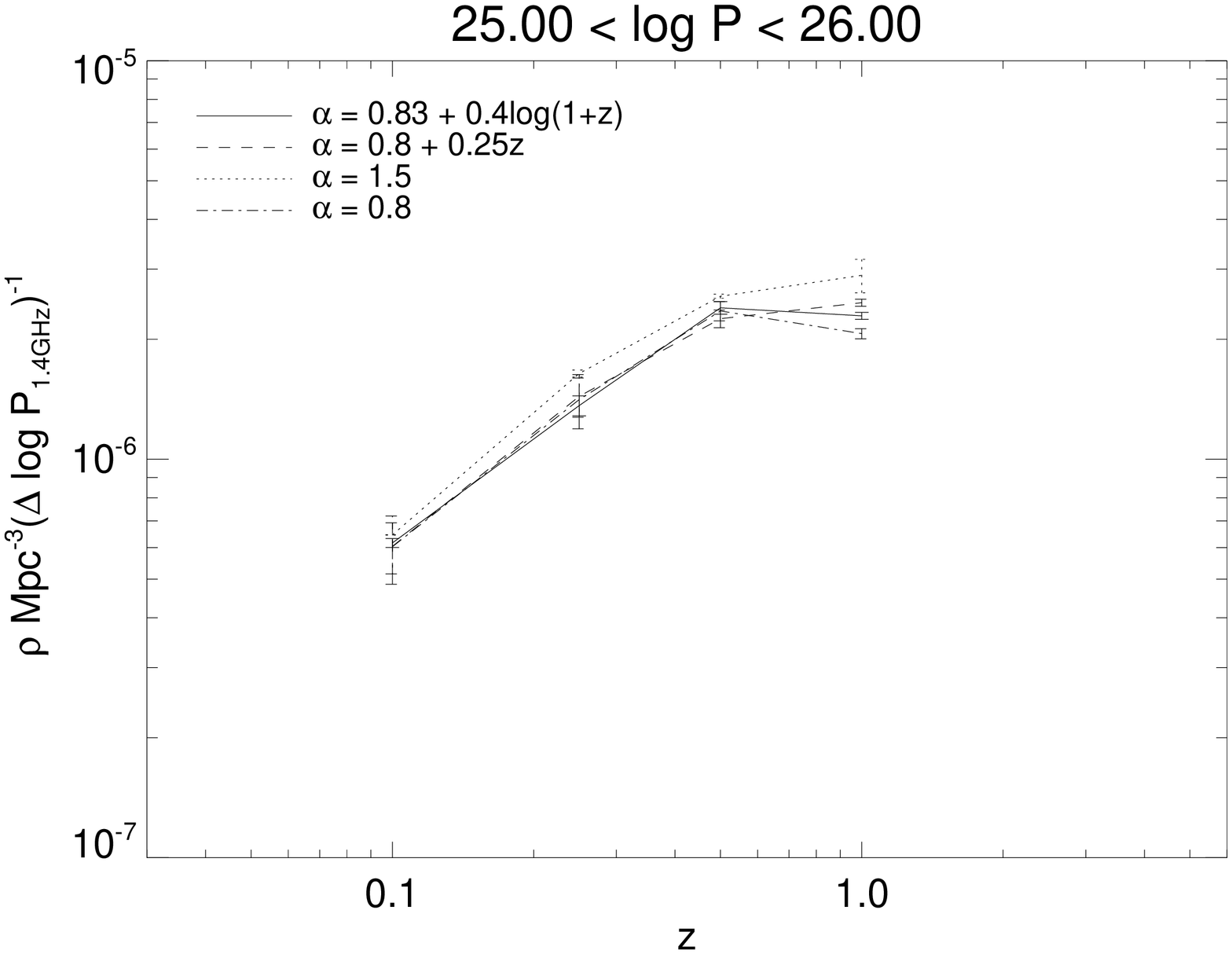}
\includegraphics[scale=0.30, angle=90]{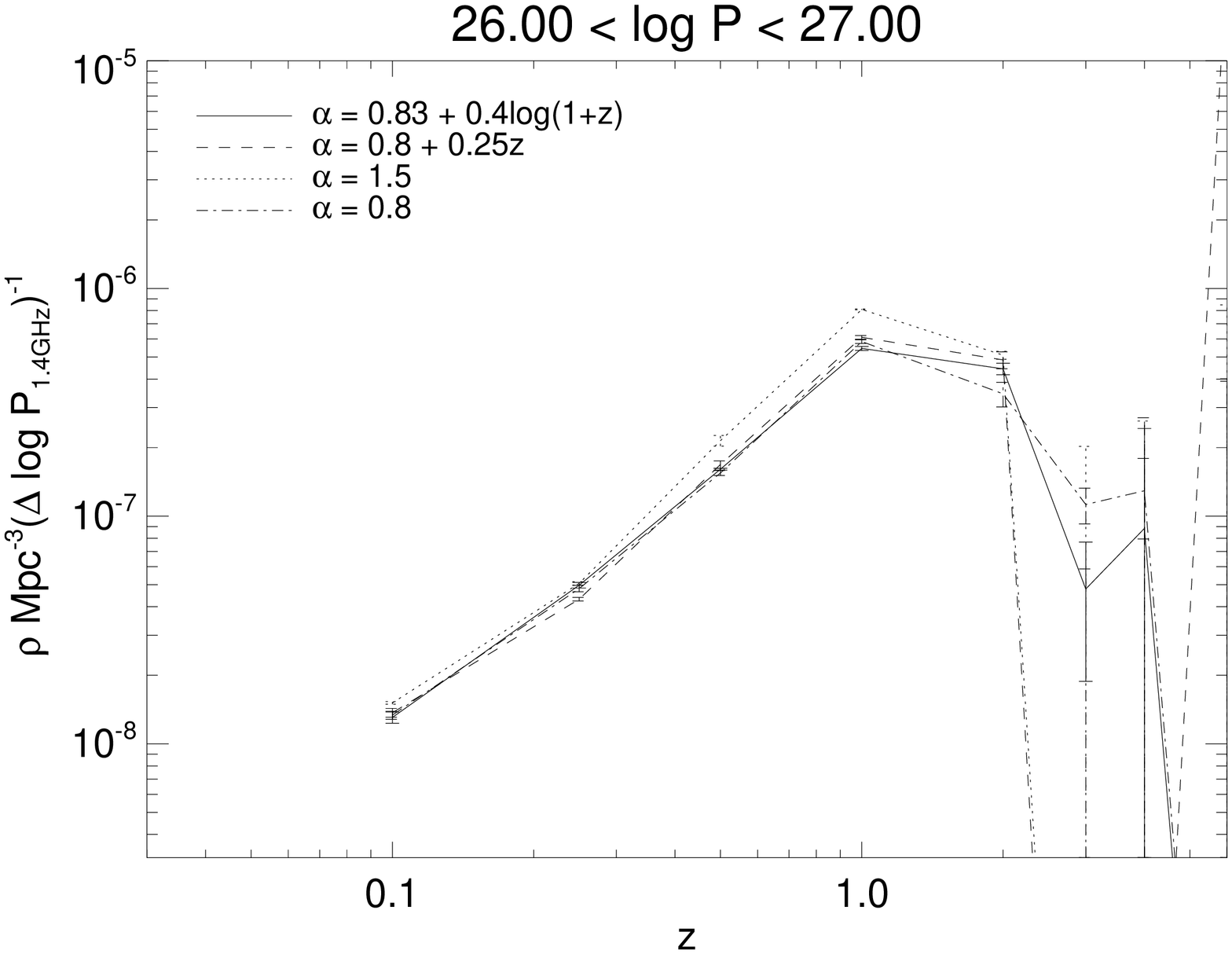}
\includegraphics[scale=0.30, angle=90]{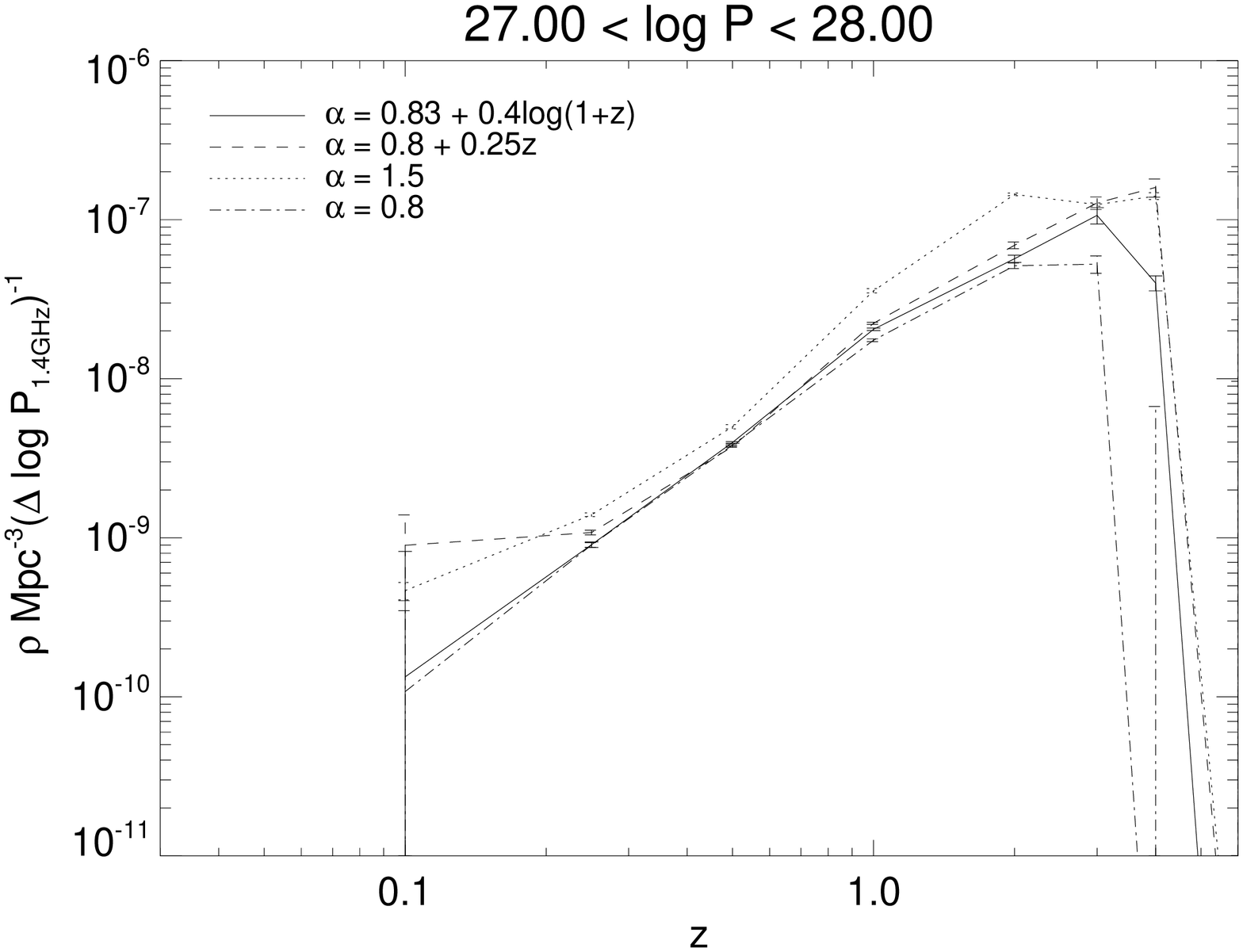}
\caption{\protect\label{alp_increase} The effect on the steep--spectrum
  model luminosity functions of changing the value of the spectral
  index, $\alpha$ for three luminosity ranges. The `$\alpha = 0.83 + 0.4\log(1+z)$' line comes from the best fitting model grid presented
  earlier. Larger bins of $\Delta\log P = 1$ are shown to better
  illustrate how the changes affect the results. In all cases a scatter 
  of $\alpha \pm 0.2$ around the mean value was also included, as previously discussed in
  Section \protect\ref{input_grids}.}  
\end{figure}

Fig. \ref{alp_increase} presents the results of using both an
extreme spectral index of 1.5, and a stronger increase with redshift
(arbitrarily modelled as $\alpha = 0.8 + 0.25z $ to give ultra--steep
spectra at $z \gsim 2$). Note that these will represent extreme cases
since some radio sources with typical $\alpha \sim 0.8$ spectra are
found at high redshift \citep[e.g.][]{jarvis09}. The cut--off strengths  are also given in Table \ref{res_table}.  
In both cases the general effect is to increase the densities at $z \gsim
0.3$, and weaken the high--redshift cut--off, as sources in the $S-z$ grid have moved into higher radio power
bins. Overall this is a good illustration of how far $\alpha$ needs to be increased to reduce the significance of the cut--offs to the $\sim3\sigma$ level.
Also shown, as a comparison, is the effect of using $\alpha = 0.8$,
the low--redshift mean value; this generally decreases the densities
but the decline is still present at high significance for all powers.

\subsubsection{The effect of the redshift incompleteness}
\label{z_change}

The other uncertainty in the model results comes from the estimated
redshifts and redshift limits present to some degree in the five input
samples. Attempts are made to take these into account in the modelling
process, but this is likely to be less successful for the $z$--limits
as their true value is less constrained. To investigate what effect this has, the model was run 50 times; in each run each limit is
assigned a new, higher, redshift, drawn randomly from a uniform 10000
element distribution, starting at the given limit, up to a maximum of $z = 6$. As a further check
the modelling was repeated with all the uncertain redshifts, both
estimates and limits, moved up and down by $1\sigma$.  
No other attempt is made to account for the uncertainty in the
estimated redshifts and limits in either of these cases.  

\begin{figure}
\includegraphics[scale=0.30, angle=90]{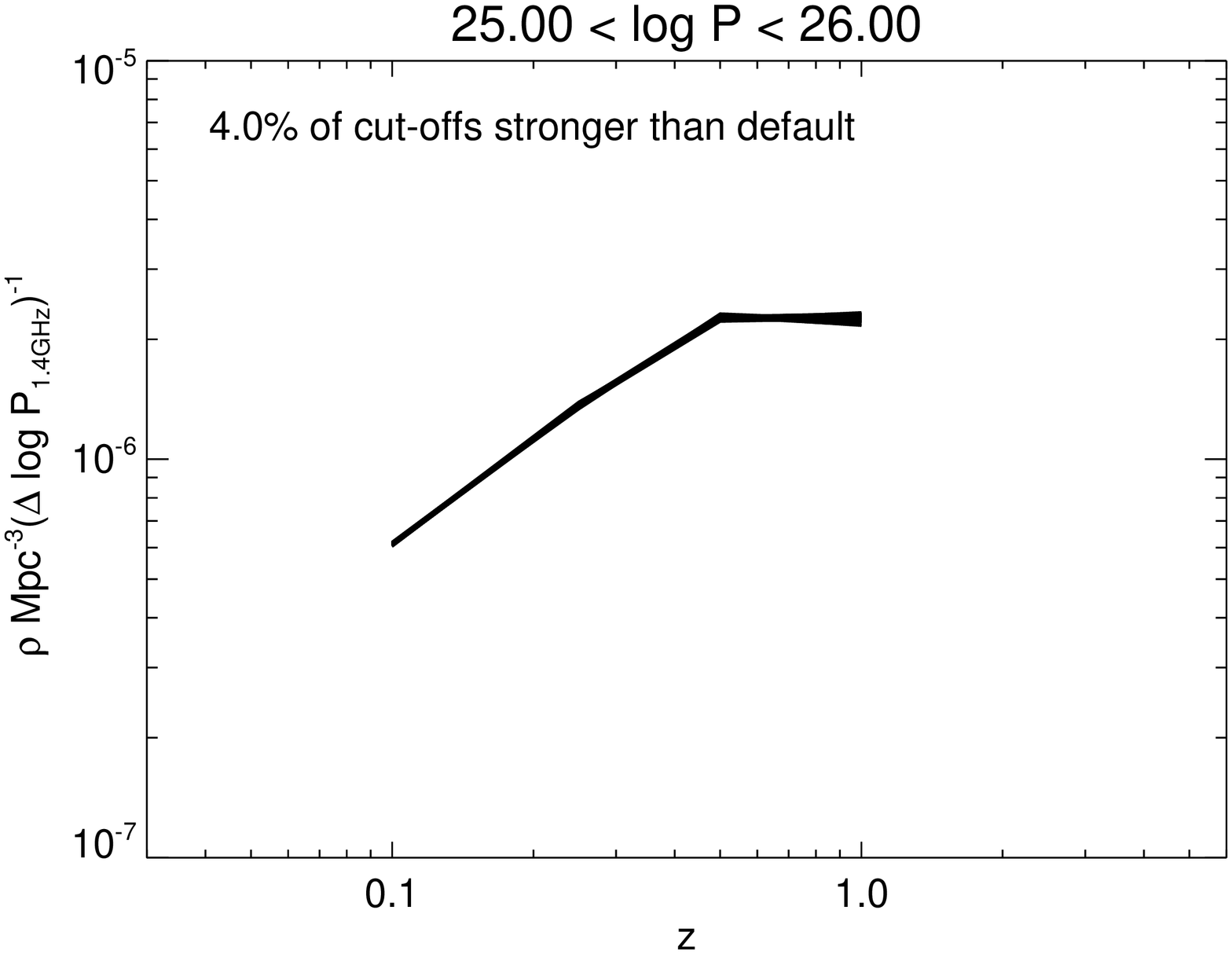}
\includegraphics[scale=0.30, angle=90]{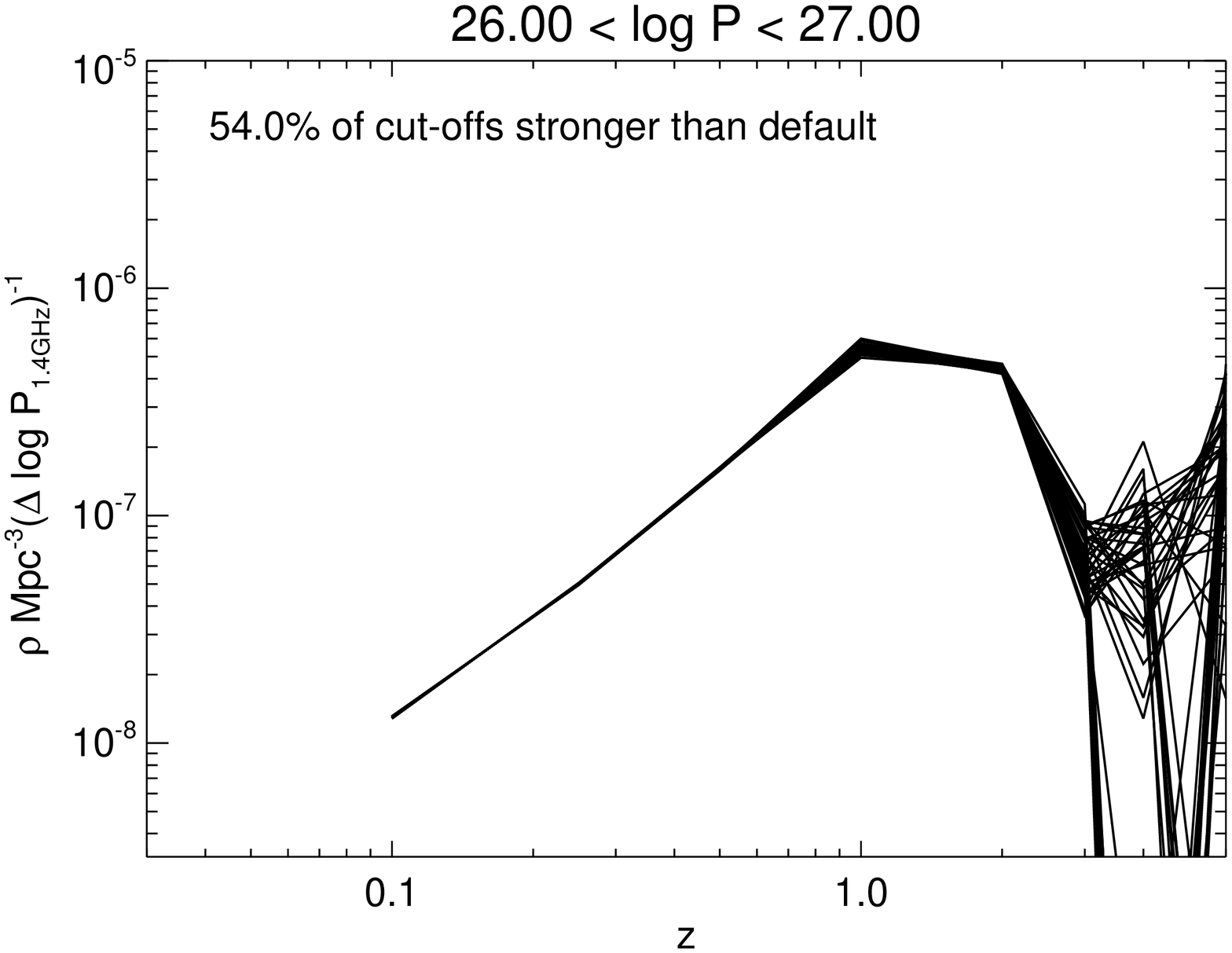}
\includegraphics[scale=0.30, angle=90]{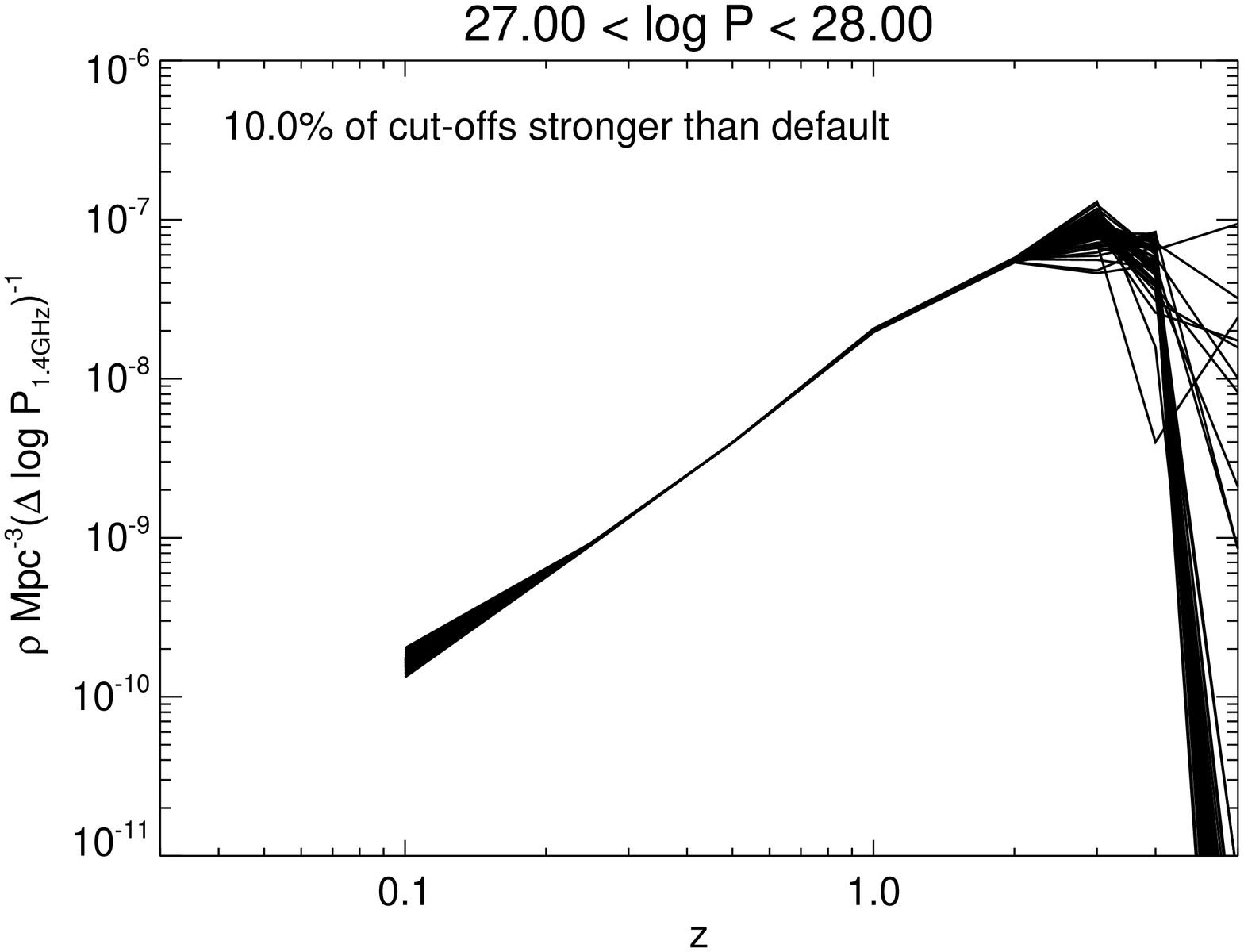}
\caption{\protect\label{z_increase} The effect on the steep--spectrum
  model luminosity functions of repeatedly randomly moving the
  redshift limits to a new, higher, value, up to a maximum of $z =
  6$. For this the estimated redshifts were kept at their given
  values, with no attempt to take their uncertainties into account as
  done previously. `default' in the Fig. labels refers to the
  original modelling results presented in Section
  \protect\ref{mod_res}. Larger bins of $\Delta\log P = 1$ are shown to better
  illustrate how the changes affect the results. In all cases an error
  of $\alpha \pm 0.2$ was also included, as previously discussed in
  Section \protect\ref{input_grids}.} 
\end{figure}

\begin{figure}
\includegraphics[scale=0.30, angle=90]{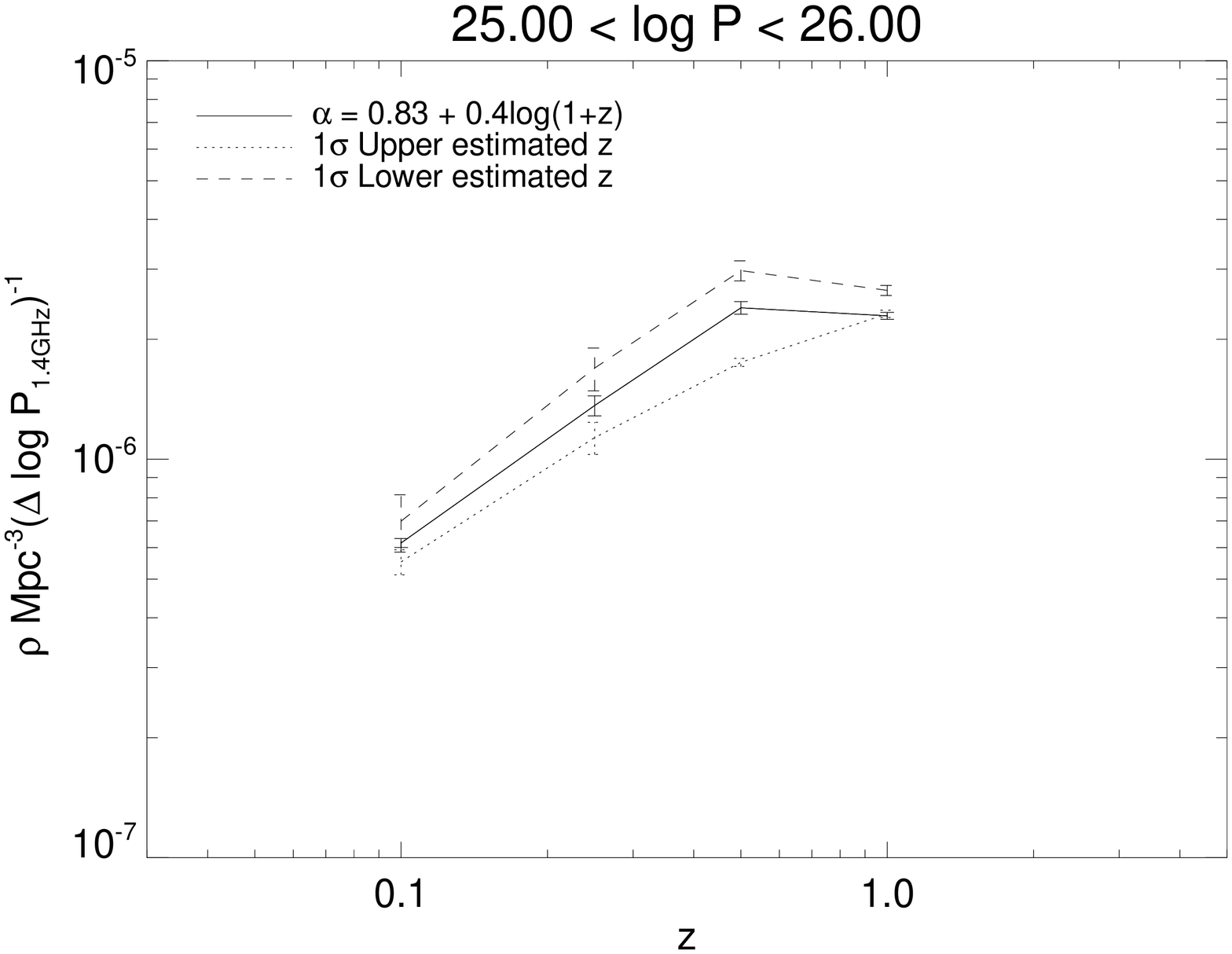}
\includegraphics[scale=0.30, angle=90]{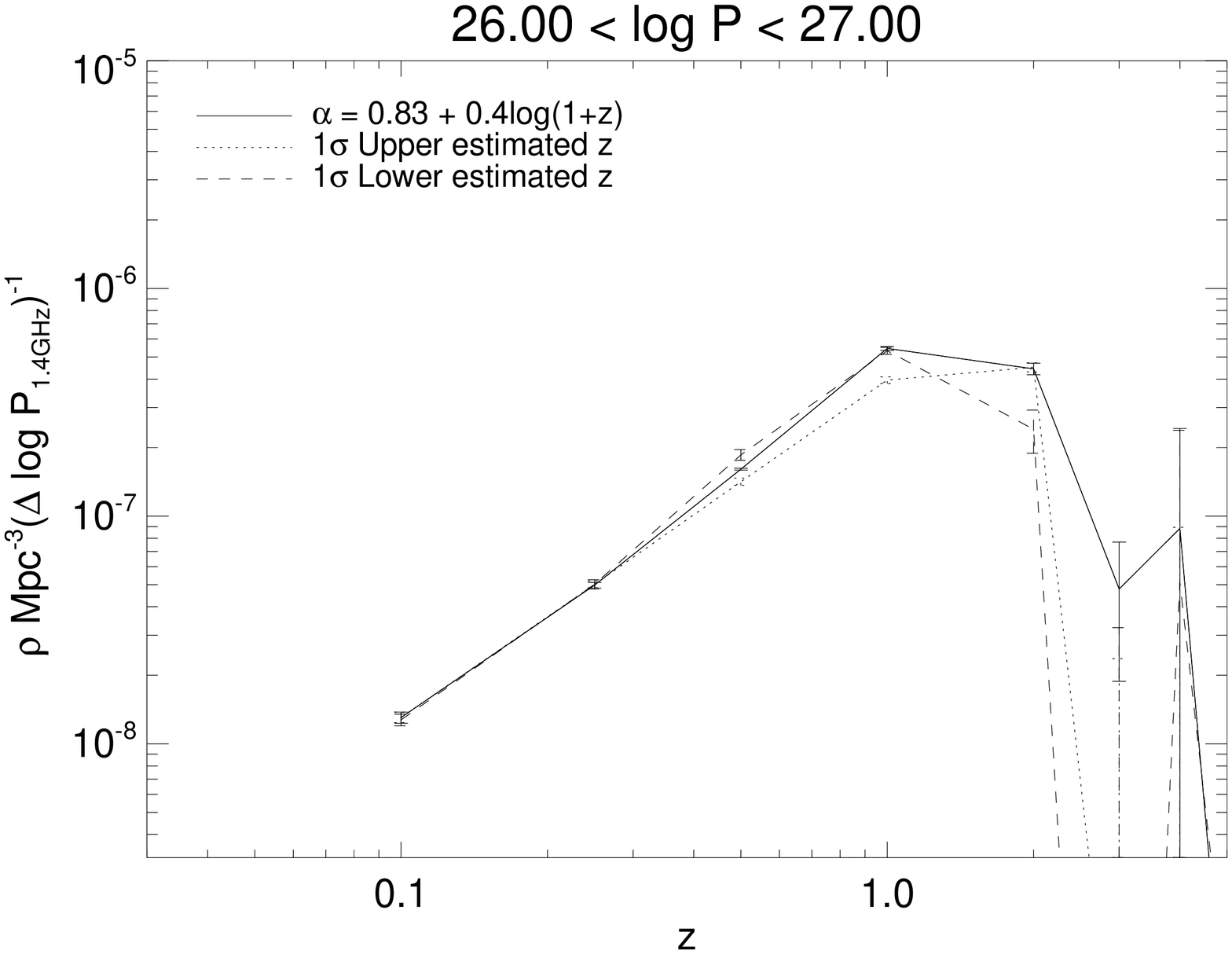}
\includegraphics[scale=0.30, angle=90]{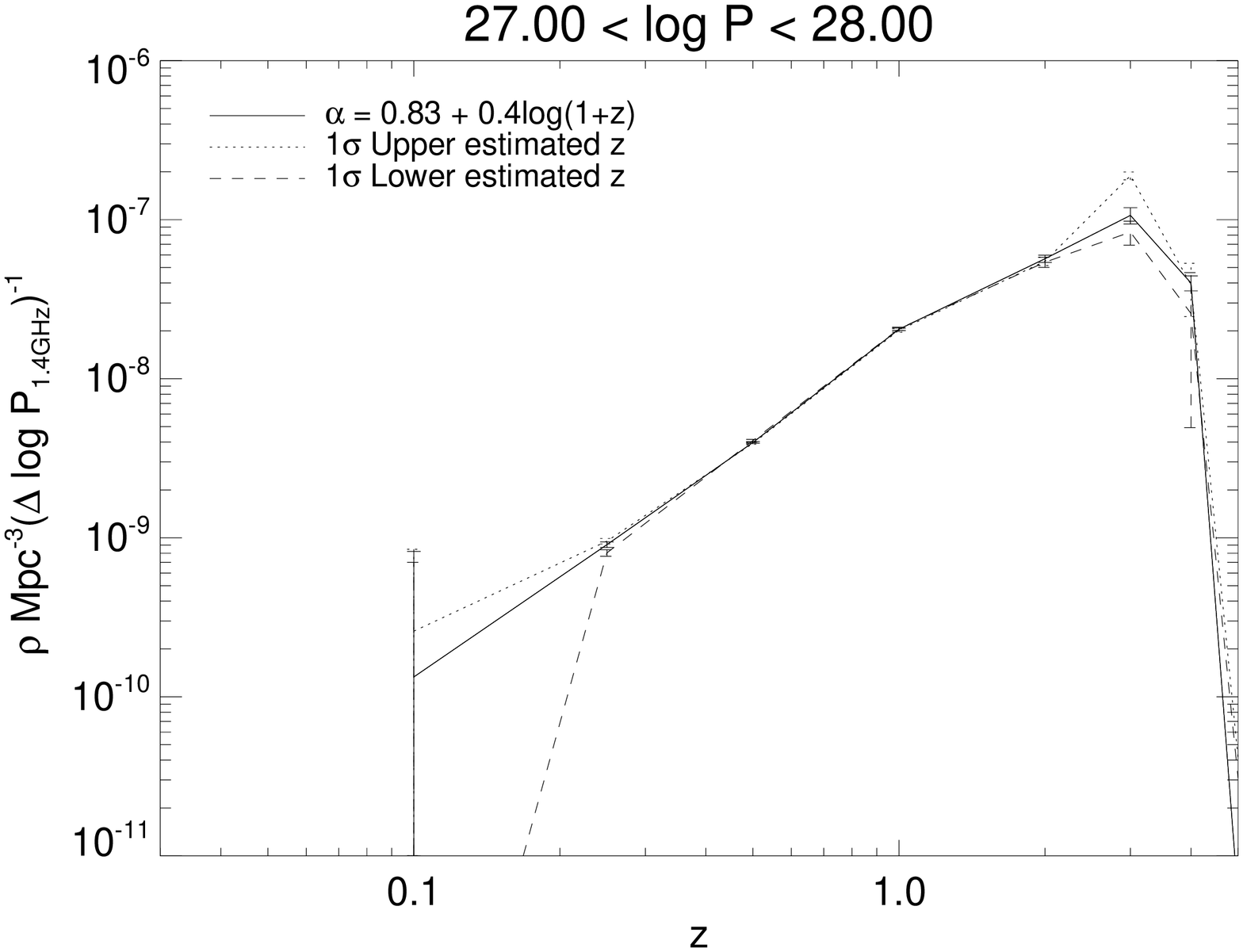}
\caption{\protect\label{z_increase2}
 The effect on the steep--spectrum
  model luminosity functions of moving the estimated redshifts and
  redshift limits up and down by their $1\sigma$ values (these were
  taken as $+$0.4 and $-$0.1 for the limits); no other attempt is made
  to take the redshift uncertainties into account. The `$\alpha = 0.83
  + 0.4\log(1+z)$' line comes from the best fitting model grid
  presented earlier. Larger bins of $\Delta\log P = 1$ are shown to
  better 
  illustrate how the changes affect the results. In all cases an error
  of $\alpha \pm 0.2$ was also included, as previously discussed in
  Section \protect\ref{input_grids}.}
\end{figure}

Figs \ref{z_increase} and \ref{z_increase2} show the spread in densities resulting from
this, and the cut--off strengths can be found in Table
\ref{res_table}. They show that whilst the turnover is preserved in both cases, it
is generally strengthened at moderate powers when all the uncertain redshifts
are shifted by $\pm 1\sigma$, compared with the original results,
because sources are shifted out of these regions to higher power and
redshift bins. When the redshift limits are randomly increased the
turnover is weakened due to sources being moved into the higher
redshift bins. 

\subsection{Testing the robustness of the redshift turnover}

The excellent coverage of the $P$--$z$ plane in the range $26 < \log P
< 28$ to $z \sim 5$, demonstrated in Fig. \ref{pz_plot} and \ref{cons_fig}, allows a
further test of how robust the redshift cut--offs seen in Fig.
\ref{indivs_lfs1} are to possible incompleteness in the radio samples.
This is done by determining the number of fake high--redshift radio
sources that need to be inserted into this luminosity range to reduce
the cut--off strength to $< 3\sigma$. In practice this was split into
two parts because of the changing position of the cut--off with luminosity and the
range of the different samples. Firstly, different numbers of new Hercules sources, with redshifts and luminosities randomly selected from $2.5<z<4$ and $26 <
\log P < 27$, were inserted and the modelling repeated. Next, Hercules was
returned to its original composition and the process repeated for
CENSORS, but this time with extra sources drawn from $3.5 < z < 6$ and
$27 < \log P < 28$. The number of real sources in these two redshift
ranges (2.2 in Hercules, 2.6 in CENSORS, incorporating the spread in estimated redshifts as described
in Section \ref{comp_pred}) is well reproduced by the polynomial fits
used to represent the data, which give 3.0 and 4.2 sources respectively, so adding fake
sources in this way should give a good indication of the number needed
to significantly affect the turnover.

Table \ref{fk_source} gives the resulting cut--off strengths for the average density following
the peak; it indicates that the number of CENSORS or Hercules sources in these ranges has to approximately double to push the significance of the cut--off below 3$\sigma$.
It should also be noted that these numbers are likely to be a lower limit, as the modelling is likely to overpredict the number of real sources at these redshifts. This is because
of the input polynomial used for the fitting, which typically underestimates at lower redshifts, thus leading to overestimation at $z \gsim 3$. 

Both the CENSORS and Hercules samples contain sources without host
galaxy identifications or with no spectroscopic redshifts, so the
extra sources required to remove the turnovers could simply be missing. However, this possibility has
already been considered in Section \ref{z_change}, where it was shown
that moving all estimated redshifts to their upper limits does not
remove the space density declines. 

An alternative determination of the number of fake sources required to
remove the redshift turnovers in these radio power and redshift
ranges can be made by freezing the relevant high--redshift space densities at
their peak values, and then calculating the total number of sources
that would have been detected in the CENSORS or Hercules samples in
the absence of a decline in density. 
These numbers -- 11.7 for Hercules and 12.4 for CENSORS -- are
comparable to the total number of sources present in these bins with the addition
of the fake sources discussed earlier in this Section. 

\begin{table}
\centering
\caption{\protect\label{fk_source} The cut--off strength for the
  average redshift point following the peak for $26 < \log P <
  27$ and $2.5<z<4$ and $27< \log P <28$ and $3.5<z<6$ for the addition of different numbers of fake sources, $N(\rm
  fake)$ into the Hercules and CENSORS sample respectively. The $N(\rm fake) = 0$ value is for the original version of the redshift
  samples.} 
\begin{tabular}{c|c|c}
\hline
& Hercules & CENSORS \\
$N(\rm fake)$  & $C_{\sigma}$ & $C_{\sigma}$ \\
\hline

0 &  6.5 & 5.0  \\
1 &  3.0 & 3.6  \\ 
2 &  2.9 & 2.5  \\
3 &  2.3 & 3.0  \\
4 &  1.1 & 2.9  \\
5 &  0.6 & 2.2  \\
6 &  1.0 & 1.5 \\
7 &  0.1 & 1.2  \\

\hline
\end{tabular}
\end{table}

\subsection{Polynomial approximation of the best--fit $P$--$z$ grid}

The usefulness of the best--fit $P$--$z$ grid (given in Appendix \ref{fullgrid}) to the reader is limited, because it only varies smoothly in regions where it is covered by the available data. 
To improve this situation the whole grid is fit four times with a
fourth order polynomial series expansion, similar to the one used for
the DP90 models. This provides an easy method to calculate the space
density values at any $(z,P)$ point, as well as allowing
extrapolations of the data to regions currently not well constrained.    

\begin{figure}
\centering
\includegraphics[scale=0.35, angle=90]{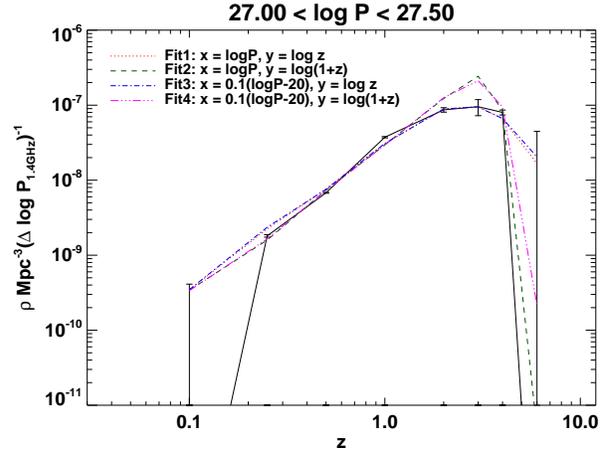}
\caption{\protect\label{sfit_example}  The results of the four smooth
  fits to the $P$--$z$ grid for one radio power range. The full set of
  plots for $\log P = 25.25$ to $\log P = 27.75$ inclusive can also be
  found in Appendix \protect\ref{fullgrid}).}
\end{figure}

The basic polynomial series used for the fit is:
\begin{equation}
\label{fit_eqn}
\log~\rho =   \Bigg (\sum_{i=0}^{4}{ {\sum^{4-i}_{j=0} A_{ij} x(P)^{i} y(z)^{j}}} \Bigg ) + Bx^{5}y^{5}  ,
\end{equation}
where $x$ and $y$ represent the radio power and redshift axes of the
grid respectively. Only points constrained by two of the input
datasets are used in this fitting. The $z=6$ points are also excluded
as the uncertainties in this region are large. The co--efficients for the 
four different versions of this fit are given in Appendix
\ref{fullgrid}. These were chosen to give several extensions into the unconstrained region, and the only difference between them is the
co--ordinates used for $x$ and $y$: 
\begin{itemize}
\item Fit 1: $x = \log P$, $y = \log z$ 
\item Fit 2: $x = \log P$, $y = \log(1+z)$  
\item Fit 3: $x = 0.1(\log P - 20)$, $y = \log z$  
\item Fit 4: $x = 0.1(\log P - 20)$, $y = \log(1+z)$  .
\end{itemize}
The results of the fits for one radio power range are shown in Fig.
\ref{sfit_example} (the full set of plots for $\log P = 25.25$ to $\log P = 27.75$ inclusive can also be found in
Appendix \ref{fullgrid}) and the agreement with the best--fit grid is
good. However, it should be stressed that this smooth version of the grid
is not a perfect representation of the model output and this is why it
was not used for the analyses in the previous subsections. 

\section{Summary and conclusions}
\label{conc}

The results presented in this paper demonstrate that the method of RLF determination described here works
well; it gives an easy means of estimating errors and hence assessing the
robustness of any evolutionary behaviour seen, such as the presence of
a redshift turnover. Examination of the best--fitting steep--spectrum $P$--$z$ grid suggests that the
turnover in the radio luminosity function occurs at $z \simeq 0.7$ for the
faintest luminosities considered here, and then moves to $z \gsim 2$  for
higher powers. These changes are consistent with those seen for
steep--spectrum radio sources by \citet{waddington01} who found
turnovers in redshift at $z \simeq 1$ for low--luminosity sources
($P_{\rm 1.4 GHz} > 10^{25.1}$ W/Hz) but at $z \simeq 2$ for the more
powerful ($P_{\rm 1.4 GHz} > 10^{26-27}$ W/Hz). Similarly a redshift
peak at $z \simeq 2$ for the brightest sources is also seen for
flat--spectrum quasars, in radio--loud, optical and X--ray selected
samples \citep[e.g.][]{zotti09, wall08, hasinger05, richards05}. They are also
in broad agreement with the assumption of a low luminosity peak at
$z=0.706$, and a high luminosity peak at $z=1.91$, in the simulations
of \citet{wilman}, which make predictions for the next generation of
radio telescopes. The results presented here nevertheless suggest that a
luminosity--dependent peak, with a high--redshift decline, would be a
better representation of the real data than the two population model,
with a flat post--peak space density,  that these simulations
currently adopt.  

Physically, the luminosity dependence of the redshift peak in the
radio galaxy RLF suggests that the most massive black holes have formed by $z
\simeq 4$ and that their lower mass counterparts formed later.
This `cosmic downsizing' may initially appear to be at variance with the hierarchical model of structure formation. However,
this discrepancy can be solved if the mode of AGN fuelling changes
with cosmic time: in the early Universe major mergers provide the cold
gas to power the accretion at high rates, but at lower redshift it is low--luminosity, radiatively
inefficient accretion from hot gas haloes that dominates \citep{fanidakis}.

The datasets available for this work mean that only a narrow range in
luminosity is constrained well enough to draw firm conclusions about
the luminosity function evolution. Better coverage of the $P$ --$z$
plane will improve this. However, the density turnovers are robust and remain present even in the unlikely scenario that
all the estimates are 1 $\sigma$ higher or lower. The turnovers seen
in this work are also in good agreement with the work of
\citet{wall05} who find a cut--off at a significance level  $>
4\sigma$ for their sample of flat spectrum quasars.  

The agreement with the DP90 results seen at high redshift for the 
brightest luminosities is not surprising; this region is dominated by 
CENSORS sources, which have already been shown to be consistent with 
two of their models (Paper III), both of which include a redshift 
cut--off. 

The modelling method presented here can be easily modified to
investigate different populations separately. The modelling results in
this paper are limited by the uncertain redshifts present in some
samples and further spectroscopic observations for the CENSORS sample
are ongoing to improve this situation. Future increases in sample size would allow independent
minimisation of all three grids, as well as subdividing the grids
further into additional populations, e.g. Fanaroff \& Riley Class I
and II galaxies \citep{fanaroff}, or high and low excitation line
radio sources \citep[e.g.][]{hardcastle}. The upcoming large, deep radio surveys from both the LOw Frequency ARray (LOFAR) and the Australian Square Kilometre Array Pathfinder (ASKAP) will be ideal for this, but complementary redshift data, using the deep multicolour optical photometry from the planned Large Synoptic Survey Telescope (LSST) for example, is essential. Such an extension of this work would yield an invaluable tool for investigating the links between the different AGN subspecies. 

\section*{Acknowledgments}

The authors are grateful to Vernesa Smol\v ci\'c for kindly
  providing the VLA--COSMOS data. PNB is grateful for support from the Leverhulme Trust. JSD
acknowledges the support of the Royal Society via a Wolfson Research
Merit award, and also the support of the European Research Council via
the award of an Advanced Grant. MHB acknowledges the support of a
PPARC studentship. 

\setlength{\bibhang}{\parindent}
\setlength\labelwidth{0.0em}

\clearpage
\onecolumn
\appendix

\section{Additional Spectroscopic Observations of Wall \& Peacock 1985 radio sources}
\label{morespecs_notcen}

\begin{figure*}
\begin{tabular}{cc}
\scriptsize{0407-65: z = 0.962}  &\scriptsize{1308-22: z = 0.005} \\
\psfig{file=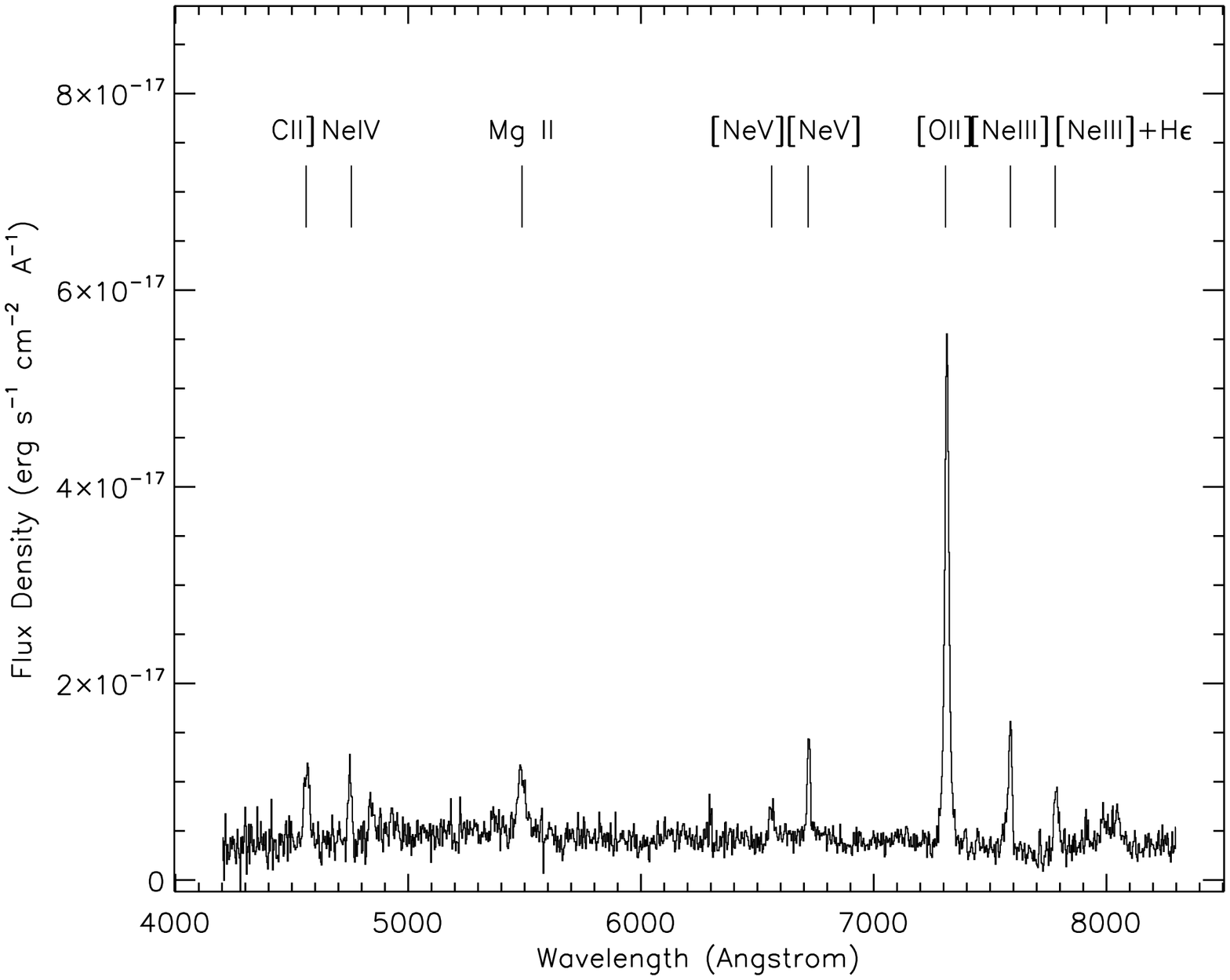,angle=90,width=6.0cm,clip=} &
\psfig{file=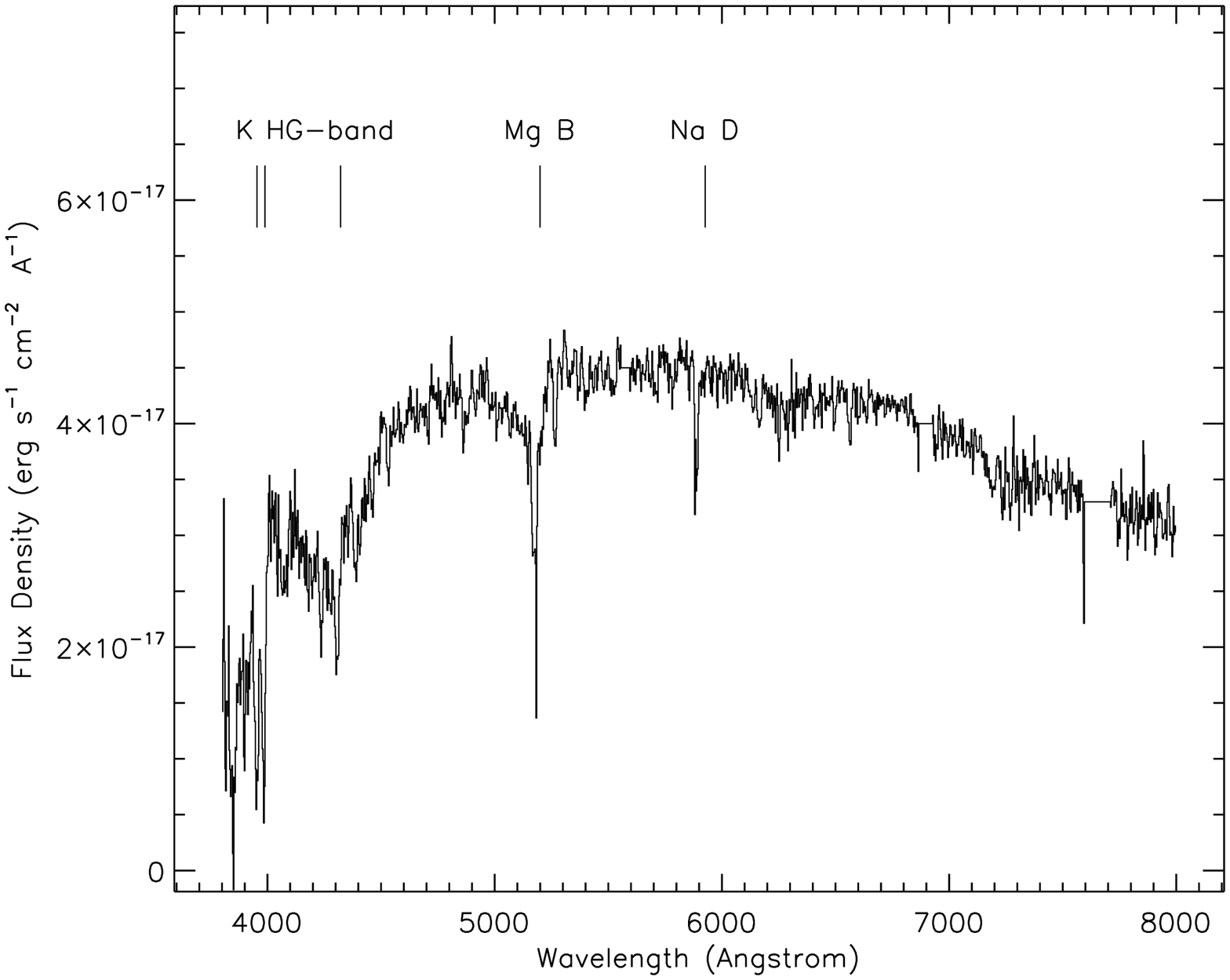,angle=90,width=6.0cm,clip=} \\
\end{tabular}
\caption{New  spectroscopic observations for WP85 sources.\label{other_spec}}
\end{figure*}

As part of the back-up program for the CENSORS observations, several
radio galaxies were targeted from the other samples used in the radio
luminosity function analysis. 
During the 2006 FORS2 run at the VLT, 0407-65 and 1308-22, from the
WP85 sample, were observed with a 1.5\arcsec slit and the 300V
grism. Redshifts of 0.962 and 0.005 were found respectively. 
The spectra are presented in Fig. \ref{other_spec} and the results are presented in Table \ref{other_spec_tab}.

\begin{table}
\centering
\caption[Results of spectroscopy observing for the WP85
  sample]{Results of spectroscopy observing for WP85
  sources.\label{other_spec_tab}} 
\begin{tabular}{|l|l|l|l|l|l|l|l|l|l|l|} \hline 
Source 	&RA ~~~~~~~~~~~~~~~DEC         	&Exp. 	 	&Slit PA
      	&$z$	&$\Delta z$ 	&Line	&$\lambda_{\rm obs}$
      	& Flux	&$\Delta v_{\rm FWHM}$	&Eq. Width
      	\\ 
      	& 				&Time (s)	&(E of N)
      	&	&	   	&	&\AA~
      	&$10^{-16}$ erg/s/cm$^{2}$		&kms$^{-1}$		&
      	\\ 
\hline
0407-65	&04 08 20.4  $-65$ 45 09	&3600		&+0
      	&0.962	&0.001		&CII]	&4564			&2.12
      	$\pm$ 0.24	&1423 $\pm$ 427		&28 $\pm$ 4	\\ 
	&				&		&
      	&	&		&NeIV 	&4749   		&0.77
      	$\pm$ 0.1   	&-       		&9 $\pm$ 1	\\
      	
	&				&		&
      	&	&		&MgII 	&5486   		&2.71
      	$\pm$ 0.29   	&2042 $\pm$ 437      	&29	$\pm$ 3	\\ 
	&				&		&
      	&	&		&[NeV] 	&6562   		&0.70
      	$\pm$ 0.1    	&564 $\pm$ 318       	&8 $\pm$ 1	\\ 
	&				&		&
      	&	&		&[NeV] 	&6721   		&1.86
      	$\pm$  0.19    	&305 $\pm$   204     	&23 $\pm$    2	\\ 
	&				&		&
      	&	&		&[OII] 	&7314   		&12.1
      	$\pm$  1.2    	&731 $\pm$   182     	&157 $\pm$  16	\\ 
	&				&		&
      	&	&		&[NeIII] &7586   		&2.59
      	$\pm$  0.27    	&489 $\pm$   182      	&40 $\pm$   4	\\ 
	&				&		&
      	&	&		&[NeIII]+ &7784   		&0.96
      	$\pm$  0.12    	&431 $\pm$   205      	&13 $\pm$   2	\\ 
1308-22	&13 11 40.1  $-22$ 17 04	&	3600	 & +0
      	&0.005	&0.001		&-	&-			&-
	&-			&-		\\ 
\hline
\end{tabular}
\end{table}

\section{The CENSORS data table}
\label{full_cen_table}

This Appendix provides an up--to--date summary of the CENSORS data. It draws together the key properties presented in Papers I, II and III, updating these where relevant. Column (1) gives the CENSORS ID number; 
(2) the redshift, estimated redshift or redshift limit; 
(3) the redshift type -- 1 $=$ spectroscopic, 2 $=$ $K$--$z$ estimate, 3 $=$ $K$--$z$ limit, 4 $=$ $I$--$z$ estimate; 
(4) the object class -- 0 $=$ AGN, 1 $=$ quasar, 2 $=$ starburst galaxy; 
(5) and (6)  Radio position from the VLA observations described in Paper I;  
(7) and (8)  1.4 GHz radio flux density, and associated error, taken from the NVSS;
(9) Radio morphology -- S$=$single, D$=$double, T$=$triple, M$=$multiple, E$=$extended diffuse;
(10) Largest angular size of the radio source;
(11) and (12) host galaxy position, taken from the $K$--band imaging if present, optical $I$--band imaging if not; 
(13) aperture radius used to measure $K$--magnitude; 
(14) and (15) $K$--magnitude, and associated error, corrected to the standard 63.9 kpc aperture; 
(16) EISD name; 
(17) and (18) $I$--band magnitude and associated error; 
(19) and (20) $V$--band magnitude and associated error; 
(21) and (22) $B$--band magnitude and associated error. 

CENSORS 58 has been removed from the sample due to its proximity to a bright star; `CENSORS 66' represents the combination of CENSORS 66 and 82; `CENSORS 84' represents the combination of CENSORS 84 and 85. 


\begin{landscape}

\begin{table*}
\scriptsize
\begin{tabular}{cccccccccccccccccccccc}
\hline 
 CENSORS & z         & T &   C &      \multicolumn{2}{c}{Radio Position}    &$S_{\rm 1.4GHz}$&  $S_{\rm err}$   & Morph. & $D_{\rm rad}$ &     \multicolumn{2}{c}{Host Position}         & ap. used&  K   & K$_{\rm err}$    &  EISD     &   I    &   I$_{\rm err}$   &    V    &   V$_{\rm err}$   &  B    &   B$_{\rm err}$    \\    
         &           &      &         &  RA               &    DEC           &  &        &  &&    RA  &     DEC   & for corr&  (ap cor)&      &  name     &           &          &            &          &          &           \\
         &           &       &         &      J2000           &    J2000           &  mJy  &  mJy   &  & \arcsec &    J2000  &     J2000    & arcsec  &  mag   & mag     &           &           &          &            &          &          &           \\       
(1)         &   (2)        &   (3)    &   (4)      &      (5)           &    (6)           &  (7)  &  (8)   & (9) & (10) &   (11) &    (12)    & (13)  & (14)   & (15)     &  (16)         & (17)          &  (18)        &   (19)         &   (20)       &   (21)       &   (22)        \\       
\hline                                                                                   
     1    & 1.155   &    1   &     0 &    09  51  29.07 &  -20 50 30.1 &  659.5 & 19.8  &      D   &   5.0    & 09 51 29.19  & -20 50 30.9  &   1.5  &       17.78  & 0.30  &  EISD  1      & 21.74   &   0.10  &     22.93   &   0.09  &   23.27   &   0.09        \\     	  
     2    & 0.913   &    1   &     0 &    09  46  50.21 &  -20 20 44.4 &  452.3 & 13.6  &      S   &   0.9    & 09 46 50.20  & -20 20 44.0  &   1.0  &       18.47  & 0.19  &  EISD  2      & 22.59   &   0.18  &             &         &   24.02   &   0.17        \\   	  
     3    & 0.790   &    1   &     0 &    09  50  31.39 &  -21 02 44.8 &  355.3 & 10.7  &      S   &   0.7    & 09 50 31.41  & -21 02 44.3  &   2.5  &       16.47  & 0.15  &  EISD  3      & 20.60   &   0.20  &     23.11   &   0.11  &   23.77   &   0.16        \\   	  
     4    & 1.013   &    1   &     0 &    09  49  53.60 &  -21 56 18.4 &  283.0 &  9.5  &      T   &   29.5   & 09 49 53.30  & -21 56 20.7  &   2.5  &       17.65  & 0.27  &  EISD  6      & 21.30   &   0.08  &     23.36   &   0.12  &   23.41   &   0.13        \\   	  
     5    & 2.588   &    1   &     0 &    09  53  44.42 &  -21 36 02.5 &  244.7 &  8.2  &      T   &   31.7   & 09 53 44.52  & -21 36 01.1  &   1.5  &       19.01  & 0.34  &  EISD  8      & 22.15   &   0.12  &     22.73   &   0.06  &   22.57   &   0.15        \\   	  
     6    & 0.547   &    1   &     1 &    09  51  43.63 &  -21 23 58.0 &  239.7 &  1.3  &      S   &   1.8    & 09 51 43.61  & -21 23 58.6  &   2.5  &       16.18  & 0.16  &  EISD  7      & 18.26   &   0.01  &     18.65   &   0.01  &   18.40   &   0.01        \\   	  
     7    & 1.437   &    1   &     1 &    09  45  56.71 &  -21 16 54.4 &  148.2 &  5.1  &      M   &   32.2   & 09 45 56.69  & -21 16 53.5  &   2.5  &       17.79  & 0.33  &  EISD 10      & 22.62   &   0.17  &     23.54   &   0.19  &   23.38   &   0.11        \\   	  
     8    & 0.271   &    1   &     0 &    09  57  30.07 &  -21 30 59.8 &  126.3 &  3.8  &      S   &   5.8    & 09 57 30.06  & -21 30 58.9  &   4.5  &       15.10  & 0.10  &  EISD 11      & 17.77   &   0.01  &     19.38   &   0.02  &   20.35   &   0.03        \\   	  
     9    & 0.242   &    1   &     0 &    09  49  35.43 &  -21 56 23.5 &  118.2 &  3.6  &      S   &  $<$0.6    & 09 49 35.54  & -21 56 24.1  &   2.5  &       15.74  & 0.25  &  EISD 12      & 18.28   &   0.01  &     20.08   &   0.01  &   20.96   &   0.02        \\   	  
    10    & 1.074   &    1   &     1 &    09  47  26.99 &  -21 26 22.6 &   79.4 &  2.9  &      T   &   66.5   & 09 47 26.99  & -21 26 33.4  &   2.5  &       15.98  & 0.07  &  EISD 16      & 18.26   &   0.01  &     18.73   &   0.01  &   19.15   &   0.01        \\   	  
    11    & 1.589   &    1   &     1 &    09  53  29.51 &  -20 02 12.5 &   78.1 &  2.4  &      S   &  $<$0.6    & 09 53 29.56  & -20 02 12.0  &   1.5  &       18.56  & 0.37  &  EISD 15      & 21.75   &   0.07  &     22.49   &   0.05  &   22.74   &   0.05        \\   	  
    12    & 0.821   &    1   &     0 &    09  46  41.13 &  -20 29 27.3 &   70.4 &  2.6  &      S   &   1.8    & 09 46 41.17  & -20 29 26.2  &   1.5  &       18.74  & 0.31  &  EISD 18      & 21.88   &   0.09  &     23.35   &   0.15  &   23.90   &   0.19        \\   	  
    13    & 2.950   &    1   &     0 &    09  54  28.97 &  -21 56 55.0 &   66.3 &  2.7  &      S   &   2.1    & 09 54 29.00  & -21 56 54.9  &   1.5  &       19.49  & 0.21  &  EISD 20      & 23.89   &   0.26  &             &         &           &               \\   	  
    14    & 1.415   &    2   &     0 &    09  54  47.66 &  -20 59 43.8 &   65.6 &  2.4  &      D   &   10.0   & 09 54 47.65  & -20 59 44.0  &   1.5  &       18.23  & 0.21  &  EISD 21      & 23.60   &   0.25  &             &         &           &               \\   	  
    15    & 1.395   &    2   &     0 &    09  46  51.12 &  -20 53 17.8 &   63.0 &  1.9  &      D   &   6.1    & 09 46 50.99  & -20 53 18.2  &   2.5  &       18.20  & 0.11  &  EISD 22      & 20.57   &   0.06  &     20.91   &   0.04  &   21.45   &   0.03        \\   	  
    16    & 3.126   &    1   &     0 &    09  57  51.42 &  -21 33 24.2 &   61.7 &  2.3  &      D   &   13.1   & 09 57 51.41  & -21 33 22.5  &   1.0  &       19.32  & 0.41  &  EISD 23      &         &         &             &         &           &               \\   	  
    17    & 0.893   &    1   &     0 &    09  52  42.95 &  -19 58 20.4 &   61.5 &  2.3  &      D   &   11.2   & 09 52 43.11  & -19 58 21.9  &   2.5  &       17.84  & 0.30  &  EISD 24      & 21.38   &   0.07  &     23.63   &   0.14  &           &               \\   	  
    18    & 0.109   &    1   &     0 &    09  55  13.60 &  -21 23 03.1 &   58.3 &  1.8  &      S   &   0.8    & 09 55 13.59  & -21 23 02.9  &   4.5  &       12.45  & 0.23  &  EISD 25      & 14.88   &   0.01  &     16.13   &   0.01  &   16.96   &   0.01        \\   	  
    19    & 1.235   &    2   &     0 &    09  53  30.69 &  -21 35 50.0 &   55.1 &  2.1  &      M   &   23.9   & 09 53 30.52  & -21 36 02.8  &   1.5  &       17.94  & 0.22  &  EISD 27      &         &         &             &         &           &               \\   	  
    20    & 1.377   &    1   &     0 &    09  46  04.75 &  -21 15 11.4 &   54.2 &  2.1  &      D   &   7.1    &              &              &   2.5  &       $>$19.6  &       &  EISD 30      &         &         &             &         &           &               \\   	  
    21    & 1.247   &    2   &     0 &    09  47  58.94 &  -21 21 50.9 &   54.0 &  1.7  &      S   &  $<$1.0    & 09 47 59.02  & -21 21 51.7  &   1.5  &       17.96  & 0.28  &  EISD 28      & 22.11   &   0.10  &     24.38   &   0.19  &   24.52   &   0.14        \\   	  
    22    & 0.984   &    2   &     0 &    09  57  30.92 &  -21 32 39.5 &   52.9 &  1.7  &      D   &   4.6    & 09 57 30.83  & -21 32 39.2  &   1.5  &       17.45  & 0.25  &  EISD 29      &         &         &             &         &           &               \\   	  
    23    & 1.734   &    2   &     0 &    09  56  30.01 &  -20 01 31.0 &   52.4 &  2.0  &      D   &   21.7   & 09 56 29.93  & -20 01 32.5  &   1.5  &       18.66  & 0.28  &  EISD 31      &         &         &             &         &           &               \\   	  
    24    & 3.431   &    1   &     0 &    09  54  38.33 &  -21 04 25.1 &   51.0 &  1.6  &      S   &   1.4    & 09 54 38.32  & -21 04 24.5  &   1.5  &       19.30  & 0.31  &  EISD 32      &         &         &             &         &           &               \\   	  
    25    & 1.793   &    2   &     0 &    09  48  04.05 &  -21 47 36.8 &   49.2 &  1.9  &      S   &  $<$0.7    & 09 48 04.06  & -21 47 36.1  &   2.5  &       18.73  & 0.24  &  EISD 34      &         &         &             &         &           &               \\   	  
    26    & 4.45    &    3   &     0 &    09  52  17.69 &  -20 08 36.2 &   44.4 &  1.4  &      S   &   2.1    &              &              &   2.5  &       $>$20.6  &       &  EISD 36      &         &         &             &         &           &               \\   	  
    27    & 0.423   &    1   &     0 &    09  51  49.78 &  -21 24 57.7 &   40.4 &  2.3  &      M   &   115.2  & 09 51 49.84  & -21 24 58.1  &   4.5  &       15.78  & 0.23  &  EISD 44      & 18.29   &   0.01  &     20.12   &   0.02  &   21.14   &   0.02        \\   	  
    28    & 0.472   &    1   &     0 &    09  46  31.32 &  -20 26 07.2 &   40.1 &  1.9  &      T   &   17.6   & 09 46 32.14  & -20 26 15.4  &   4.5  &       15.91  & 0.11  &  EISD 38      & 19.02   &   0.02  &     20.86   &   0.04  &   21.59   &   0.03        \\   	  
    29    & 0.965   &    1   &     1 &    09  48  15.71 &  -21 40 06.3 &   38.2 &  1.6  &      M   &   27.6   & 09 48 15.81  & -21 40 07.0  &   2.5  &       17.29  & 0.23  &  EISD 39      & 18.73   &   0.01  &     19.02   &   0.01  &   19.23   &   0.01        \\   	  
    30    & 0.108   &    1   &     0 &    09  45  55.86 &  -20 28 30.2 &   37.8 &  2.0  &      T   &   50.1   & 09 45 55.92  & -20 28 29.7  &   2.0  &       13.20  & 0.10  &  EISD 40      & 16.41   &   0.00  &     18.58   &   0.01  &   19.50   &   0.05        \\   	  
    31    & 2.47    &    3   &     0 &    09  45  19.60 &  -21 42 43.8 &   37.3 &  1.5  &      D   &   28.5   &              &              &   2.5  &       $>$19.4  &       &  EISD 41      &         &         &             &         &           &               \\   	  
    32    & 1.151   &    1   &     0 &    09  51  41.02 &  -20 11 18.4 &   35.3 &  1.5  &      D   &   36.3   & 09 51 40.85  & -20 11 16.1  &   1.5  &       17.56  & 0.23  &  EISD 43      & 22.25   &   0.12  &             &         &           &               \\   	  
    33    & 1.203   &    1   &     0 &    09  53  04.71 &  -20 44 09.8 &   34.3 &  1.1  &      D   &   23.2   & 09 53 05.00  & -20 44 13.9  &   2.5  &       18.75  & 0.27  &  EISD 45      &         &         &             &         &           &               \\   	  
    34    & 1.325   &    2   &     0 &    09  47  53.55 &  -21 47 19.6 &   34.2 &  1.1  &      S   &  $<$0.9    & 09 47 53.59  & -21 47 19.3  &   1.0  &       18.09  & 0.32  &  EISD 47      & 24.95   &   0.63  &             &         &           &               \\   	  
    35    & 0.473   &    1   &     0 &    09  54  52.43 &  -21 19 29.0 &   34.1 &  1.4  &      D   &   12.2   & 09 54 52.47  & -21 19 29.5  &   4.5  &       16.46  & 0.20  &  EISD 48      & 18.71   &   0.01  &     21.11   &   0.03  &   22.12   &   0.04        \\   	  
    36    & 1.485   &    1   &     0 &    09  49  33.23 &  -21 27 08.3 &   32.3 &  1.1  &      S   &   0.7    & 09 49 33.32  & -21 27 06.8  &   1.5  &       18.54  & 0.27  &  EISD 51      & 22.90   &   0.17  &             &         &   24.74   &   0.16        \\   	  
    37    & 0.511   &    1   &     1 &    09  49  19.44 &  -21 51 35.4 &   31.8 &  1.4  &      T   &   21.3   & 09 49 19.55  & -21 51 33.9  &   1.5  &       19.45  & 0.26  &  EISD 52      & 23.67   &   0.19  &     24.55   &   0.26  &   24.52   &   0.20        \\   	  
    38    & 2.116   &    1   &     1 &    09  51  16.77 &  -20 56 38.4 &   31.7 &  1.1  &      S   &   3.4    & 09 51 16.89  & -20 56 37.0  &   2.5  &       17.25  & 0.15  &  EISD 53      & 19.12   &   0.02  &     19.56   &   0.01  &   19.57   &   0.01        \\   	  
    39    & 1.572   &    1   &     1 &    09  48  35.99 &  -21 06 22.6 &   31.5 &  1.1  &      D   &   6.4    & 09 48 36.18  & -21 06 22.4  &   2.5  &       17.63  & 0.33  &  EISD 54      & 20.60   &   0.03  &     21.73   &   0.03  &   21.61   &   0.02        \\   	  
    40    & 1.158   &    1   &     0 &    09  50  58.63 &  -21 14 20.3 &   30.9 &  1.3  &      D   &   11.3   & 09 50 58.98  & -21 14 23.8  &   4.5  &       18.06  & 0.30  &  EISD 55      & 21.54   &   0.10  &     23.18   &   0.13  &   23.66   &   0.12        \\   	  
    41    & 0.295   &    1   &     0 &    09  49  18.18 &  -20 54 45.4 &   27.5 &  1.7  &      T   &   42.2   & 09 49 18.23  & -20 54 46.1  &   4.5  &       14.89  & 0.21  &  EISD 58      & 17.10   &   0.01  &     18.92   &   0.03  &   20.09   &   0.01        \\   	  
    42    & 1.254   &    1   &     0 &    09  52  01.86 &  -21 15 52.3 &   26.5 &  0.9  &      D   &   18.2   & 09 52 01.59  & -21 15 53.0  &   4.5  &       19.30  & 0.20  &  EISD 60      & 23.70   &   0.19  &             &         &   25.06   &   0.28        \\   	  
    43    & 0.778   &    1   &     0 &    09  52  59.17 &  -21 48 42.4 &   26.4 &  0.9  &      D   &   5.7    & 09 52 59.15  & -21 48 41.7  &   1.5  &       17.15  & 0.24  &  EISD 64      & 20.93   &   0.05  &     24.37   &   0.14  &           &               \\   	  
    44    & 0.790   &    1   &     1 &    09  54  27.06 &  -20 29 46.5 &   26.1 &  0.9  &      S   &  $<$1.1    & 09 54 27.08  & -20 29 46.5  &   2.5  &       17.98  & 0.30  &  EISD 62      & 19.85   &   0.02  &     20.48   &   0.01  &   20.80   &   0.02        \\   	  
    45    & 0.796   &    1   &     0 &    09  57  42.91 &  -20 06 36.1 &   25.5 &  1.2  &      S   &   6.2    & 09 57 42.98  & -20 06 36.8  &   2.5  &       16.84  & 0.13  &  EISD 66      & 20.50   &   0.04  &     23.33   &   0.10  &   24.16   &   0.15        \\   	  
    46    & 0.718   &    1   &     0 &    09  54  03.02 &  -20 25 13.2 &   25.2 &  0.9  &      S   &  $<$1.1    & 09 54 03.06  & -20 25 12.9  &   2.5  &       16.94  & 0.13  &  EISD 65      & 20.39   &   0.05  &     22.70   &   0.10  &   23.53   &   0.13        \\   	  
    47    & 0.508   &    1   &     0 &    09  47  03.32 &  -20 50 02.2 &   25.2 &  0.9  &      D   &   9.0    & 09 47 03.36  & -20 50 00.7  &   4.5  &       16.45  & 0.21  &  EISD 63      & 19.34   &   0.02  &     21.08   &   0.02  &   21.82   &   0.04        \\   	  
    48    & 1.606   &    1   &     1 &    09  54  28.28 &  -20 39 26.6 &   24.2 &  0.9  &      S   &   1.4    & 09 54 28.38  & -20 39 28.1  &   2.5  &       17.52  & 0.26  &  EISD 68      & 19.48   &   0.02  &     20.44   &   0.01  &   20.41   &   0.01        \\   	  
    49    & 0.410   &    1   &     0 &    09  53  23.18 &  -20 13 43.5 &   23.8 &  0.9  &      S   &  $<$1.0    & 09 53 23.25  & -20 13 44.8  &   4.5  &       15.78  & 0.25  &  EISD 67      & 19.20   &   0.02  &     20.57   &   0.16  &   20.55   &   0.01        \\   	  
    50    & 1.529   &    1   &     0 &    09  52  12.71 &  -21 02 36.3 &   22.3 &  0.8  &      D   &   5.0    & 09 52 12.71  & -21 02 36.5  &   1.5  &       18.39  & 0.27  &  EISD 69      &         &         &             &         &           &               \\   	  
    51    & 2.267   &    2   &     0 &    09  51  22.89 &  -21 51 55.1 &   21.7 &  0.8  &      D   &   5.8    & 09 51 22.98  & -21 51 53.4  &   1.0  &       19.22  & 0.43  &  EISD 75      &         &         &             &         &           &               \\   	  
    52    & 1.625   &    1   &     0 &    09  45  42.64 &  -21 15 44.9 &   21.7 &  0.8  &      S   &  $<$1.2    & 09 45 42.60  & -21 15 43.8  &   1.0  &       18.43  & 0.22  &  EISD 72      &         &         &             &         &           &               \\   	  
    53    & 0.426   &    1   &     0 &    09  51  32.40 &  -21 00 29.6 &   21.6 &  1.1  &      D   &   10.9   & 09 51 32.44  & -21 00 29.1  &   2.5  &       15.24  & 0.22  &  EISD 76      & 17.50   &   0.01  &     19.89   &   0.02  &   20.95   &   0.05        \\   	  
    54    & 0.410   &    1   &     0 &    09  53  20.56 &  -21 43 59.2 &   21.4 &  0.8  &      S   &  $<$2.3    & 09 53 20.67  & -21 43 59.2  &   4.5  &       14.15  & 0.11  &  EISD 74      & 15.75   &   0.01  &     17.43   &   0.01  &   18.59   &   0.01        \\   	  
    55    & 0.557   &    1   &     0 &    09  49  30.56 &  -20 23 34.2 &   21.4 &  0.8  &      D   &   14.0   & 09 49 30.80  & -20 23 34.5  &   4.5  &       16.64  & 0.15  &  EISD 71      & 19.61   &   0.03  &     22.00   &   0.18  &   22.94   &   0.11        \\   	  
    56    & 1.483   &    1   &     0 &    09  50  43.20 &  -21 26 40.7 &   20.8 &  1.1  &      D   &   20.4   & 09 50 43.20  & -21 26 42.6  &   2.5  &       17.84  & 0.31  &  EISD 78      & 22.58   &   0.14  &             &         &           &               \\   	  
\hline
\end{tabular}
\end{table*} 

\begin{table*}
\scriptsize
\begin{tabular}{cccccccccccccccccccccc}
\hline 
 CENSORS & z         & T &   C &      \multicolumn{2}{c}{Radio Position}    &$S_{\rm 1.4GHz}$&  $S_{\rm err}$   & Morph. & $D_{\rm rad}$ &     \multicolumn{2}{c}{Host Position}         & ap. used&  K   & K$_{\rm err}$    &  EISD     &   I    &   I$_{\rm err}$   &    V    &   V$_{\rm err}$   &  B    &   B$_{\rm err}$    \\    
         &           &      &         &  RA               &    DEC           &  &        &  &&    RA  &     DEC   & for corr&  (ap cor)&      &  name     &           &          &            &          &          &           \\
         &           &       &         &      J2000           &    J2000           &  mJy  &  mJy   &  & \arcsec &    J2000  &     J2000    & arcsec  &  mag   & mag     &           &           &          &            &          &          &           \\       
(1)         &   (2)        &   (3)    &   (4)      &      (5)           &    (6)           &  (7)  &  (8)   & (9) & (10) &   (11) &    (12)    & (13)  & (14)   & (15)     &  (16)         & (17)          &  (18)        &   (19)         &   (20)       &   (21)       &   (22)        \\       
\hline                                                                                   
    57    & 1.223   &    2   &     0 &    09  51  21.02 &  -21 29 55.4 &   20.7 &  1.1  &      D   &   22.5   & 09 51 21.08  & -21 29 54.4  &   4.5  &       17.93  & 0.26  &  EISD 80      & 22.19   &   0.13  &             &         &           &               \\   	  
    59    & 1.070   &    1   &     0 &    09  48  42.44 &  -21 52 24.8 &   19.1 &  1.1  &      T   &   33.1   & 09 48 42.49  & -21 52 24.6  &   1.5  &       17.91  & 0.28  &  EISD 81      & 21.69   &   0.10  &             &         &           &               \\   	  
    60    & 1.540   &    2   &     0 &    09  51  48.66 &  -20 31 52.9 &   18.9 &  0.7  &      S   &  $<$0.8    & 09 51 48.71  & -20 31 53.2  &   2.5  &       18.41  & 0.27  &  EISD 83      & 23.48   &   0.26  &             &         &           &               \\   	  
    61    & 1.422   &    2   &     0 &    09  48  01.87 &  -20 09 11.4 &   18.5 &  0.7  &      D   &   21.3   & 09 48 01.98  & -20 09 11.9  &   1.5  &       18.24  & 0.26  &  EISD 82      &         &         &             &         &           &               \\   	  
    62    & 0.574   &    1   &     0 &    09  49  45.67 &  -21 50 06.2 &   18.4 &  0.7  &      D   &   17.7   & 09 49 45.90  & -21 50 06.3  &   1.5  &       16.81  & 0.28  &  EISD 84      & 20.25   &   0.03  &     22.94   &   0.07  &   24.30   &   0.14        \\   	  
    63    & 0.314   &    1   &     0 &    09  45  29.51 &  -21 18 50.5 &   18.3 &  0.7  &      D   &   6.3    & 09 45 29.64  & -21 18 51.5  &   4.5  &       16.08  & 0.25  &  EISD 88      & 17.96   &   0.01  &     19.96   &   0.01  &   20.89   &   0.01        \\   	  
    64    & 0.403   &    1   &     0 &    09  49  00.07 &  -20 50 08.4 &   18.1 &  1.0  &      D   &   54.1   & 09 49 01.60  & -20 50 00.7  &   6.0  &       15.04  & 0.08  &  EISD 85      & 17.44   &   0.02  &             &         &           &               \\   	  
    65    & 0.549   &    1   &     0 &    09  57  26.04 &  -20 13 05.7 &   17.9 &  1.0  &      S   &   7.3    & 09 57 26.04  & -20 13 04.5  &   2.5  &       16.65  & 0.26  &  EISD 87      & 19.87   &   0.03  &     22.25   &   0.05  &   23.06   &   0.07        \\   	  
    66+82 & 1.395   &    2   &     0 &    09  50  46.38 &  -21 32 55.1 &   31.0 &  1.1  &      M   &   103.3  & 09 50 48.97  & -21 32 55.8  &   4.5  &       18.2   & 0.1   &  EISD 90+113  & 21.7    &   0.1   &             &         &           &               \\   
    67    & 0.428   &    1   &     0 &    09  57  31.87 &  -21 20 26.7 &   17.3 &  0.7  &      T   &   38.8   & 09 57 31.81  & -21 20 30.5  &   4.5  &       15.96  & 0.18  &  EISD 89      & 18.62   &   0.01  &     20.87   &   0.03  &   21.95   &   0.04        \\   	  
    68    & 0.514   &    1   &     0 &    09  54  51.96 &  -21 30 16.1 &   17.2 &  0.7  &      S   &  $<$0.9    & 09 54 51.97  & -21 30 16.6  &   1.5  &       16.28  & 0.22  &  EISD 91      & 19.88   &   0.20  &     21.54   &   0.20  &   22.55   &   0.20        \\   	  
    69    & 4.01    &    1   &     0 &    09  56  02.36 &  -21 56 04.2 &   17.0 &  0.7  &      S   &  $<$1.4    & 09 56 02.45  & -21 56 03.8  &   2.0  &       19.6   & 0.3   &  EISD 92      & 22.82   &   0.15  &             &         &           &               \\   	  
    70    & 0.645   &    1   &     0 &    09  48  10.91 &  -20 00 59.9 &   17.0 &  2.0  &      M   &   154.3  & 09 48 10.60  & -20 00 58.6  &   4.5  &       17.19  & 0.25  &  EISD124      & 20.87   &   0.06  &     23.11   &   0.11  &   23.82   &   0.21        \\   	  
    71    & 2.857   &    1   &     0 &    09  55  41.89 &  -20 39 39.2 &   16.7 &  0.7  &      S   &  $<$3.5    & 09 55 41.88  & -20 39 38.2  &   1.5  &       19.62  & 0.25  &  EISD 93      &         &         &             &         &           &               \\   	  
    72    & 2.427   &    1   &     0 &    09  49  25.99 &  -20 37 24.2 &   16.5 &  0.7  &      S   &  $<$0.7    & 09 49 26.00  & -20 37 23.7  &   2.5  &       17.88  & 0.21  &  EISD 97      &         &         &             &         &           &               \\   	  
    73    & 1.357   &    2   &     0 &    09  56  28.10 &  -20 48 45.3 &   16.2 &  0.7  &      T   &   15.8   & 09 56 28.09  & -20 48 44.8  &   2.5  &       18.14  & 0.27  &  EISD 94      &         &         &             &         &           &               \\   	  
    74    & 0.667   &    1   &     0 &    09  49  29.75 &  -21 29 38.6 &   16.0 &  0.7  &      S   &   2.5    & 09 49 30.11  & -21 29 39.9  &   4.5  &       17.00  & 0.23  &  EISD 96      & 21.79   &   0.09  &     23.23   &   0.15  &   23.44   &   0.16        \\   	  
    75    & 0.265   &    1   &     0 &    09  45  26.97 &  -20 33 55.0 &   15.7 &  1.0  &      D   &   10.9   & 09 45 26.95  & -20 33 53.3  &   2.5  &       14.80  & 0.20  &  EISD 98      & 16.73   &   0.01  &     18.37   &   0.01  &   19.88   &   0.01        \\   	  
    76    & 0.282   &    1   &     0 &    09  57  45.89 &  -21 23 23.6 &   15.3 &  0.7  &      D   &   10.9   & 09 57 46.11  & -21 23 27.9  &   4.5  &       15.09  & 0.31  &  EISD102      & 17.23   &   0.01  &     18.99   &   0.01  &   20.16   &   0.01        \\   	  
    77    & 1.512   &    1   &     0 &    09  49  42.98 &  -20 37 45.5 &   15.0 &  0.7  &      S   &  $<$2.2    & 09 49 42.95  & -20 37 45.0  &   1.5  &       18.67  & 0.30  &  EISD104      & 23.60   &   0.20  &             &         &   24.99   &   0.31        \\   	  
    78    & 0.413   &    1   &     0 &    09  55  59.23 &  -20 42 51.6 &   14.6 &  0.7  &      D   &   6.9    & 09 55 59.31  & -20 42 53.2  &   2.5  &       16.48  & 0.35  &  EISD107      & 19.24   &   0.02  &     21.05   &   0.03  &   22.27   &   0.05        \\   	  
    79    & 1.255   &    1   &     0 &    09  45  48.48 &  -21 59 06.1 &   14.6 &  1.1  &      D   &   16.9   & 09 45 48.55  & -21 59 06.5  &   4.5  &       17.44  & 0.25  &  EISD106      & 22.57   &   0.15  &             &         &   24.27   &   0.19        \\   	  
    80    & 0.366   &    1   &     0 &    09  54  53.26 &  -21 15 12.9 &   14.5 &  0.6  &      S   &   10.9   & 09 54 53.25  & -21 15 13.3  &   4.5  &       14.65  & 0.15  &  EISD110      & 20.01   &   0.02  &     21.25   &   0.03  &   23.94   &   0.05        \\   	  
    81    & 0.462   &    1   &     0 &    09  54  16.43 &  -21 29 01.6 &   14.5 &  1.4  &      D   &   40.1   & 09 54 16.45  & -21 29 04.3  &   2.5  &       18.80  & 0.20  &  EISD105      & 21.16   &   0.06  &     22.44   &   0.08  &   22.87   &   0.08        \\   	  
    83    & 0.521   &    1   &     0 &    09  51  29.69 &  -20 16 42.8 &   13.5 &  0.6  &      S   &  $<$1.2    & 09 51 29.71  & -20 16 42.2  &   1.5  &       16.17  & 0.22  &  EISD116      & 20.13   &   0.03  &     21.72   &   0.08  &   22.77   &   0.10        \\ 
    84+85 & 0.148   &    2   &     0 &    09  55  45.19 &  -21 25 23.0 &   92.4 &  3.8  &      M   &   425.1  & 09 55 36.87  & -21 27 12.5  &  12.0  &       13.10  & 0.20  &  EISD103+112  &         &         &     16.81   &   0.04  &   17.72   &   0.02        \\   	  
    86    & 0.902   &    2   &     0 &    09  48  04.20 &  -20 34 34.8 &   13.2 &  0.6  &      S   &  $<$1.3    & 09 48 04.28  & -20 34 35.2  &   1.5  &       17.26  & 0.23  &  EISD120      & 23.00   &   0.30  &             &         &           &               \\   
    87    & 2.72    &    3   &     0 &    09  45  56.03 &  -21 20 51.0 &   13.2 &  0.6  &      D   &   9.4    &              &              &   2.5  &       $>$19.6  &       &  EISD111      &         &         &             &         &           &               \\   	  
    88    & 1.064   &    1   &     0 &    09  45  20.95 &  -22 01 22.2 &   13.1 &  0.6  &      S   &   2.3    & 09 45 20.95  & -22 01 21.0  &        &              &       &  EISD119      & 22.62   &   0.16  &             &         &           &               \\   	  
    89    & 0.909   &    1   &     0 &    09  53  09.24 &  -20 01 21.3 &   13.0 &  1.0  &      T   &   18.9   & 09 53 09.89  & -20 01 17.6  &   1.5  &       19.16  & 0.32  &  EISD117      &         &         &             &         &           &               \\   	  
    90    & 2.62    &    3   &     0 &    09  47  34.47 &  -21 26 58.0 &   12.8 &  0.6  &      S   &  $<$3.1    &              &              &   2.5  &       $>$19.5  &       &  EISD114      &         &         &             &         &           &               \\   	  
    91    & 1.265   &    2   &     0 &    09  48  22.16 &  -21 05 08.9 &   12.7 &  0.6  &      S   &  $<$1.3    & 09 48 22.25  & -21 05 08.5  &   1.5  &       17.99  & 0.34  &  EISD127      & 22.43   &   0.12  &             &         &           &               \\   	  
    92    & 0.743   &    1   &     1 &    09  52  55.92 &  -20 51 45.4 &   12.6 &  1.1  &      T   &   94.1   & 09 52 55.98  & -20 51 44.6  &   2.5  &       16.47  & 0.25  &  EISD122      & 18.68   &   0.01  &     19.29   &   0.01  &   19.41   &   0.01        \\   	  
    93    & 0.183   &    1   &     0 &    09  46  18.86 &  -20 37 57.4 &   12.2 &  0.6  &      D   &   14.5   & 09 46 19.16  & -20 37 58.5  &   1.5  &       15.13  & 0.31  &  EISD132      & 17.41   &   0.01  &     18.86   &   0.01  &   19.84   &   0.01        \\   	  
    94    & 1.555   &    2   &     0 &    09  45  21.12 &  -20 43 21.4 &   12.2 &  0.6  &      S   &   8.5    & 09 45 20.96  & -20 43 19.2  &   0.5  &       18.43  & 0.58  &  EISD125      &         &         &             &         &           &               \\     
    95    & 0.045   &    1   &     2 &    09  54  21.48 &  -21 48 07.2 &   12.2 &  1.2  &      S   &   1.6    & 09 54 21.59  & -21 48 06.6  &   4.5  &       12.08  & 0.27  &  EISD123      &         &         &             &         &           &               \\   	  
    96    & 2.706   &    1   &     0 &    09  49  25.99 &  -20 05 20.2 &   12.0 &  0.6  &      S   &  $<$1.0    & 09 49 26.06  & -20 05 19.9  &   1.0  &       20.07  & 0.27  &  EISD131      &         &         &             &         &           &               \\   	  
    97    & 1.548   &    2   &     0 &    09  54  36.32 &  -21 44 26.6 &   12.0 &  1.2  &      D   &   51.8   & 09 54 36.24  & -21 44 31.0  &   1.0  &       18.42  & 0.28  &  EISD126      & 22.54   &   0.14  &             &         &           &               \\   	  
    98    & 1.669   &    2   &     0 &    09  49  35.13 &  -21 58 10.5 &   11.8 &  0.6  &      S   &  $<$1.1    & 09 49 35.19  & -21 58 10.5  &   1.5  &       18.58  & 0.34  &  EISD130      &         &         &             &         &           &               \\   	  
    99    & 0.738   &    1   &     0 &    09  57  02.25 &  -21 56 51.8 &   11.6 &  0.6  &      S   &   4.0    & 09 57 02.40  & -21 56 50.6  &   4.5  &       16.20  & 0.21  &  EISD133      & 19.49   &   0.03  &     22.22   &   0.07  &   24.15   &   0.14        \\   	  
   100    & 1.701   &    2   &     0 &    09  50  48.57 &  -21 54 57.1 &   11.5 &  0.6  &      S   &   4.7    & 09 50 48.52  & -21 54 55.5  &   2.5  &       18.62  & 0.15  &  EISD136      & 23.77   &   0.25  &     24.69   &   0.18  &   24.60   &   0.22        \\   	  
   101    & 1.043   &    1   &     0 &    09  52  50.38 &  -21 31 48.0 &   11.4 &  0.6  &      S   &  $<$3.0    & 09 52 50.45  & -21 31 48.2  &   2.5  &       17.86  & 0.35  &  EISD139      & 22.35   &   0.14  &     24.33   &   0.25  &           &               \\   	  
   102    & 0.468   &    1   &     0 &    09  46  49.27 &  -21 16 48.7 &   11.1 &  1.1  &      D   &   12.0   & 09 46 49.49  & -21 16 46.8  &   2.5  &       15.63  & 0.28  &  EISD134      & 18.36   &   0.01  &     20.83   &   0.02  &   21.82   &   0.04        \\   	  
   103    & 2.50    &    3   &     0 &    09  47  28.14 &  -21 28 57.9 &   10.7 &  0.6  &      D   &   12.6   &              &              &   2.5  &       $>$19.4  &       &  EISD 56      &         &         &             &         &           &               \\   	  
   104    & 0.962   &    2   &     0 &    09  57  39.51 &  -20 03 22.6 &   10.7 &  0.6  &      D   &   31.8   & 09 57 39.08  & -20 03 12.0  &   4.5  &       17.40  & 0.21  &  EISD145      & 23.33   &   0.19  &             &         &           &               \\   	  
   105    & 3.377   &    1   &     0 &    09  47  24.38 &  -21 05 02.3 &   10.6 &  0.6  &      S   &  $<$6.8    & 09 47 24.54  & -21 05 02.5  &   1.0  &       20.16  & 0.36  &  EISD138      &         &         &             &         &           &               \\   	  
   106    & 1.300   &    2   &     0 &    09  56  06.94 &  -20 05 43.8 &   10.5 &  0.6  &      S   &   5.6    & 09 56 07.13  & -20 05 44.0  &   1.5  &       18.05  & 0.34  &  EISD142      & 21.96   &   0.10  &     24.37   &   0.18  &           &               \\   	  
   107    & 0.512   &    1   &     0 &    09  45  37.77 &  -21 11 14.2 &   10.3 &  1.0  &      D   &   7.0    & 09 45 38.10  & -21 11 13.6  &   2.5  &       16.01  & 0.17  &  EISD148      & 18.92   &   0.02  &     21.38   &   0.03  &   22.23   &   0.05        \\   	  
   108    & 0.230   &    1   &     0 &    09  56  49.76 &  -20 35 25.9 &   10.2 &  0.6  &      S   &  $<$1.3    & 09 56 49.86  & -20 35 26.2  &   4.5  &       14.68  & 0.23  &  EISD153      & 17.09   &   0.01  &     18.63   &   0.01  &   19.70   &   0.01        \\   	  
   109    & 0.804   &    2   &     0 &    09  52  10.91 &  -20 50 11.2 &   10.1 &  0.6  &      S   &  $<$3.9    & 09 52 10.86  & -20 50 08.9  &   4.5  &       17.01  & 0.27  &  EISD154      & 20.51   &   0.05  &     24.03   &   0.18  &           &               \\   	  
   110    & 0.282   &    1   &     0 &    09  55  11.49 &  -20 30 18.7 &   10.1 &  1.3  &      T   &   83.4   & 09 55 11.46  & -20 30 19.2  &   4.5  &       14.60  & 0.10  &  EISD141      & 17.52   &   0.01  &     19.17   &   0.01  &   20.25   &   0.10        \\   	  
   111    & 0.411   &    1   &     0 &    09  47  44.76 &  -21 12 23.6 &   10.0 &  0.6  &      S   &   2.3    & 09 47 44.79  & -21 12 23.3  &   4.5  &       15.84  & 0.24  &  EISD149      & 18.34   &   0.01  &     20.77   &   0.02  &   21.73   &   0.03        \\   	  
   112    & 1.75    &    4   &     0 &    09  56  42.31 &  -21 19 44.6 &    9.8 &  0.6  &      S   &  $<$1.1    & 09 56 42.30  & -21 19 44.3  &        &              &       &  EISD146      & 23.50   &   0.23  &     23.90   &   0.17  &   22.95   &   0.07        \\   	  
\hline
\end{tabular}
\end{table*} 

\begin{table*}
\scriptsize
\begin{tabular}{cccccccccccccccccccccc}
\hline 
 CENSORS & z         & T &   C &      \multicolumn{2}{c}{Radio Position}    &$S_{\rm 1.4GHz}$&  $S_{\rm err}$   & Morph. & $D_{\rm rad}$ &     \multicolumn{2}{c}{Host Position}         & ap. used&  K   & K$_{\rm err}$    &  EISD     &   I    &   I$_{\rm err}$   &    V    &   V$_{\rm err}$   &  B    &   B$_{\rm err}$    \\    
         &           &      &         &  RA               &    DEC           &  &        &  &&    RA  &     DEC   & for corr&  (ap cor)&      &  name     &           &          &            &          &          &           \\
         &           &       &         &      J2000           &    J2000           &  mJy  &  mJy   &  & \arcsec &    J2000  &     J2000    & arcsec  &  mag   & mag     &           &           &          &            &          &          &           \\       
(1)         &   (2)        &   (3)    &   (4)      &      (5)           &    (6)           &  (7)  &  (8)   & (9) & (10) &   (11) &    (12)    & (13)  & (14)   & (15)     &  (16)         & (17)          &  (18)        &   (19)         &   (20)       &   (21)       &   (22)        \\       
\hline                                                                                   
   113    & 1.017   &    2   &     0 &    09  47  10.01 &  -20 35 52.8 &    9.7 &  0.6  &      D   &   19.7   & 09 47 10.36  & -20 35 52.2  &   2.5  &       17.52  & 0.28  &  EISD150      & 22.10   &   0.11  &     24.30   &   0.24  &           &               \\   	  
   114    & 1.426   &    1   &     1 &    09  56  04.45 &  -21 44 36.7 &    9.6 &  0.6  &      S   &  $<$1.8    & 09 56 04.52  & -21 44 36.7  &   1.5  &       19.23  & 0.34  &  EISD166      & 22.73   &   0.22  &     23.62   &   0.16  &   23.68   &   0.10        \\   	  
   115    & 0.545   &    1   &     0 &    09  57  24.93 &  -20 22 48.0 &    9.6 &  1.0  &      D   &   13.1   & 09 57 24.89  & -20 22 42.5  &   4.5  &       15.18  & 0.23  &  EISD155      & 17.84   &   0.01  &     20.10   &   0.02  &   21.09   &   0.02        \\   	  
   116    & 2.637   &    1   &     1 &    09  57  35.35 &  -20 29 35.4 &    9.6 &  0.6  &      S   &  $<$1.3    & 09 57 35.46  & -20 29 35.5  &   2.5  &       18.23  & 0.27  &  EISD143      & 19.97   &   0.02  &     20.57   &   0.03  &   20.90   &   0.02        \\   	  
   117    & 1.204   &    1   &     0 &    09  54  10.54 &  -21 58 00.9 &    9.5 &  0.6  &      D   &   5.6    & 09 54 10.58  & -21 58 01.1  &   2.5  &       18.15  & 0.33  &  EISD165      & 22.73   &   0.15  &             &         &           &               \\   	  
   118    & 2.294   &    1   &     0 &    09  47  48.55 &  -20 48 34.0 &    9.4 &  0.6  &      S   &   3.7    & 09 47 48.46  & -20 48 35.3  &   1.0  &       19.31  & 0.22  &  EISD161      &         &         &             &         &           &               \\   	  
   119    & 1.484   &    1   &     0 &    09  49  02.22 &  -21 15 05.5 &    9.4 &  0.6  &      S   &   7.9    & 09 49 02.22  & -21 15 04.8  &   4.5  &       17.92  & 0.32  &  EISD157      & 23.58   &   0.26  &             &         &   25.42   &   0.19        \\   	  
   120    & 2.829   &    1   &     0 &    09  53  57.38 &  -20 36 51.3 &    9.1 &  0.6  &      S   &  $<$1.2    & 09 53 57.51  & -20 36 50.7  &   1.5  &       17.31  & 0.24  &  EISD159      & 21.10   &   0.06  &     22.28   &   0.06  &   22.41   &   0.05        \\   	  
   121    & 0.246   &    1   &     0 &    09  52  01.20 &  -20 24 56.5 &    9.0 &  0.5  &      S   &  $<$0.9    & 09 52 01.26  & -20 24 56.5  &   4.5  &       14.64  & 0.20  &  EISD164      & 17.15   &   0.01  &     18.58   &   0.01  &   19.57   &   0.01        \\   	  
   122    & 0.250   &    1   &     0 &    09  56  37.11 &  -20 19 05.5 &    9.0 &  0.6  &      T   &   28.4   & 09 56 37.20  & -20 19 05.7  &   4.5  &       14.40  & 0.22  &  EISD156      & 16.94   &   0.01  &     18.38   &   0.01  &   19.60   &   0.01        \\   	  
   123    & 0.906   &    2   &     0 &    09  54  31.06 &  -20 35 38.0 &    8.7 &  0.5  &      S   &  $<$1.2    & 09 54 31.08  & -20 35 37.1  &   2.5  &       17.27  & 0.22  &  EISD173      & 21.05   &   0.07  &     22.96   &   0.14  &           &               \\   	  
   124    & 0.0156  &    1   &     2 &    09  49  10.88 &  -20 21 53.0 &    8.7 &  0.6  &      E   &   24.8   & 09 49 10.80  & -20 21 53.0  &        &              &       &  EISD163      &         &         &             &         &           &               \\     
   125    & 0.701   &    1   &     0 &    09  49  22.31 &  -21 18 19.4 &    8.4 &  0.5  &      D   &   11.8   & 09 49 22.34  & -21 18 17.7  &   4.5  &       15.80  & 0.10  &  EISD175      & 18.95   &   0.02  &     21.83   &   0.05  &   23.00   &   0.25        \\   	  
   126    & 0.445   &    2   &     0 &    09  47  50.58 &  -21 42 08.2 &    8.4 &  1.3  &      D   &   38.3   & 09 47 50.69  & -21 42 11.7  &   1.5  &       15.69  & 0.26  &  EISD171      & 17.43   &   0.01  &     18.84   &   0.01  &   19.34   &   0.01        \\   	  
   127    & 0.922   &    1   &     0 &    09  49  24.64 &  -21 11 12.0 &    8.3 &  0.5  &      S   &  $<$1.0    & 09 49 24.73  & -21 11 11.8  &   2.5  &       17.10  & 0.24  &  EISD186      & 20.77   &   0.04  &     22.83   &   0.07  &   23.10   &   0.10        \\   	  
   128    & 3.153   &    2   &     0 &    09  49  02.78 &  -20 16 11.5 &    8.3 &  0.5  &      S   &  $<$1.3    & 09 49 02.78  & -20 16 10.9  &   2.5  &       19.90  & 0.30  &  EISD174      & 22.50   &   0.13  &     22.97   &   0.12  &   23.99   &   0.12        \\   	  
   129    & 2.421   &    1   &     0 &    09  52  26.51 &  -20 01 07.1 &    8.3 &  0.6  &      S   &  $<$2.1    & 09 52 26.41  & -20 01 07.1  &   1.0  &       19.00  & 0.36  &  EISD170      &         &         &             &         &           &               \\   	  
   130    & 2.234   &    2   &     0 &    09  57  22.18 &  -21 01 06.0 &    8.2 &  0.5  &      S   &  $<$1.2    & 09 57 22.17  & -21 01 05.4  &   2.5  &       19.19  & 0.34  &  EISD172      &         &         &             &         &           &               \\   	  
   131    & 0.470   &    1   &     0 &    09  51  48.94 &  -21 33 41.6 &    8.2 &  0.6  &      D   &   9.5    & 09 51 49.00  & -21 33 39.7  &   1.0  &       15.87  & 0.33  &  EISD169      & 17.15   &   0.01  &     19.50   &   0.01  &   19.97   &   0.02        \\   	  
   132    & 2.069   &    2   &     0 &    09  46  02.36 &  -21 51 44.2 &    7.9 &  0.6  &      S   &  $<$2.7    & 09 46 02.37  & -21 51 44.2  &   1.5  &       19.03  & 0.28  &  EISD167      &         &         &             &         &           &               \\   	  
   133    & 1.335   &    1   &     0 &    09  51  29.36 &  -20 25 34.6 &    7.8 &  1.2  &      D   &   11.4   & 09 51 29.42  & -20 25 35.4  &   4.5  &       17.77  & 0.10  &  EISD183      &         &         &             &         &           &               \\   	  
   134    & 2.354   &    1   &     0 &    09  49  49.00 &  -21 34 33.7 &    7.8 &  0.6  &      D   &   22.4   & 09 49 48.77  & -21 34 28.2  &   1.5  &       19.93  & 0.14  &  EISD182      &         &         &             &         &           &               \\   	  
   135    & 1.316   &    1   &     0 &    09  47  48.33 &  -21 00 40.4 &    7.8 &  0.6  &      D   &   10.4   & 09 47 47.91  & -21 00 45.2  &   4.5  &       18.78  & 0.38  &  EISD178      & 21.42   &   0.08  &     22.63   &   0.10  &   23.12   &   0.06        \\   	  
   136    & 0.629   &    1   &     0 &    09  54  41.85 &  -20 49 43.0 &    7.5 &  0.6  &      S   &  $<$3.8    & 09 54 41.88  & -20 49 43.4  &   2.5  &       $>$19.4  &       &  EISD181      & 24.85   &   0.35  &             &         &           &               \\   	  
   137    & 0.526   &    1   &     0 &    09  50  38.80 &  -21 41 08.4 &    7.4 &  1.2  &      D   &   33.0   & 09 50 38.70  & -21 41 12.2  &   2.5  &       16.53  & 0.26  &  EISD187      & 18.73   &   0.02  &     21.49   &   0.05  &   22.43   &   0.08        \\   	  
   138    & 0.508   &    1   &     0 &    09  55  26.95 &  -20 46 06.0 &   14.7 &  0.5  &      S   &   121.3  & 09 55 26.95  & -20 46 06.2  &   2.5  &       17.03  & 0.35  &  EISD177      & 19.99   &   0.04  &     22.63   &   0.06  &   23.45   &   0.13        \\   	  
   139    & 0.344   &    1   &     0 &    09  49  12.72 &  -22 00 23.4 &    6.9 &  0.6  &      S   &  $<$2.8    & 09 49 12.74  & -22 00 23.4  &   4.0  &       15.00  & 0.10  &  EISD180      & 17.61   &   0.01  &     19.85   &   0.02  &   20.81   &   0.02        \\   	  
   140    & 0.265   &    1   &     0 &    09  45  26.34 &  -21 55 00.4 &    6.8 &  0.5  &      S   &  $<$1.1    & 09 45 26.34  & -21 55 00.2  &   2.5  &       15.67  & 0.33  &  EISD199      & 17.99   &   0.01  &     19.64   &   0.01  &   20.71   &   0.02        \\   	  
   141    & 2.829   &    1   &     0 &    09  45  51.03 &  -20 14 46.9 &    6.6 &  0.6  &      S   &  $<$1.2    & 09 45 50.99  & -20 14 46.4  &   1.0  &       19.05  & 0.27  &  EISD189      &         &         &             &         &           &               \\   	  
   142    & 2.192   &    2   &     0 &    09  57  15.56 &  -20 30 34.8 &    6.3 &  0.6  &      S   &  $<$1.9    & 09 57 15.55  & -20 30 34.6  &   1.0  &       19.15  & 0.35  &  EISD195      &         &         &             &         &           &               \\   	  
   143    & 1.701   &    2   &     0 &    09  47  46.12 &  -21 27 51.2 &    6.1 &  0.6  &      S   &  $<$1.7    & 09 47 46.07  & -21 27 50.4  &   1.5  &       18.62  & 0.42  &  EISD188      &         &         &             &         &           &               \\   	  
   144    & 0.696   &    1   &     0 &    09  49  59.72 &  -21 27 19.0 &    6.0 &  0.6  &      S   &  $<$1.0    & 09 49 59.73  & -21 27 19.0  &   2.5  &       17.28  & 0.32  &  EISD179      & 19.95   &   0.03  &     22.29   &   0.07  &   22.60   &   0.15        \\   	  
   145    & 0.400   &    1   &     0 &    09  48  14.15 &  -19 59 56.0 &    5.8 &  0.3  &      S   &  $<$6.8    & 09 48 14.22  & -19 59 56.5  &   4.5  &       15.87  & 0.11  &  EISD137      & 18.93   &   0.02  &     20.84   &   0.02  &   21.92   &   0.04        \\   	  
   146    & 0.0294  &    1   &     2 &    09  50  27.68 &  -21 48 08.7 &    5.4 &  0.6  &      E   &   0.0    & 09 50 27.69  & -21 48 09.2  &        &              &       &  EISD191      &         &         &             &         &           &               \\   	  
   147    & 1.338   &    2   &     0 &    09  45  21.73 &  -20 36 00.3 &    4.2 &  0.7  &      S   &  $<$1.7    & 09 45 21.72  & -20 35 59.5  &   1.5  &       18.11  & 0.30  &  EISD197      &         &         &             &         &           &               \\   	  
   148    & 0.758   &    2   &     0 &    09  56  39.20 &  -20 10 43.6 &    4.1 &  0.8  &      S   &  $<$3.8    & 09 56 39.22  & -20 10 44.3  &   1.0  &       16.88  & 0.32  &  EISD162      & 19.20   &   0.30  &             &         &           &               \\   	  
   149    & 0.0290  &    1   &     2 &    09  52  14.34 &  -21 40 19.0 &    4.0 &  0.7  &      S   &  $<$2.9    & 09 52 14.41  & -21 40 18.6  &   2.5  &       12.58  & 0.35  &  EISD185      & 14.84   &   0.01  &     15.87   &   0.01  &   16.15   &   0.01        \\   	  
   150    & 0.146   &    2   &     0 &    09  45  27.69 &  -20 57 35.2 &    3.8 &  0.7  &      D   &   23.1   & 09 45 27.77  & -20 57 48.2  &   2.5  &       13.08  & 0.35  &  EISD194      & 14.67   &   0.01  &     15.44   &   0.01  &   15.87   &   0.01        \\   	  
 																																					
\hline
\end{tabular}
\end{table*} 

\end{landscape}



\section{Additional datasets}
\label{extra_data}

\begin{table}

\subsection{The Hercules sample}

The 1.4 GHz flux densities, spectral indices (measured between 1.4 GHz and 0.6 GHz) and redshifts for the Hercules sample. 
`z type' is defined as:
(1) spectroscopic, 
(2) photo-z from Waddington et al. (2001), 
(3) $K$--$z$ limit from $K$--band magnitude in either Rigby et al. (2007) or Waddington et al. (2000), 
(4) $K$--$z$ value from Rigby et al. (2007) for 53W054B, whose host galaxy was misidentified by Waddington et al. (2000) 

\begin{center}
\begin{tabular}{ccccc}
\hline
Name     &   z     &  z type  &   $S_{\rm 1.4 GHz}$  (mJy)  &   $\alpha$ \\ 
\hline
53W002   &  2.390   &   1    &    50.1   &    1.10 \\
53W004   &  1.12    &   2    &    54.5   &    0.20 \\
53W005   &  0.95    &   1    &     7.6   &    1.09 \\
53W008   &  0.733   &   1    &   306.6   &    0.79 \\
53W009   &  1.090   &   1    &    92.7   &    0.38 \\
53W010   &  0.48    &   1    &     8.1   &    0.73 \\
53W011   &  0.61    &   2    &     3.5   &    0.28 \\
53W012   &  1.328   &   1    &    47.6   &    0.41 \\
53W013   &  1.49    &   2    &     3.7   &   -0.39 \\
53W014   &  1.28    &   2    &     5.3   &   -0.81 \\
53W015   &  1.129   &   1    &   184.6   &    0.78 \\
53W019   &  0.542   &   1    &     6.8   &    0.72 \\
53W020   &  0.100   &   1    &     6.7   &    1.07 \\
53W021   &  1.12    &   2    &     4.7   &    1.07 \\
53W022   &  0.528   &   1    &    11.8   &    0.43 \\
53W023   &  0.57    &   1    &   109.9   &    0.87 \\
53W024   &  1.961   &   1    &    10.3   &    0.55 \\
53W026   &  0.55    &   1    &    21.1   &    0.74 \\
53W027   &  0.403   &   1    &     8.3   &    0.80 \\
53W029   &  1.23    &   2    &    22.2   &   -0.23 \\
53W031   &  0.628   &   1    &   116.5   &    0.70 \\
53W032   &  0.37    &   1    &    10.5   &    0.80 \\
53W034   &  0.281   &   1    &    10.9   &    1.00 \\
53W035   &  1.41    &   2    &     4.4   &   -0.44 \\
53W036   &  1.50    &   2    &     3.2   &    1.24 \\
53W037   &  4.20    &   3    &     6.6   &    1.07 \\
53W039   &  0.402   &   1    &     3.4   &    0.82 \\
53W041   &  0.59    &   2    &     9.4   &    0.88 \\
53W042   &  1.58    &   2    &     6.6   &    1.07 \\
53W046   &  0.528   &   1    &    63.1   &    0.69 \\
53W047   &  0.534   &   1    &    23.9   &    0.67 \\
53W048   &  0.676   &   1    &    11.5   &    0.81 \\
53W049   &  0.23    &   1    &    95.1   &    0.81 \\
53W051   &  1.01    &   2    &   141.6   &    0.87 \\
53W052   &  0.46    &   1    &     8.6   &    0.74 \\
53W054A  &  1.25    &   2    &     3.9   &   -0.39 \\
53W054B  &  3.50    &   4    &     3.0   &   -0.42 \\
53W057   &  1.53    &   2    &     2.9   &   -0.36 \\
53W059   &  1.42    &   2    &    18.7   &    0.90 \\
53W060   &  0.62    &   2    &     9.7   &    0.93 \\
53W061   &  2.88    &   2    &     2.6   &   -0.15 \\
53W065   &  1.185   &   1    &     5.3   &    1.21 \\
53W066   &  1.82    &   2    &     4.1   &    0.91 \\
53W067   &  0.759   &   1    &    23.2   &    0.81 \\
53W068   &  0.25    &   2    &     3.9   &    0.33 \\
53W069   &  1.432   &   1    &     3.7   &    0.87 \\
53W070   &  1.315   &   1    &     2.6   &   -0.04 \\
53W071   &  0.287   &   1    &     2.8   &    1.43 \\
53W072   &  0.019   &   1    &     6.6   &    0.17 \\
53W075   &  2.150   &   1    &    96.1   &    0.78 \\
53W077   &  0.80    &   1    &     7.8   &    0.87 \\
53W078   &  0.27    &   1    &     2.0   &    0.53 \\
53W079   &  0.548   &   1    &    13.3   &    0.05 \\
53W080   &  0.546   &   1    &    27.6   &    0.80 \\
53W081   &  2.060   &   1    &    12.2   &    0.84 \\
53W082   &  1.19    &   2    &     2.0   &    1.41 \\
\hline
\end{tabular}
\end{center}
\end{table}


\begin{table}
\begin{center}
\begin{tabular}{ccccc}
\hline
Name     &   z     &  z type  &   $S_{\rm 1.4 GHz}$  (mJy)  &   $\alpha$ \\ 
\hline
53W083   &  0.628   &   1    &     5.0   &    0.70 \\
53W085   &  1.35    &   1    &     4.3   &    1.29 \\
53W086   &  0.40    &   1    &     4.9   &    0.35 \\
53W087   &  2.57    &   3    &     5.8   &    1.18 \\
53W088   &  1.773   &   1    &    14.9   &   -0.10 \\
53W089   &  0.635   &   1    &     2.5   &    1.29 \\
53W090   &  0.094   &   1    &     2.1   &    0.83 \\
53W091   &  1.552   &   1    &    22.1   &    1.30 \\
\hline
\end{tabular}
\end{center}

\subsection{The WP85 sample}

The 1.4 GHz flux densities, converted from the original 2.7 GHz values, spectral indices (determined between 2.7 GHz and 5 GHz) and redshifts for the $z \geq 0.1$ WP85 sample. 
`ztype' is defined as:
(1) spectroscopic, 
(2) estimated z taken from the WP85 paper, 
(3) estimated z taken from \citet{burgess}.

\begin{center}
\begin{tabular}{ccccc}
\hline
  Name       &   z      &   z type  &  $ S_{\rm 1.4 GHz}$  (Jy) & $\alpha$   \\
\hline
   0003-00   &  1.040   &     1 &      4.22    &   0.86 \\
   0008-42   &  1.600   &     2 &      4.86    &   1.03 \\
   0022-42   &  0.937   &     1 &      4.71    &   0.77 \\
   0023-26   &  0.322   &     1 &      9.19    &   0.70 \\
   0035-02   &  0.220   &     1 &      6.48    &   0.72 \\
   0038+09   &  0.190   &     1 &      5.79    &   1.00 \\
   0040+51   &  0.170   &     1 &     10.45    &   0.72 \\
   0105-16   &  0.400   &     1 &      4.63    &   1.10 \\
   0114-21   &  1.410   &     1 &      4.16    &   0.95 \\
   0117-15   &  0.565   &     1 &      4.91    &   0.90 \\
   0134+32   &  0.370   &     1 &     15.87    &   0.85 \\
   0157-31   &  0.680   &     1 &      4.03    &   0.81 \\
   0210+86   &  0.190   &     1 &      8.91    &   1.31 \\
   0213-13   &  0.140   &     1 &      4.54    &   0.74 \\
   0235-19   &  0.620   &     1 &      4.27    &   0.87 \\
   0237-23   &  2.220   &     1 &      7.46    &   0.64 \\
   0252-71   &  0.568   &     1 &      6.55    &   1.14 \\
   0307+16   &  0.260   &     1 &      4.60    &   0.93 \\
   0316+16   &  0.907   &     1 &      8.01    &   0.79 \\
   0404+76   &  0.599   &     1 &      6.01    &   0.60 \\
   0407-65   &  0.962   &     1 &     13.47    &   1.11 \\
   0409-75   &  0.693   &     1 &     12.72    &   0.86 \\
   0428+20   &  0.220   &     1 &      4.50    &   0.53 \\
   0433+29   &  0.220   &     1 &     48.50    &   0.86 \\
   0442-28   &  0.147   &     1 &      7.07    &   0.93 \\
   0453+22   &  0.210   &     1 &      4.08    &   1.01 \\
   0518+16   &  0.760   &     1 &     12.99    &   0.92 \\
   0538+49   &  0.550   &     1 &     21.79    &   0.77 \\
   0605+48   &  0.280   &     1 &      4.18    &   0.89 \\
   0743-67   &  0.400   &     1 &      5.18    &   0.97 \\
   0809+48   &  0.870   &     1 &     14.37    &   0.94 \\
   0834-19   &  1.032   &     1 &      4.28    &   0.82 \\
   0859-25   &  0.305   &     1 &      6.71    &   1.08 \\
   0917+45   &  0.170   &     1 &      8.83    &   1.06 \\
   0958+29   &  0.180   &     1 &      5.94    &   1.06 \\
   1005+07   &  0.880   &     1 &      6.62    &   0.97 \\
   1017-42   &  1.280   &     1 &      4.74    &   1.08 \\
   1136-13   &  0.550   &     1 &      4.29    &   0.65 \\
   1151-34   &  0.260   &     1 &      6.58    &   0.69 \\
   1157+73   &  0.970   &     1 &      6.41    &   0.70 \\
   1245-19   &  1.275   &     1 &      6.49    &   0.76 \\
   1254+47   &  1.000   &     1 &      5.59    &   1.02 \\
   1306-09   &  0.464   &     1 &      4.29    &   0.65 \\
   1323+32   &  0.360   &     1 &      4.97    &   0.60 \\
   1328+25   &  1.050   &     1 &      7.05    &   0.65 \\
   1328+30   &  0.850   &     1 &     14.70    &   0.53 \\
   1355-41   &  0.310   &     1 &      4.59    &   0.93 \\
   1358+62   &  0.430   &     1 &      4.20    &   0.68 \\
\hline
\end{tabular}
\end{center}
\end{table}

\begin{table}
\begin{center}
\begin{tabular}{ccccc}
\hline
Name     &   z     &  z type  &   $S_{\rm 1.4 GHz}$  (mJy)  &   $\alpha$ \\ 
\hline
   1409+52   &  0.460   &     1 &     22.88    &   0.99 \\
   1416+06   &  1.440   &     1 &      4.97    &   0.93 \\
   1453-10   &  0.940   &     1 &      4.60    &   0.93 \\
   1458+71   &  0.900   &     1 &      8.89    &   0.77 \\
   1518+04   &  1.296   &     1 &      5.10    &   1.28 \\
   1529+24   &  0.100   &     1 &      4.65    &   1.14 \\
   1559+02   &  0.100   &     1 &      9.41    &   0.95 \\
   1602+01   &  0.460   &     1 &      4.32    &   1.07 \\
   1607+26   &  0.473   &     1 &      6.18    &   1.08 \\
   1609+66   &  0.550   &     1 &      6.19    &   0.76 \\
   1634+62   &  0.990   &     1 &      5.09    &   0.96 \\
   1637+62   &  0.750   &     1 &      4.45    &   1.03 \\
   1648+05   &  0.150   &     1 &     51.00    &   1.11 \\
   1740-51   &  0.350   &     2 &      7.38    &   0.72 \\
   1828+48   &  0.690   &     1 &     16.69    &   0.78 \\
   1832+47   &  0.160   &     1 &      4.38    &   0.96 \\
   1932-46   &  0.231   &     1 &     12.86    &   1.03 \\
   1934-63   &  0.180   &     1 &     19.78    &   0.88 \\
   1938-15   &  0.452   &     1 &      6.51    &   0.82 \\
   1939+60   &  0.200   &     1 &      5.35    &   0.99 \\
   2032-35   &  0.631   &     1 &      7.62    &   1.10 \\
   2121+24   &  0.100   &     1 &     14.14    &   1.07 \\
   2128+04   &  0.990   &     1 &      4.84    &   0.67 \\
   2135-20   &  0.635   &     1 &      4.27    &   0.82 \\
   2153+37   &  0.290   &     1 &      7.26    &   1.22 \\
   2211-17   &  0.150   &     1 &     10.34    &   1.26 \\
   2230+11   &  1.040   &     1 &      8.23    &   0.67 \\
   2250-41   &  0.310   &     1 &      4.48    &   0.99 \\
   2314+03   &  0.220   &     1 &      4.46    &   0.97 \\
   2331-41   &  0.907   &     1 &      4.84    &   0.91 \\
   2342+82   &  0.730   &     1 &      4.35    &   0.95 \\
   2356-61   &  0.100   &     1 &     24.97    &   1.36 \\
\hline
\multicolumn{5}{c}{WP85 additional sources } \\
\hline
     3c325   &  1.135   &     1 &      4.29    &   1.29 \\
  1526-423   &  0.500   &     3 &      5.08    &   1.02 \\
  1827-360   &  0.120   &     3 &      6.49    &   1.12 \\
\hline
\end{tabular}
\end{center}

\subsection{The PSR sample}

The 1.4 GHz flux densities, converted from the original 2.7 GHz values, spectral indices (determined between 1.4 GHz and 5 GHz) and redshifts for the PSR sample. 
`ztype' is defined as:
(1) spectroscopic, 
(2) $K$--$z$ estimated redshift, 
(3) \citet{DP93} estimated z derived from spectral fitting, 
(4) $K$--$z$ using the $B-R$ colour to get the $K$ magnitude, 
(5) $K$--$z$ limit


\begin{center}
\begin{tabular}{ccccc}
\hline
  Name       &   z      &   z type  &  $ S_{\rm 1.4 GHz}$  (Jy) &  $\alpha$     \\
\hline
  0000+035  &    0.61   &     4 &      0.37   &     1.41 \\
  0003+006  &    0.92   &     3 &      0.47   &     0.97 \\
  0003-003  &   1.037   &     1 &      3.87   &     0.72 \\
  0010+005  &   0.606   &     1 &      1.74   &     0.92 \\
  0011-023  &   2.080   &     1 &      0.40   &     0.74 \\
  0038-019  &   1.679   &     1 &      1.45   &     1.22 \\
  0041+007  &   0.112   &     1 &      0.48   &     1.24 \\
  0043+000  &    0.60   &     2 &      0.53   &     1.03 \\
  0043-010  &    1.07   &     2 &      0.32   &     0.82 \\
  0045-009  &    0.60   &     2 &      0.30   &     1.13 \\
  0053-016  &   0.044   &     1 &      1.15   &     0.74 \\
  0053-015  &   0.044   &     1 &      1.33   &     0.81 \\
  0054+018  &   0.291   &     1 &      0.51   &     0.95 \\
  0055-016  &   0.045   &     1 &      5.08   &     0.63 \\
  0059+017  &    0.52   &     4 &      0.80   &     1.06 \\
  0059+027  &    1.48   &     3 &      0.30   &     1.06 \\
  0101-025  &   2.050   &     1 &      0.30   &     0.76 \\
  0222-008  &   0.687   &     1 &      1.11   &     0.79 \\

\hline
\end{tabular}
\end{center}
\end{table}


\begin{table}
\begin{center}
\begin{tabular}{ccccc}
\hline
Name     &   z     &  z type  &   $S_{\rm 1.4 GHz}$  (mJy)  &   $\alpha$ \\ 
\hline

  0223-023  &    0.93   &     2 &      0.41   &     0.93 \\
  0223+012  &   1.369   &     1 &      0.40   &     0.76 \\
  0225+002  &    1.64   &     2 &      0.30   &     1.15 \\
  0225-014  &   2.037   &     1 &      0.49   &     0.75 \\
  0230-027  &   0.239   &     1 &      0.57   &     0.83 \\
  0233-025  &   1.321   &     1 &      0.97   &     0.78 \\
  0235-019  &   0.840   &     1 &     0.40   &     0.89 \\
  0235+023  &   0.209   &     1 &     0.47   &     0.72 \\
  0240-002  &   0.004   &     1 &     5.40   &     0.83 \\
  0242+028  &   0.767   &     1 &     0.53   &     0.95 \\
  1155-029  &    0.35   &     4 &     0.34   &     0.56 \\
  1159-023  &    1.13   &     4 &     0.85   &     1.03 \\
  1201-002  &    0.28   &     2 &     0.32   &     0.79 \\
  1207-013  &    0.33   &     4 &     0.59   &     0.70 \\
  1211+000  &   0.321   &     1 &     0.47   &     1.08 \\
  1212-007  &   1.600   &     1 &     0.98   &     1.00 \\
  1212+005  &    0.39   &     2 &     0.51   &     0.89 \\
  1329+012  &    0.84   &     3 &     0.39   &     0.95 \\
  1330+022  &   0.216   &     1 &     2.82   &     0.59 \\
  1331+004  &   1.400   &     1 &     0.36   &     1.44 \\
  1331+025  &   1.228   &     1 &     0.32   &     1.19 \\
  1336+020  &   0.567   &     1 &     0.74   &     1.02 \\
  1337-033  &    0.79   &     4 &     1.01   &     0.84 \\
  1339+015  &   3.510   &     1 &     0.30   &     0.86 \\
  1340+022  &    0.49   &     2 &     0.98   &     0.91 \\
  1342-016  &   0.167   &     1 &     0.40   &     0.87 \\
  1343-007  &    0.45   &     2 &     0.98   &     0.69 \\
  1343-026  &    0.36   &     2 &     0.45   &     0.93 \\
  1345+008  &   1.500   &     1 &     0.31   &     1.21 \\
  1349-017  &   0.167   &     1 &     0.55   &     0.88 \\
  1352+008  &    0.80   &     2 &     0.70   &     0.85 \\
  2150-202  &   1.330   &     1 &     0.58   &     0.95 \\
  2152-218  &   0.306   &     1 &     0.71   &     1.48 \\
  2154-184  &   0.668   &     1 &     2.39   &     1.17 \\
  2154-183  &   1.423   &     1 &     1.68   &     0.79 \\
  2155-202  &    0.55   &     2 &     0.49   &     1.24 \\
  2157-214  &    0.73   &     2 &     0.31   &     0.96 \\
  2158-206  &   0.370   &     1 &     0.36   &     0.67 \\
  2158-170  &    1.56   &     2 &     0.37   &     1.12 \\
  2158-177  &    0.81   &     2 &     0.54   &     0.93 \\
  2159-187  &   0.334   &     1 &     0.44   &     1.33 \\
  2159-192  &    1.24   &     2 &     0.34   &     1.07 \\
  2159-201  &    0.75   &     4 &     0.60   &     1.66 \\
  2202-179  &   1.350   &     1 &     0.70   &     1.11 \\
  2204-182  &    2.04   &     2 &     0.59   &     0.83 \\
  2204-203  &   1.620   &     1 &     0.77   &     1.20 \\
  2205-178  &    2.04   &     2 &     0.34   &     1.21 \\
  2211-172  &   0.153   &     1 &     9.74   &     1.17 \\
  2213-167  &   0.074   &     1 &     0.35   &     0.85 \\
  2213-156  &    0.81   &     2 &     0.48   &     0.80 \\
  2215-179  &    0.49   &     2 &     0.54   &     1.26 \\
  2353-003  &   0.198   &     1 &     0.36   &     1.20 \\
  2354+008  &    0.73   &     3 &     0.31   &     1.04 \\
  2355-010  &    0.76   &     3 &     0.83   &     1.02 \\
  2356+033  &    0.57   &     2 &     0.38   &     0.90 \\
  2357+004  &   0.084   &     1 &     0.39   &     0.54 \\

\hline
\end{tabular}
\end{center}
\end{table}

\newpage

\subsection{The Local RLF}

The complete dataset used in this work for the local RLF comparison. N.B. The data presented here have been converted to the cosmology used here where necessary. $\rho$ is in units of Mpc$^{-3} (\Delta \log P_{\rm 1.4 GHz})^{-1}$

\begin{table}
\begin{center}
\begin{tabular}{ccccccccc}
\hline
\multicolumn{3}{c}{Sadler et al. 2002} &    \multicolumn{3}{c}{Best et al. {\it in prep}} & \multicolumn{3}{c}{Mauch et al. 2007} \\                
\hline                                                                                                                         
$\log P_{\rm 1.4 GHz}$ & $\log \rho$ & $+1\sigma$ & $\log P_{\rm 1.4 GHz}$ & $\log \rho$ & $+1\sigma$ & $\log P_{\rm 1.4 GHz}$ & $\log \rho$ & $+1\sigma$\\
\hline
     21.91 &      -3.68 &       0.13  &           22.15 &      -4.35 &       0.14 &   20.40 &      -3.23 &       0.23 \\ 
     22.31 &      -3.89 &       0.08  &           22.45 &      -4.39 &       0.05 &   20.80 &      -3.37 &       0.11 \\ 
     22.71 &      -4.14 &       0.06  &           22.75 &      -4.44 &       0.04 &   21.20 &      -3.61 &       0.09 \\ 
     23.11 &      -4.48 &       0.05  &           23.05 &      -4.47 &       0.02 &   21.60 &      -3.64 &       0.05 \\ 
     23.51 &      -4.65 &       0.04  &           23.35 &      -4.68 &       0.02 &   22.00 &      -3.78 &       0.04 \\ 
     23.91 &      -4.90 &       0.05  &           23.65 &      -4.80 &       0.01 &   22.40 &      -3.95 &       0.03 \\ 
     24.31 &      -5.03 &       0.05  &           23.95 &      -4.86 &       0.07 &   22.80 &      -4.22 &       0.02 \\ 
     24.71 &      -5.54 &       0.09  &           24.25 &      -5.11 &       0.01 &   23.20 &      -4.51 &       0.03 \\ 
     25.11 &      -5.73 &       0.10  &           24.55 &      -5.27 &       0.02 &   23.60 &      -4.69 &       0.03 \\ 
     25.51 &      -6.02 &       0.13  &           24.85 &      -5.55 &       0.02 &   24.00 &      -4.96 &       0.03 \\ 
     25.91 &      -7.17 &       0.30  &           25.15 &      -5.85 &       0.03 &   24.40 &      -5.17 &       0.04 \\ 
                                                       &&&      25.45 &      -6.33 &       0.05 &    24.80 &      -5.63 &       0.06 \\ 
                                                       &&&      25.75 &      -6.61 &       0.07 &    25.20 &      -5.93 &       0.08 \\ 
                                                       &&&      26.05 &      -7.20 &       0.12 &    25.60 &      -6.34 &       0.16 \\ 
                                                       &&&      26.35 &      -7.79 &       0.21 &    26.00 &      -6.90 &       0.25 \\ 
                                                       &&&                 &                 &                &    26.40 &      -7.72 &       0.27 \\ 

\hline
\end{tabular}
\end{center}
\end{table}


\subsection{The source counts}

The complete dataset used for the source count comparison here.  Count values are expressed as $\frac{\rm dN}{\rm dS} S^{2.5}$sr$^{-1}$Jy$^{1.5}$

\begin{table}
\begin{center}
\begin{tabular}{ccccccccccccccc}
\hline
\multicolumn{3}{c}{Bondi et al. 2008} & \multicolumn{3}{c}{Seymour et al. 2004} &\multicolumn{3}{c}{Windhorst et al. 1984} &\multicolumn{3}{c}{White et al. 1997} &\multicolumn{3}{c}{Kellerman \& Wall 1987} \\
\hline

$S_{\rm 1.4 GHz}$ & Cnts & $\pm$ & $S_{\rm 1.4 GHz}$& Cnts & $\pm$ & $S_{\rm 1.4 GHz}$ & Cnts & $\pm$ & $S_{\rm 1.4 GHz}$ ) & Cnts & $\pm$ & $S_{\rm 1.4 GHz}$  & Cnts & $\pm$ \\
(mJy) & & & (mJy) & & & (mJy) & & & (mJy) & & & (mJy) & & \\
\hline

      0.07 &       3.32 &       0.17 &      0.047 &       3.26 &       0.34 &       1.10 &       9.59 &       2.61 &       2.00 &      10.04 &       0.12 &     51.60 &     142.56 &      14.04 \\ 
      0.08 &       3.84 &       0.21 &      0.071 &       3.74 &       0.33 &       1.85 &      16.05 &       3.09 &       2.24 &      12.27 &       0.15 &     65.70 &     168.30 &      22.75 \\ 
      0.10 &       3.26 &       0.23 &      0.107 &       4.15 &       0.43 &       2.62 &      16.97 &       3.34 &       2.51 &      12.68 &       0.17 &     98.00 &     202.84 &      30.82 \\ 
      0.12 &       3.09 &       0.26 &      0.162 &       4.02 &       0.58 &       3.70 &      20.65 &       3.75 &       2.82 &      14.16 &       0.19 &    140.50 &     282.67 &      43.22 \\ 
      0.15 &       3.99 &       0.34 &      0.245 &       3.90 &       0.78 &       5.23 &      32.44 &       6.33 &       3.16 &      15.41 &       0.22 &    218.30 &     282.67 &      27.85 \\ 
      0.18 &       4.36 &       0.42 &      0.371 &       4.65 &       1.16 &       7.40 &      35.42 &       6.55 &       3.55 &      17.10 &       0.25 &    332.50 &     320.13 &      31.55 \\ 
      0.22 &       4.02 &       0.47 &      0.861 &       5.18 &       1.44 &      12.40 &      47.26 &       7.36 &       3.98 &      18.99 &       0.29 &    656.70 &     347.83 &      20.97 \\ 
      0.28 &       5.30 &       0.63 &            &            &            &      24.80 &      77.39 &      15.01 &       4.47 &      20.54 &       0.33 &    904.70 &     340.70 &      27.14 \\ 
      0.34 &       4.49 &       0.67 &            &            &            &      49.60 &     127.66 &      31.78 &       5.01 &      22.89 &       0.37 &   1522.70 &     300.83 &      29.63 \\ 
      0.41 &       4.47 &       0.78 &            &            &            &     116.80 &     234.05 &      67.60 &       5.62 &      25.26 &       0.43 &   2318.60 &     313.56 &      30.89 \\ 
      0.50 &       5.14 &       0.97 &            &            &            &            &            &            &       6.31 &      27.28 &       0.49 &   2948.30 &     249.59 &      15.05 \\ 
      0.62 &       5.97 &       1.28 &            &            &            &            &            &            &       7.08 &      31.38 &       0.57 &   3674.70 &     207.09 &      27.99 \\ 
      0.76 &       4.05 &       1.17 &            &            &            &            &            &            &       7.94 &      34.54 &       0.65 &   5269.10 &     202.84 &      34.54 \\ 
      0.93 &       5.49 &       1.58 &            &            &            &            &            &            &       8.91 &      37.24 &       0.74 &   8023.20 &     136.76 &      18.47 \\ 
           &            &            &            &            &            &            &            &            &      10.00  &      42.04 &       0.85 &  23185.50 &     145.55 &      22.27 \\ 
           &            &            &            &            &            &            &            &            &      11.22 &      46.01 &       0.97 & 94170.00 &     175.50 &      90.00     \\ 
           &            &            &            &            &            &            &            &            &      12.59 &      50.67 &       1.11 &           &            &            \\ 
           &            &            &            &            &            &            &            &            &      14.13 &      54.59 &       1.26 &           &            &            \\ 
           &            &            &            &            &            &            &            &            &      15.85 &      59.78 &       1.43 &           &            &            \\ 
           &            &            &            &            &            &            &            &            &      17.78 &      66.18 &       1.65 &           &            &            \\ 
           &            &            &            &            &            &            &            &            &      19.95 &      77.71 &       1.94 &           &            &            \\ 
           &            &            &            &            &            &            &            &            &      22.39 &      85.57 &       2.22 &           &            &            \\ 
           &            &            &            &            &            &            &            &            &      25.12 &      80.12 &       2.35 &           &            &            \\ 
           &            &            &            &            &            &            &            &            &      28.18 &     100.88 &       2.87 &           &            &            \\ 
           &            &            &            &            &            &            &            &            &      31.62 &     114.91 &       3.34 &           &            &            \\ 
           &            &            &            &            &            &            &            &            &      35.48 &     102.91 &       3.45 &           &            &            \\ 
           &            &            &            &            &            &            &            &            &      39.81 &     134.73 &       4.30 &           &            &            \\ 
           &            &            &            &            &            &            &            &            &      44.67 &     119.97 &       4.42 &           &            &            \\ 
           &            &            &            &            &            &            &            &            &      50.12 &     138.13 &       5.17 &           &            &            \\ 
           &            &            &            &            &            &            &            &            &      56.23 &     160.17 &       6.07 &           &            &            \\ 
           &            &            &            &            &            &            &            &            &      63.10 &     179.24 &       7.00 &           &            &            \\ 
           &            &            &            &            &            &            &            &            &      70.79 &     172.49 &       7.49 &           &            &            \\ 
           &            &            &            &            &            &            &            &            &      79.43 &     208.60 &       8.98 &           &            &            \\ 
           &            &            &            &            &            &            &            &            &      89.13 &     230.73 &      10.29 &           &            &            \\ 
           &            &            &            &            &            &            &            &            &     100.00 &     223.84 &      11.05 &           &            &            \\ 
           &            &            &            &            &            &            &            &            &     112.20 &     231.56 &      12.26 &           &            &            \\ 
           &            &            &            &            &            &            &            &            &     125.89 &     258.50 &      14.12 &           &            &            \\ 
           &            &            &            &            &            &            &            &            &     141.25 &     246.72 &      15.03 &           &            &            \\ 
           &            &            &            &            &            &            &            &            &     158.49 &     328.63 &      18.92 &           &            &            \\ 
           &            &            &            &            &            &            &            &            &     177.83 &     260.16 &      18.35 &           &            &            \\ 
           &            &            &            &            &            &            &            &            &     199.53 &     308.06 &      21.77 &           &            &            \\ 
           &            &            &            &            &            &            &            &            &     223.87 &     321.10 &      24.23 &           &            &            \\ 
           &            &            &            &            &            &            &            &            &     251.19 &     371.50 &      28.41 &           &            &            \\ 
           &            &            &            &            &            &            &            &            &     281.84 &     375.95 &      31.16 &           &            &            \\ 
           &            &            &            &            &            &            &            &            &     316.23 &     305.91 &      30.64 &           &            &            \\ 
           &            &            &            &            &            &            &            &            &     354.81 &     296.63 &      32.89 &           &            &            \\ 
           &            &            &            &            &            &            &            &            &     398.11 &     340.95 &      38.44 &           &            &            \\ 
           &            &            &            &            &            &            &            &            &     446.68 &     351.93 &      42.58 &           &            &            \\ 
           &            &            &            &            &            &            &            &            &     501.19 &     364.21 &      47.22 &           &            &            \\ 
           &            &            &            &            &            &            &            &            &     562.34 &     328.27 &      48.88 &           &            &            \\ 
           &            &            &            &            &            &            &            &            &     630.96 &     632.16 &      73.94 &           &            &            \\ 
           &            &            &            &            &            &            &            &            &     707.95 &     482.54 &      70.43 &           &            &            \\ 
           &            &            &            &            &            &            &            &            &     794.33 &     273.39 &      57.79 &           &            &            \\ 
           &            &            &            &            &            &            &            &            &     891.25 &     490.07 &      84.35 &           &            &            \\ 
           &            &            &            &            &            &            &            &            &    1000.00&     288.70 &      74.38 &           &            &            \\

\hline
\end{tabular}
\end{center}
\end{table}

\clearpage

\section{Predicted $P$--$z$ Grid}
\label{fullgrid}

\begin{table}
\caption{The space densities, $\rho$ (Mpc$^{-3}(\Delta \log P_{\rm 1.4GHz})^{-1}$), given by the best--fitting steep--spectrum $P$--$z$ grid. The most reliable points (defined as those which have values two times higher than their fitted error and are constrained by at least two of the input datasets) are highlighted in bold.}
\begin{center}
\begin{tabular}{ccccccccc}
\hline
$\log P$ & \multicolumn{8}{c}{$z$} \\
      &     0.10  &    0.25  &    0.50  &    1.00  &    2.00 &     3.00 &     4.00 &     6.00 \\
\hline
19.25 &     &      &      &      &      &      &      &      \\ 
19.75 &     &      &      &      &      &      &      &      \\ 
20.25 &     &      &      &      &      &      &      &      \\ 
20.75 &     &      &      &      &      &      &      &      \\ 
21.25 &{\bf 1.38e-4}  &      &      &      &      &      &      &      \\ 
21.75 &{\bf 1.33e-4} &      &      &      &      &      &      &      \\ 
22.25 &{\bf 5.63e-5} &      4.54e-4 &      &      &      &      &      &      \\ 
22.75 &{\bf 4.47e-5} & {\bf 1.22e-4} &      2.59e-4  &      &      &      &      &      \\ 
23.25 &{\bf 3.00e-5} & {\bf 1.49e-5} & {\bf 7.60e-5} &      1.85e-4  &      &      &      &      \\ 
23.75 &{\bf 1.66e-5} & {\bf 2.13e-5} & {\bf 1.49e-5} & {\bf 3.20e-5} &      &      &      &      \\ 
24.25 &{\bf 8.60e-6} & {\bf 9.64e-6} & {\bf 1.27e-5} & {\bf 1.33e-5} & 3.82e-10 &     4.39e-4     &      &      \\ 
24.75 &{\bf 3.50e-6} & {\bf 5.38e-6} & {\bf 5.06e-6} & {\bf 6.76e-6} & 1.51e-10 & 3.40e-9 & 6.86e-5 &  4.11e-4    \\ 
25.25 &{\bf 1.09e-6} & {\bf 2.18e-6} & {\bf 3.74e-6} & {\bf 3.05e-6} &      4.21e-6  & 6.05e-11 & 2.68e-5  & 9.36e-10 \\ 
25.75 &{\bf 1.47e-7} & {\bf 5.49e-7} & {\bf 1.06e-6} & {\bf 1.53e-6} & {\bf 3.09e-7} & 2.20e-12 & 1.17e-11 & 2.04e-5  \\ 
26.25 &{\bf 2.61e-8} & {\bf 8.43e-8} & {\bf 2.73e-7} & {\bf 9.51e-7} & {\bf 8.37e-7} & 6.42e-13 & 4.79e-13 & 6.84e-12 \\ 
26.75 &1.68e-14 & {\bf 1.52e-8} & {\bf 4.84e-8} & {\bf 1.40e-7} & {\bf 4.97e-8} & {\bf 9.57e-8} & {\bf 1.77e-7} & 3.15e-12 \\ 
27.25 &2.82e-14 & {\bf 1.81e-9} & {\bf 6.79e-9} & {\bf 3.74e-8} & {\bf 8.66e-8} & {\bf 9.55e-8} & {\bf 8.00e-8} & 6.15e-15 \\ 
27.75 &2.67e-10 & 2.19e-17 & {\bf 1.17e-9} & {\bf 3.55e-9} & {\bf 2.71e-8} & {\bf 1.17e-7} & 8.81e-14 & 1.42e-15 \\ 
28.25 &5.79e-13 & 4.19e-16 & 7.30e-16 & {\bf 4.93e-10} & {\bf 1.06e-9} & {\bf 4.13e-9} & 2.02e-14 & 3.04e-14 \\ 
28.75 &3.08e-13 & 5.10e-17 & 1.78e-16 & {\bf 7.69e-11} & {\bf 1.27e-10} & 8.51e-18 & 3.40e-16 & 3.19e-9 \\ 
29.25 &1.16e-12 & 2.32e-3  & 2.58e-11 & 1.61e-17 & 3.01e-18 & 4.48e-19 & {\bf 2.66e-12} & 7.69e-20 \\

\hline
\end{tabular}

\caption{The corresponding errors in $\rho$  for the best--fitting steep--spectrum grid above.}

\begin{tabular}{ccccccccc}
\hline
$\log P$ & \multicolumn{8}{c}{$z$} \\
      &     0.10  &    0.25  &    0.50  &    1.00  &    2.00 &     3.00 &     4.00 &     6.00 \\
\hline
19.25 &     &      &      &      &      &      &      &      \\ 
19.75 &     &      &      &      &      &      &      &      \\ 
20.25 &     &      &      &      &      &      &      &      \\ 
20.75 &     &      &      &      &      &      &      &      \\ 
21.25 &9.84e-5 &      &      &      &      &      &      &      \\ 
21.75 &3.29e-5 &      &      &      &      &      &      &      \\ 
22.25 &1.07e-5 & 1.87e-5 &      &      &      &      &      &      \\ 
22.75 &5.55e-6 & 7.05e-6 & 1.11e-5 &      &      &      &      &      \\ 
23.25 &2.65e-6 & 4.28e-6 & 2.06e-6 & 3.62e-6 &      &      &      &      \\ 
23.75 &1.18e-6 & 2.53e-6 & 1.61e-6 & 4.31e-7 &      &      &      &      \\ 
24.25 &5.59e-7 & 1.12e-6 & 3.65e-7 & 5.86e-7 & 1.79e-6 & 1.51e-4  &      &      \\ 
24.75 &1.72e-7 & 5.69e-7 & 2.75e-7 & 3.65e-7 & 1.80e-6 & 6.20e-6 & 4.23e-5 &  2.35e-3    \\ 
25.25 &2.68e-8 & 1.56e-7 & 1.58e-7 & 8.26e-8 & 6.08e-7 & 4.51e-7 & 3.49e-6 & 1.63e-5 \\ 
25.75 &1.71e-8 & 2.67e-8 & 7.61e-8 & 4.11e-8 & 1.05e-8 & 2.74e-7 & 5.15e-7 & 2.75e-6 \\ 
26.25 &1.42e-9 & 2.81e-9 & 2.97e-9 & 2.18e-8 & 4.70e-8 & 4.52e-8 & 3.05e-7 & 7.03e-7 \\ 
26.75 &3.53e-10 & 3.95e-10 & 8.46e-10 & 4.29e-9 & 2.06e-8 & 3.65e-8 & 4.98e-8 & 3.44e-7 \\ 
27.25 &4.11e-10 & 6.99e-11 & 8.77e-11 & 7.64e-10 & 5.88e-9 & 2.31e-8 & 6.10e-9 & 4.47e-8 \\ 
27.75 &1.31e-9 & 1.60e-11 & 1.96e-11 & 3.80e-11 & 8.83e-10 & 1.00e-8 & 6.06e-9 & 1.43e-8 \\ 
28.25 &2.55e-9 & 2.81e-11 & 7.77e-12 & 8.91e-12 & 9.99e-11 & 8.68e-10 & 3.15e-9 & 8.39e-9 \\ 
28.75 &2.55e-9 & 3.63e-10 & 8.36e-12 & 2.39e-12 & 7.01e-12 & 1.71e-10 & 2.31e-10 & 2.73e-9 \\ 
29.25 &2.55e-9 & 2.32e-1 & 4.25e-11 & 1.03e-12 & 2.18e-13 & 4.79e-13 & 9.29e-13 & 3.00e-13 \\

\hline
\end{tabular}
\end{center}
\end{table}

\begin{table}
\begin{center}
\caption{The expansion co--efficients for Equation \protect\ref{fit_eqn} from the four different fits to
  the $P$--$z$ grid}
\begin{tabular}{ccrrrr}
\hline
\multicolumn{2}{c}{Order of series term} & \multicolumn{4}{c}{$A_{i,j}$} \\
\hline
x & y & Fit1 & Fit2 & Fit3 & Fit4 \\
\hline
      0 &       0 &        11458.51 &        9291.00 &        29.80 &        4.77 \\ 
      0 &       1 &        2295.69 &        8437.41 &        23.02 &        57.26 \\ 
      0 &       2 &       -24.97 &        1130.46 &       -9.26 &        122.73 \\ 
      0 &       3 &       -41.51 &       -1118.96 &       -10.11 &       -150.82 \\ 
      0 &       4 &       -1.23 &        127.83 &      -0.90 &       -44.34 \\ 
      1 &       0 &       -1747.07 &       -1511.30 &       -235.35 &       -87.84 \\ 
      1 &       1 &       -268.81 &       -1001.76 &       -144.61 &       -401.59 \\ 
      1 &       2 &      -0.38 &       -63.64 &       0.59 &       -311.17 \\ 
      1 &       3 &        1.45 &        40.56 &        10.56 &        332.54 \\ 
      2 &       0 &        99.61 &        91.86 &        590.30 &        312.23 \\ 
      2 &       1 &        10.43 &        39.37 &        268.26 &        843.97 \\ 
      2 &       2 &      0.047 &       0.77 &        15.36 &        88.06 \\ 
      3 &       0 &       -2.52 &       -2.47 &       -644.97 &       -481.38 \\ 
      3 &       1 &      -0.13 &      -0.51 &       -150.05 &       -500.62 \\ 
      4 &       0 &      0.024 &      0.025 &        249.12 &        246.39 \\ 

\hline
& & \multicolumn{4}{c}{B} \\
\hline
5 & 5 &    1.44e-08 &   -7.81e-06 &        3.74 &       -80.72 \\ 
\hline
\end{tabular}
\end{center}
\end{table}

\begin{figure*}
\centering
\includegraphics[scale=0.2, angle=90]{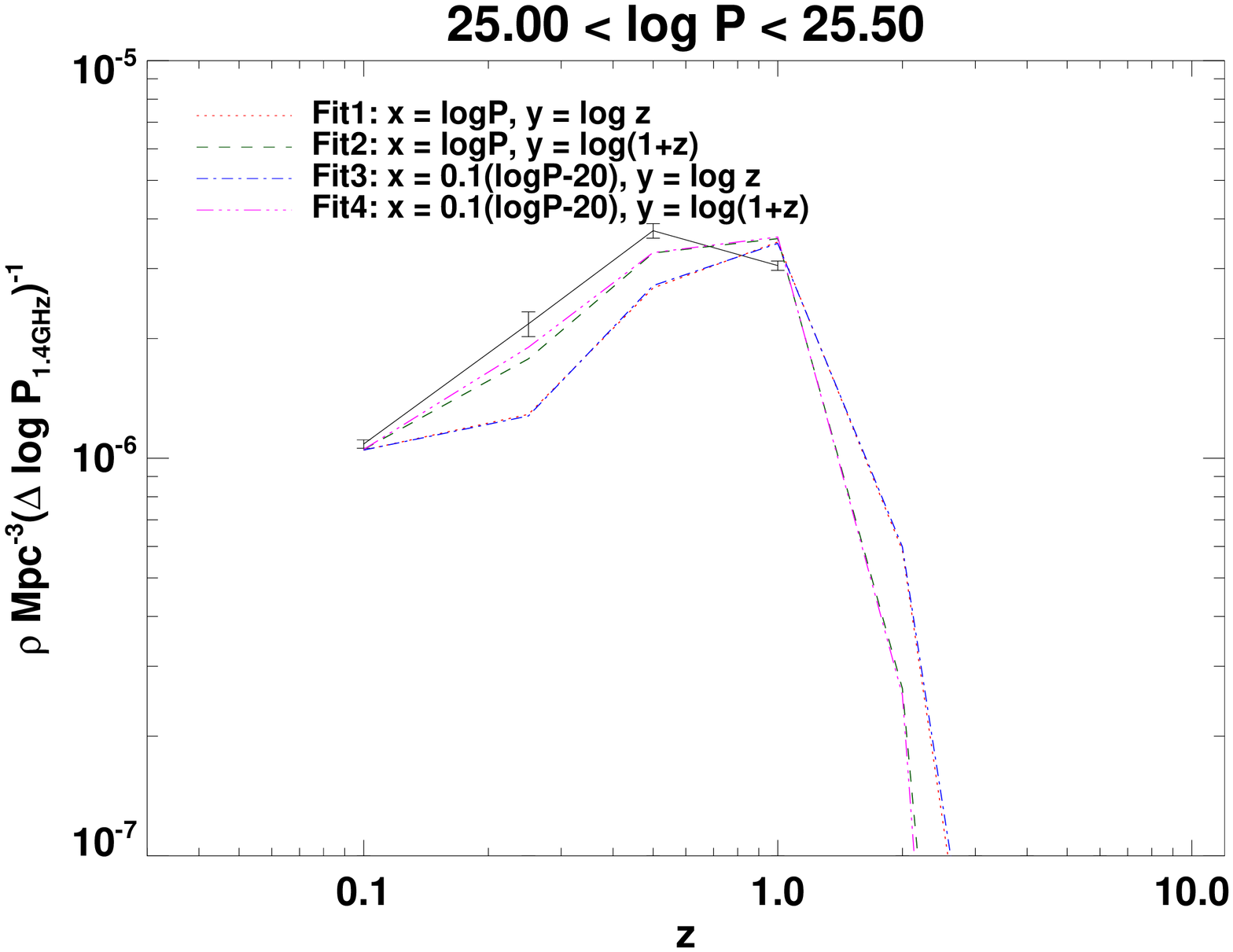}
\includegraphics[scale=0.2, angle=90]{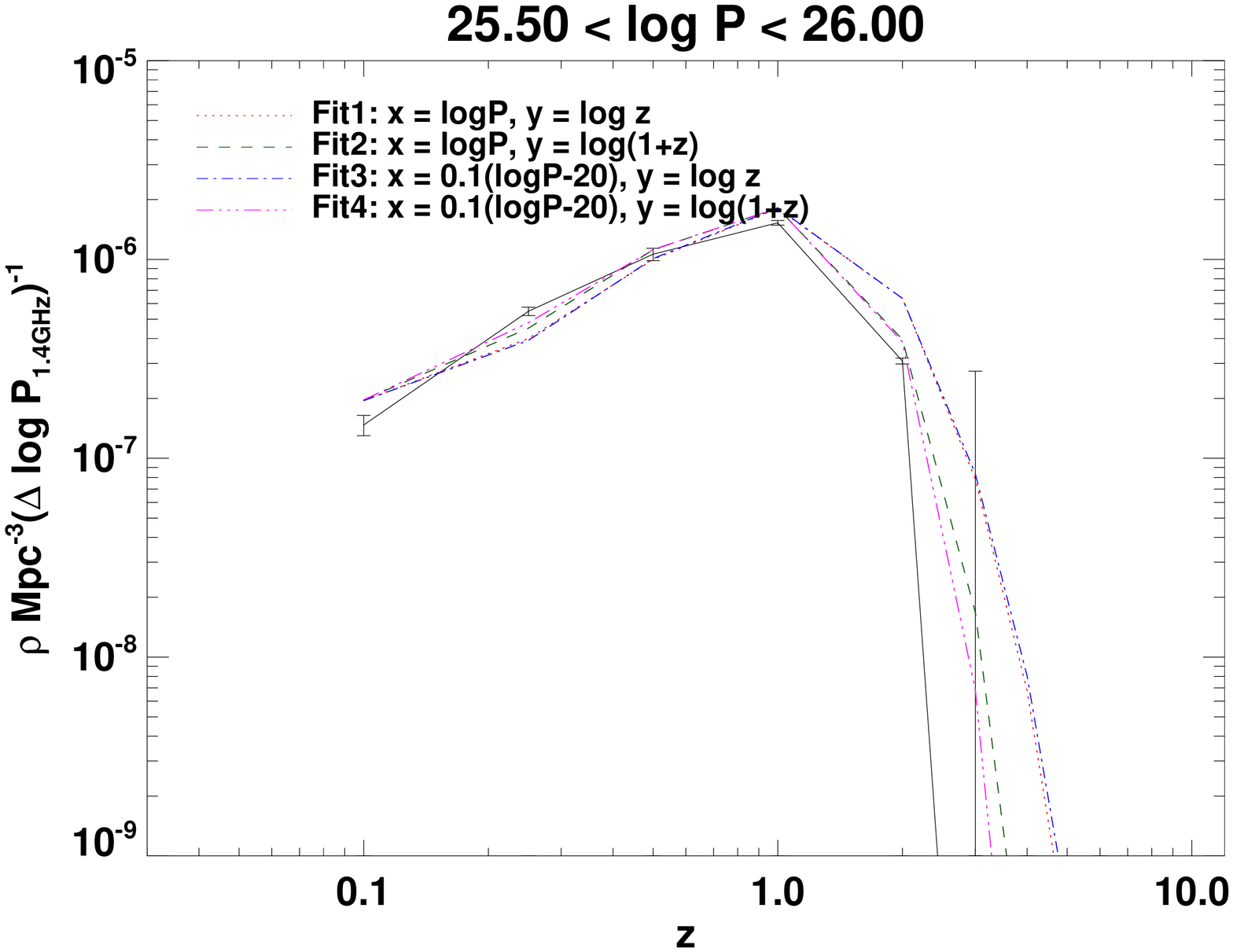}
\includegraphics[scale=0.2, angle=90]{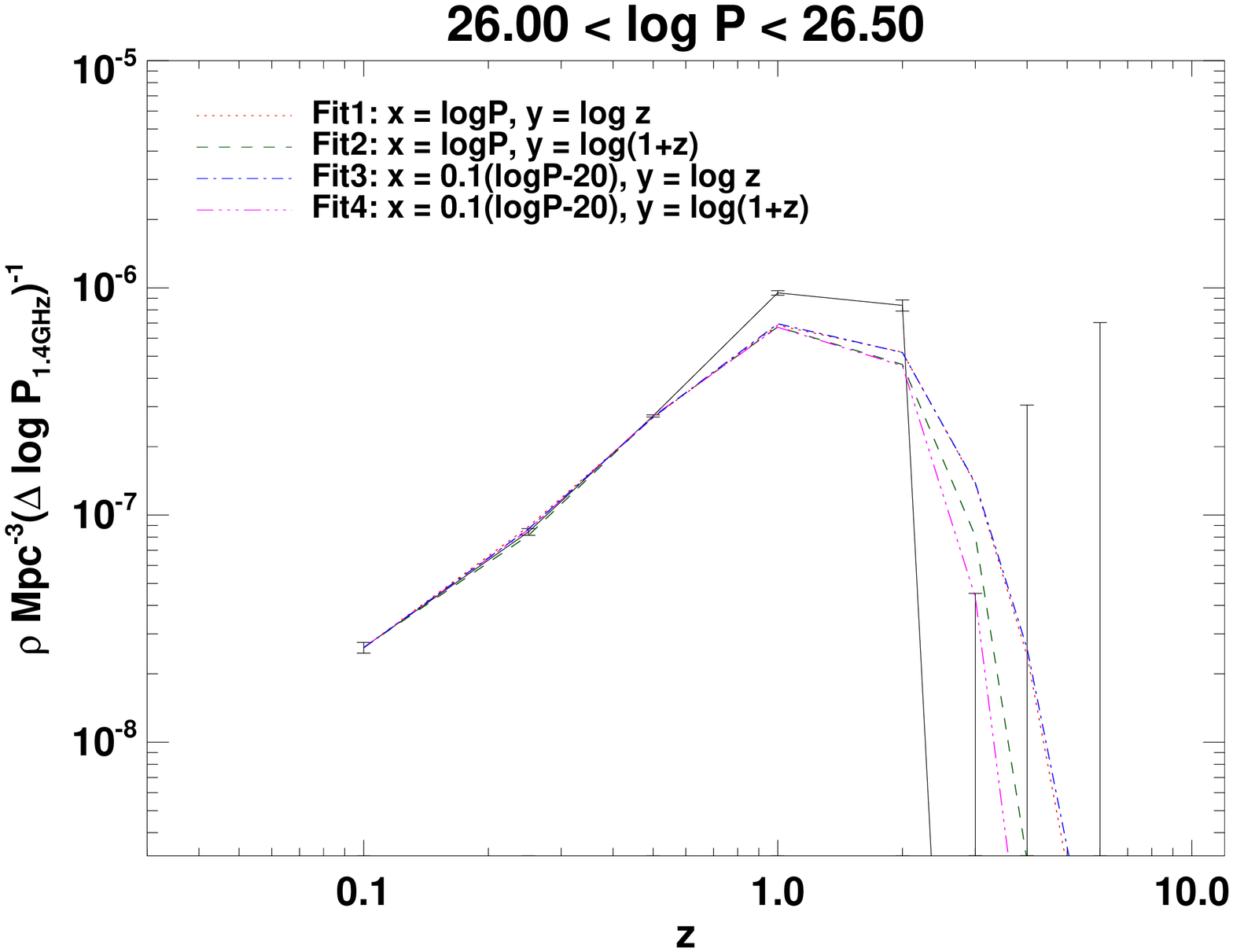}
\includegraphics[scale=0.2, angle=90]{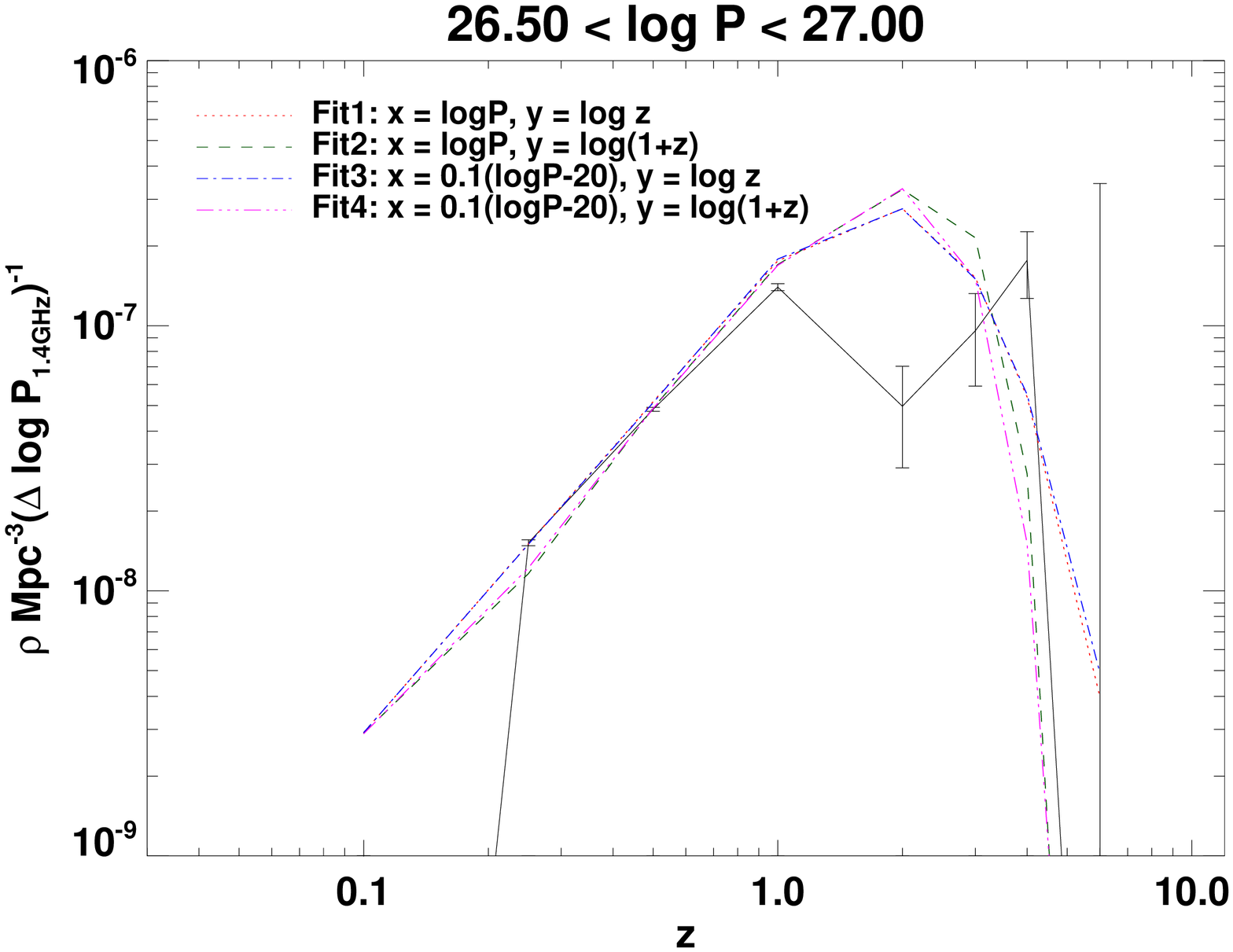}
\includegraphics[scale=0.2, angle=90]{fit6_2725.ps}
\includegraphics[scale=0.2, angle=90]{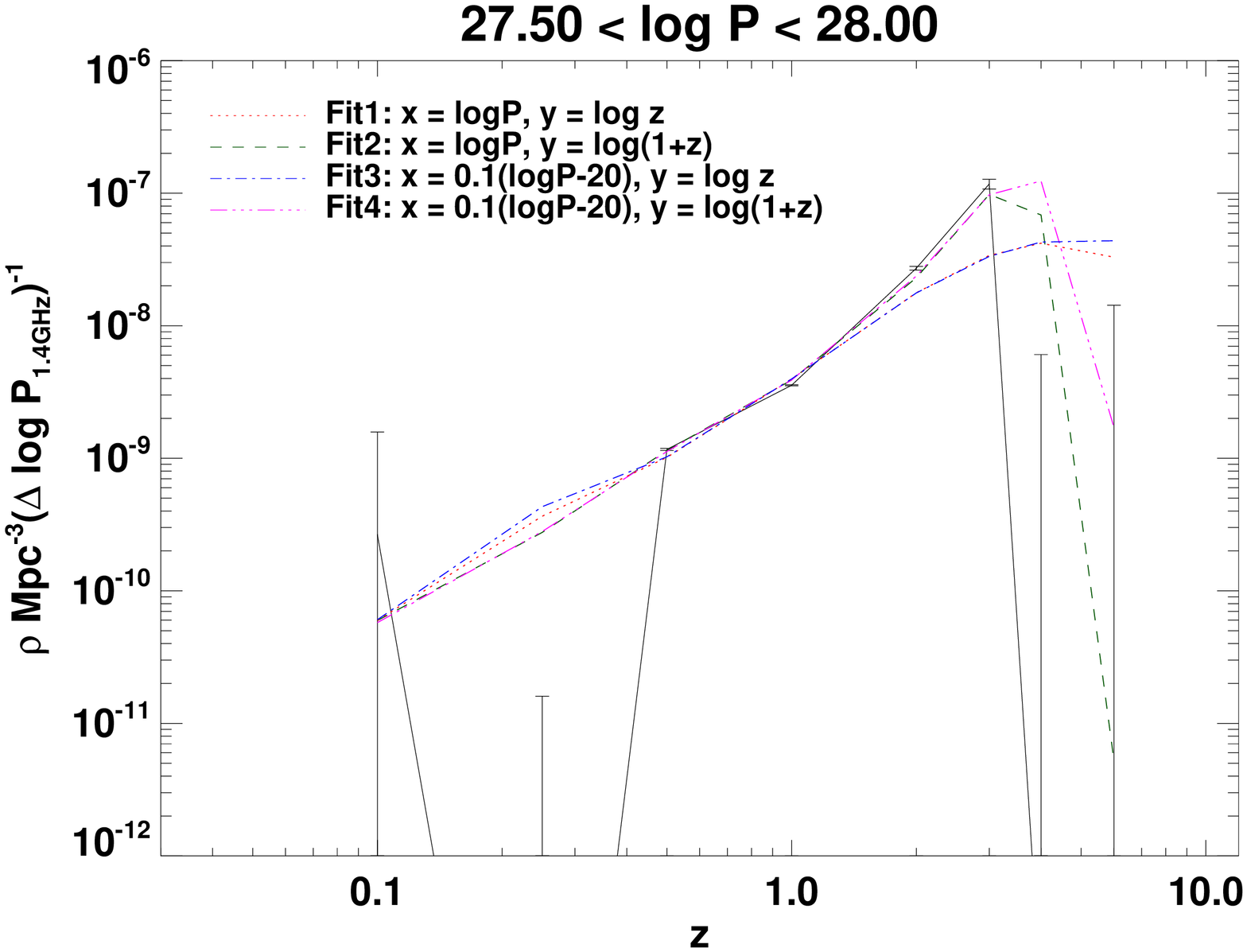}
\caption{\protect\label{sfit_full} The results of the four smooth
  fits to the $P$--$z$ grid. }
\end{figure*}

\bsp

\label{lastpage}

\end{document}